\documentclass[11pt,a4paper,reqno]{amsart}

\usepackage{graphicx}%
\usepackage{multirow}%
\usepackage{amsmath,amssymb,amsfonts,mathtools,bm}%
\usepackage{amsthm}%
\usepackage{mathrsfs}%
\usepackage[title]{appendix}%
\usepackage{xcolor}%
\usepackage{textcomp}%
\usepackage{setspace}
\usepackage{manyfoot}%
\usepackage{booktabs}%
\usepackage{algorithm}%
\usepackage{algorithmicx}%
\usepackage{algpseudocode}%
\usepackage{listings}%
\usepackage{booktabs}
\usepackage{pifont}
\usepackage{subcaption}%
\usepackage{makecell}
  \usepackage[top=1.5in, bottom=1.2in,
  left=1in, right=1in]{geometry}
\usepackage{hyperref}
\hypersetup{colorlinks = true, urlcolor = green, linkcolor = red,
  citecolor = blue }
  \usepackage[square,sort,comma,numbers]{natbib}
\newcommand{\cmark}{\ding{51}} 
\newcommand{\xmark}{\ding{55}} 
\newcommand{\Lev}{\ensuremath{\textnormal{lev}}}
\newcommand{\RV}{{\tt RV}}
\newcommand{\Dphi}{\mathcal B_{\phi,\delta}}

\newcommand{\MDA}{{\tt MDA}}

\newcommand{\R}{{\mathbb{R}}}

\newcommand{\mv}[1]{\boldsymbol{#1}}

\newcommand{\xx}{\boldsymbol{x}}
\newcommand{\XX}{\boldsymbol{\xi}}

\newcommand{\zz}{\boldsymbol{z}}

\newtheorem{assumption}{Assumption}
\newtheorem{illustration}{Numerical Illustration}
\newtheorem{theorem}{Theorem}[section]
\newtheorem{definition}{Definition}
\newtheorem{example}{Example}
\newtheorem{lemma}{Lemma}[section]
\newtheorem{proposition}{Proposition}[section]
\newtheorem{corollary}{Corollary}[section]
\newtheorem{remark}{Remark}
\renewcommand{\arraystretch}{1.5}

\theoremstyle{thmstyleone}%

\theoremstyle{thmstyletwo}%

\theoremstyle{thmstylethree}%

\title[Distributional Robustness for Tail Risk Functionals]{EVT-Based Rate-Preserving Distributional Robustness for Tail Risk Functionals}
\author{{\large A\MakeLowercase{nand}
    D\MakeLowercase{eo}}}
  \address{Indian Institute of Management, Bannerghatta Road, Bangalore 560076} \email{anand.deo@iimb.ac.in}
\begin{document}


\begin{abstract}
Risk measures such as Conditional Value-at-Risk (CVaR) focus on extreme losses, where scarce tail data makes model error unavoidable. To hedge misspecification, one evaluates worst-case tail risk over an ambiguity set. Using Extreme Value Theory (EVT), we derive first-order asymptotics for worst-case tail risk for a broad class of tail-risk measures under standard ambiguity sets, including Wasserstein balls and $\phi$-divergence neighborhoods. We show that robustification can alter the nominal tail asymptotic scaling as the tail level $\beta\to0$, leading to excess risk inflation. Motivated by this diagnostic, we propose a tail-calibrated ambiguity design that preserves the nominal tail asymptotic scaling while still guarding against misspecification. Under standard domain of attraction assumptions, we prove that the resulting worst-case risk preserves the baseline first-order scaling as $\beta\to0$, uniformly over key tuning parameters, and that a plug-in implementation based on consistent tail-index estimation inherits these guarantees. Synthetic and real-data experiments show that the proposed design avoids the severe inflation often induced by standard ambiguity sets.

\end{abstract}

\maketitle
\vspace{-20pt}
\section{Introduction}\label{sec1}
Distributional uncertainty is unavoidable in risk management whenever one wishes to evaluate a risk functional under an unknown data generating law. 
Distributionally Robust Optimization (DRO) addresses this by replacing evaluation at a single model with a worst-case evaluation over an \emph{ambiguity set}, i.e., a collection of distributions deemed plausible perturbations of a nominal law (see for instance \cite{delage2010distributionally,zymler2013distributionally,ben2013robust,lam2019recovering,duchi2021statistics,blanchet2019quantifying,gao2023distributionally}). 

In many applications, the functional emphasizes tail risk; for instance, in an insurance setting, one may wish to quantify the impact of excessively large claims.
Tail samples are scarce, so direct estimation is challenging.  A common fallback is to fit a parametric family (e.g., Gaussian) and evaluate the functional under the fitted model, risking systematic tail underestimation if the family is misspecified. DRO is a natural way to mitigate such misspecification, but for extreme tail functionals, if the ambiguity set is too permissive, worst-case evaluations become overly conservative \cite{kruse2021toolkit,jin2024constructing}. 
The challenge, therefore, is to balance robustness against unnecessary risk inflation.

This paper studies distributional robustness for tail risk functionals (e.g., CVaR at level \(\beta\)). Our contributions are as follows: \emph{First}, as a diagnostic, we characterize worst-case tail scaling as \(\beta\downarrow0\) for standard ambiguity families, revealing sources of undue conservativeness. \emph{Second}, motivated by these findings, our main contribution is the development of a data-driven, EVT-calibrated DRO (rate-preserving DRO or RPEV-DRO) whose worst-case value is rate-preserving relative to the truth: as \(\beta\to0\), the DRO worst-case tail risk and the true tail risk have the same asymptotic growth rate under standard domain of attraction assumptions, uniformly over choices of the ambiguity radius $\delta$ and a baseline intermediate EVT threshold level $\beta_0 \gg \beta$.
\emph{Finally}, we show that these results extend to multivariate losses.\\

\noindent \textbf{Measures of tail risk:}
We study tail risk functionals $\rho_{1-\beta}(P)$ that depend primarily on the $\beta$-tail of the loss distribution.
A central example is CVaR (expected shortfall), which averages losses beyond a high quantile and is coherent and law-invariant \cite{artzner1999coherent,acerbi2002spectral}.
More generally, to model heterogeneous aversion within the tail, one can also consider tail-weighted (spectral) generalizations of CVaR; see \eqref{eqn:drm_tails}. Appropriate choices of the weight recover a large class of coherent law-invariant risk measures \cite{kusuoka2001law,acerbi2002spectral}. Such tail risk measures are widely used in applications where safety and reliability are paramount (see \cite{Krokhmal,acerbi2002coherence,Ban,Tamar,bienstock2014chance,summers2015stochastic}).

Despite these modeling advantages, estimating tail dependent risks is statistically challenging due to scarce tail samples, especially under heavy tails \cite{Lim,Caccioli2018}.
Distributionally robust formulations are therefore a natural recourse.
Distributionally Robust CVaR has been studied for portfolio evaluation \cite{zhu2009worst,gotoh2013robust}, regret minimization \cite{natarajan2014probabilistic}, and CVaR-constrained programs with tractable reformulations \cite{zymler2013distributionally,chen2023approximations}.
More broadly, tractable evaluation of robust tail-risk measures has been developed in \cite{pesenti2024optimizing,bernard2023robust,cai2023distributionally,wu2025generalization}.

\subsection{Main Contributions} 
While previous works provide reformulations to ensure computational tractability, the question of how the choice of ambiguity set affects the worst-case risk evaluation remains largely unexplored. For instance, a regulator using DRO to hedge model risk needs to understand the capital overhead induced purely by robustification. Given an ambiguity set $\mathcal P$ around a baseline $P_0$, they may ask the following: 

\begin{itemize}
\item[(Q1)] \textbf{Asymptotic inflation (Diagnostic).} How does the worst-case tail risk evaluation
\[
\rho_{1-\beta}(\mathcal P):=\sup_{P\in\mathcal P}\rho_{1-\beta}(P)
\]
scale relative to $\rho_{1-\beta}(P_0)$ as $\beta\downarrow 0$? In other words, how much conservativeness is caused by the choice of ambiguity set itself?

\item[(Q2)] \textbf{Design for parsimony (Constructive).} Can one carefully tailor $\mathcal P$ using information from an intermediate tail level so that the worst-case value remains robust \emph{and} preserves the nominal tail rate (i.e., avoids excess risk inflation)?
\end{itemize}
It is also natural to ask whether such designs  admit:
(i) data-driven implementations at  practically relevant rare-event levels, and
(ii) extensions to multivariate losses. 
Our main contributions below address these questions.

\noindent \textbf{i) Characterization of Worst Case Tail Risk: } We show the existence of a scaling function $g_\beta$, depending on (i) the tail behavior of $P_0$ and (ii) the structure of $\mathcal P$, such that, as $\beta\downarrow 0$,
\begin{equation}\label{eqn:Robust_CVaR_scaling}
    \rho_{1-\beta}(\mathcal P) \;\sim\; g_\beta
\end{equation}
where $a_\beta\sim b_\beta$ denotes $a_\beta/b_\beta\to 1$. 
Using \eqref{eqn:Robust_CVaR_scaling}, we characterize conservativeness relative to the baseline $\rho_{1-\beta}(P_0)$ for two popular ambiguity set families.

\noindent\textbf{a) Wasserstein ambiguity.} 
If $\mathcal P$ is a $p$-Wasserstein ball around $P_0$, there exists a constant $c\in(1,\infty]$ such that
\[
\frac{\log g_\beta}{\log \rho_{1-\beta}(P_0)} \;\to\; c \quad \text{as }\beta\downarrow 0.
\]
When $P_0$ has a power-law tail, one obtains $c<\infty$. 
If instead the tails of $P_0$ decay faster than a power law, we show that $c=\infty$.
Thus, while computationally attractive, Wasserstein ambiguity sets can be highly conservative, with the resulting risk inflation worsening as the tail of the nominal distribution  becomes lighter.

\noindent\textbf{b) $\boldsymbol{\phi}$-divergence with polynomial growth.} 
Suppose $\mathcal P$ is a ball defined by a $\phi$-divergence whose growth at infinity is polynomial (e.g., $\chi^2$ and Cressie-Reed divergences).
To showcase the conservativeness of divergence-based ambiguity sets, we demonstrate the following structural lower bound: if $P_0$ is heavy-tailed the worst-case tail risk is necessarily asymptotically more conservative than $\rho_{1-\beta}(P_0)$, with
\[
\liminf_{\beta\to 0} \frac{\log g_\beta}{\log \rho_{1-\beta}(P_0)} \geq c,
\]
where $c \in (1,\infty)$ depends \emph{only} on the divergence parameter.
Hence, under heavy tails, polynomial divergences are conservative, but with a finite "conservativeness exponent".
If $P_0$ has tails that decay faster than any power-law (Weibull, Gaussian, Exponential, etc.), we show that
\[
\liminf_{\beta\to 0}\frac{g_\beta}{\rho_{1-\beta}(P_0)} \geq K, 
\]
with $K \in (1,\infty)$ depending  on 
\emph{both} 
the divergence parameter and the tail strength of $P_0$. Therefore, under light tails, one still incurs an unavoidable multiplicative risk inflation.


\noindent \textbf{ii) Design of tail-aligned, rate-preserving DRO: }The conservativeness of Wasserstein or $\phi$-divergence-based DRO arises because the ambiguity set contains distributions with tails heavier than that of $P_0$.
To address this, we impose two design principles (made precise in Proposition~\ref{prop:rate_preserving_suffcient}): (i) construct \emph{tail-aligned} ambiguity sets that exclude models with tails heavier than the baseline (made precise below), and (ii) choose a nominal model whose tail matches the data-generating law $Q$. The first controls conservativeness; the second aligns the nominal tail with the observed extremes. To enforce these, we choose

\begin{equation}\label{eqn:f-div-sets}
   \mathcal Q_\beta =\Big\{P: \int \phi\left(\frac{dP}{dQ_{\beta}}\right)\,dQ_{\beta} \leq \delta \Big\},
\end{equation}
with the convention that the integral in \eqref{eqn:f-div-sets} is $+\infty$ if $P\not\ll Q_\beta$, and where $Q_\beta$ may
 depend on the tail probability $\beta$.
Unlike common choices that center $\mathcal Q_\beta$ at the empirical law \(\hat P_n\) or a possibly misspecified parametric fit, our $Q_\beta$ is built by extrapolating from a baseline tail level $\beta_0\gg\beta$ at which tail quantities can still be estimated reliably (standard in EVT literature; see \cite{deHaan}). 

To ensure tail alignment we choose a convex, lower semi-continuous $\phi$ with $\phi(1)=0$ and \emph{super-polynomial} growth at infinity (see Section~\ref{sec:calib_uncert} for details), which rules out distributions whose tails are heavier than those of \(Q_\beta\). With this baseline and choice of $\phi$, the worst-case value is (first-order) rate-preserving relative to the truth. Formally, for \(\beta_0=\beta^{\theta}\), 
\[
\begin{aligned}
\log \rho_{1-\beta}(\mathcal Q_\beta) &\sim \log \rho_{1-\beta}(Q) \quad &&\text{when \(Q\) is heavy-tailed (Fréchet)},\\
\rho_{1-\beta}(\mathcal Q_\beta) &\sim \rho_{1-\beta}(Q) \quad &&\text{when \(Q\) is light-tailed (Gumbel)},
\end{aligned}
\]
uniformly over a range of radii $\delta$ and intermediate levels $\theta\in (0,1)$ (Theorem~\ref{thm:scale_preserving_dro}).

Practically, starting with $n$ samples one may take $\beta_0(n)=n^{-\kappa}$ for $\kappa\in(0,1)$ and obtain reliable evaluation at tail levels $\beta(n)=c_0 n^{-q}$ with $q\ge 1$ (a regime where sample-average methods fail due to a lack of tail observations). Consequently, RPEV-DRO is robust without undue conservativeness: it aligns to the true tail while avoiding the risk inflation typical of Wasserstein or polynomial-growth $\phi$-divergence balls.

\noindent\textbf{iii) Multivariate extension.}
We extend the framework to vector-valued risk factors: for a measurable loss map $L:\mathbb R^d\!\to\!\mathbb R$ and unknown law $\tilde Q$ of $\XX\in \R^d$, the objective is to estimate $\rho_{1-\beta}(Q)$ with $Q=\tilde Q\circ L^{-1}$. 
For any law-invariant risk measure $\rho$ and convex $\phi$, we show that worst-case evaluation is invariant under the pushforward by $L$:
\[
\sup_{\tilde P\in\mathcal \Dphi(\tilde Q)} \rho(\tilde P\circ L^{-1})
\;=\;
\sup_{P\in\mathcal \Dphi(Q)} \rho(P),
\]
where $\Dphi(P)$ is as defined in \eqref{eqn:ambiguity_sets_considered} (see  Lemma~\ref{lem:lift_to_univariate}). 
We also demonstrate that under mild regularity conditions on $L(\cdot)$, $Q$ inherits the tail class of $\XX$ (Proposition~\ref{prop:mv_cvar}): if $\XX$ is heavy (resp. light) tailed, then so is $L(\XX)$.
These two observations reduce multivariate problems to the uni-variate setting of Section~\ref{sec:calib_uncert}, so the same rate-preserving guarantees and algorithm apply without additional tuning.

\noindent\textbf{iv) Empirical validation and case studies.}
We validate our approach with synthetic experiments and two real-data applications (Danish fire losses; Fama-French equity factors). Across heavy and light-tailed settings, the proposed DRO yields worst-case evaluations that are markedly less conservative than Wasserstein and standard $\phi$-divergence based baselines while avoiding risk understatement. Further, the performance remains qualitatively stable across a range of intermediate levels $\beta_0$ and ambiguity radii $\delta$. As a decision making illustration, a numerical study of discrete delta hedging shows that the rate-preserving worst-case risk provides an effective proxy for choosing the re-balancing frequency, achieving lower tail hedging error than standard worst-case calibrations.

\subsection{DRO formulations involving tail risks}  We conclude by positioning our contributions within the literature on DRO for distribution tails.
\cite{lam2017tail,bai2023distributionally} derive tractable robust formulations with shape constraints on tail densities for rare-event probabilities and related expectations.
Kullback-Leibler divergences have been used to measure model risk in various financial contexts by \cite{glasserman2014robust,schneider2015robust}. Leveraging extreme-value theory, \cite{blanchet2020distributionally,birghila2021distributionally,yuen2020distributionally} derive asymptotics for worst-case tail probabilities and elucidate how nominal tail strength and the ambiguity specification govern the robust valuation. In particular, \cite{blanchet2020distributionally} arrive at a practical one-dimensional line-search for worst-case quantiles, and \cite{yuen2020distributionally} provide bounds on worst-case {Value-at-Risk} (VaR) under dependence uncertainty in multivariate settings. In a related  direction, \cite{engelke2017robust} analyze worst-case exceedance probabilities for divergence-based ambiguity sets augmented with moment constraints. Scaling limits for distributionally robust chance constraints are developed in \cite{deo2025scaling} 
with the aim of understanding how the choice of ambiguity set affects operational cost.

A complementary line of work characterizes tail behaviors that 
are admitted by $\phi$-divergence neighborhoods: \cite{kruse2019joint,kruse2021toolkit} give conditions on the growth of $\phi$ and properties of the nominal law under which the neighborhood can include extremely heavy-tailed models (e.g., admitting infinite mean). Building on these insights, \cite{jin2024constructing} propose growth conditions on $\phi$ that ensure a finite worst-case risk evaluation. 
Our contribution is complementary in focus to existing literature: (i) we characterize how the worst-case tail-risk scales as $\beta\downarrow0$, and (ii) we design an EVT calibrated ambiguity family that preserves the risk’s tail rate under the unknown data-generating law, uniformly over the calibration choices. This uniformity is essential: it rules out $\beta$- and tail-dependent tuning of $\delta$ as a mechanism for rate recovery, and it yields an implementable, data-driven procedure.
In contrast to existing work that either derives worst-case bounds for a fixed ambiguity specification or focuses on tractable reformulations, we explicitly link tail asymptotics, ambiguity set design, and implementable procedures for tail-risk evaluation (see 
Table~\ref{tab:rw_min_compare}).

\begin{table}[htpb]
\centering
\small
\caption{Comparison with related work (\cmark\ = provided for that paper's target functional; \xmark\ = not provided).}
\begin{tabular}{lccc}
\toprule
\textbf{} &
\shortstack{\textbf{This}\\\textbf{paper}} &
\shortstack{\textbf{Jin et al.}\\\textbf{(2024)}} &
\shortstack{\textbf{BHM (2020) / Birghila}\\\textbf{et al. (2021)}} \\
\midrule
(i) Target functional
& \shortstack{CVaR / spectral}
& \shortstack{Robust risk\\measures (general)}
& \shortstack{VaR /\\tail probability} \\
\addlinespace[3pt]
(ii) Asymptotic rates ($\beta\!\downarrow\!0$)
& \cmark
& \xmark
& \cmark\\
\addlinespace[3pt]
(iii) EVT tailored nominal?
& \cmark & \xmark & \xmark \\
\addlinespace[2pt]
(iv) Data-driven procedure?
& \cmark & \xmark & \xmark \\
\bottomrule
\end{tabular}
\label{tab:rw_min_compare}
\end{table}

\noindent{\textbf{Structure of Paper:}} In Section~\ref{sec:model}, we outline our problem formulation and state and explain the modeling assumptions under which we establish our main results. We derive asymptotics  for worst case risk under Wasserstein ambiguity in Section~\ref{sec:wass_dro} and  polynomial divergence ambiguity in Section \ref{sec:divergence_dro}. 
These constitute our first contribution: a precise quantification of risk inflation due to DRO.
In Section~\ref{sec:calib_uncert} we discuss calibration of the ambiguity set $\mathcal P$, so as to reduce the conservativeness of the resulting DRO formulation.
We present the multivariate extension in Section~\ref{sec:application}. We conclude with numerical experiments on simulated and real data in Sections~\ref{sec:simulation} and \ref{sec:real_data} respectively. 

\noindent \textbf{Notation:}
We use 
$a_n\xrightarrow{a.s.}a$, $a_n\xrightarrow{p}a$, and $a_n\xrightarrow{d}a$ to denote almost-sure, in probability, and in distribution convergence; for sequences $x_t,y_t$, $x_t\sim y_t$ means $x_t/y_t\to 1$.
For a measurable space $\mathcal X$, let $\mathcal M(\mathcal X)$ be the set of probability measures on $\mathcal X$, and write $E_P[X]=\int X\,dP$.
For $P\in\mathcal M(\R)$, let $F_P(x)=P(X\le x)$ be its cdf, $\bar F_P(x)=1-F_P(x)$ its tail cdf, and $\Lambda_P(x)=-\log \bar F_P(x)$ its (cumulative) hazard.
For a function $f$, $f^{\leftarrow}$ denotes its left-inverse.
For $(P,Q)\in\mathcal M(\mathcal X)$, let $\Pi(P,Q)\subset \mathcal M(\mathcal X\times\mathcal X)$ be the set of couplings with marginals $P$ and $Q$.
For a measurable map $T:\mathcal X\to\mathcal Y$ and $P_0\in\mathcal M(\mathcal X)$, the push-forward is $P_0\circ T^{-1}$, i.e., $(P_0\circ T^{-1})(B)=P_0(T^{-1}(B))$.
Given $\mathcal P\subset \mathcal M(\mathcal X)$ and a risk functional $\rho:\mathcal M(\mathcal X)\to\R_+$, write $\rho(\mathcal P)=\sup_{P\in\mathcal P}\rho(P)$ for the worst-case value. We use $P\ll Q$ to denote that $P$ is absolutely continuous with respect to $Q$, that is $Q(A)>0$ whenever $P(A)>0$.

\section{Problem Setup and Modeling Assumptions}\label{sec:model}
The notion of regular variation is central to our capturing regularity in tails of distributions:
\begin{definition}\label{def:RV}
    A function $f:(0,\infty) \to (0,\infty)$ is regularly varying (at infinity) with index $\rho$ if for any $x>0$,
    \[
    \lim_{t\to\infty}\frac{f(tx)}{f(t)} = x^{\rho}.
    \]
   We denote this by $f\in \RV(\rho)$ or  $f\in \RV$  if there is no explicit need to specify the index of regular variation. If $f\in \RV(0)$, we  say that $f$ is slowly varying.
\end{definition}
Regular variation offers a structured approach to studying functions with polynomial-like growth/decay. 
Examples of regularly varying functions include $x^2$, $x^k\log^p(x)$ and $x^k \exp(\log^q(x))$ for $q\in(0,1)$. 
In what follows, regular variation will serve as our main language for describing both heavy-tail decay and Weibull-type light-tail behavior (via either survival tails or hazard growth), and for expressing the scaling of tail-risk functionals as $\beta\downarrow 0$.
We refer to \cite[Proposition B.1.9]{deHaan} for background and standard properties.

\subsection{Problem Formulation}\label{sec:formulation}
Consider finding the worst case evaluation of a risk measure $\rho_{1-\beta}: \mathcal M(\R) \to [0,\infty]$: 
\[
 \rho_{1-\beta}(\mathcal P) = \sup_{P\in \mathcal P}\rho_{1-\beta}(P)
\]
where $\mathcal P$ consists of perturbations
of a nominal model $P_0$. We assume that $\rho_{1-\beta}(\cdot)$ is a law-invariant risk measure. We focus on two popular ambiguity families.


\noindent \textbf{(i) Divergence based discrepancy:} Let \(\phi:[0,\infty)\to[0,\infty]\) be lower semi-continuous and convex with $\phi(1)=0$ and $\phi^\prime(1) =0$. Let  $\mathcal D_\phi(P,R)$ denote the $\phi-$divergence between $P$ and $R$
: 
\begin{equation*}
  \mathcal D_\phi(P,R) =  \int \phi\left(\frac{dP}{dR}\right) dR.
\end{equation*}
Note that $\mathcal D_\phi(P,R)$ is finite only if $P\ll R$. 

\noindent \textbf{(ii) Wasserstein discrepancy:}  
Given measures $(P,R)\in \mathcal M(\mathbb R)$, and $p\geq 1$, define the $p-$th order Wasserstein distance between $P$ and $R$ as
\begin{equation*}
 \mathcal D_{W}^{(p)}(P,R) = \left(\inf_{\pi\in\Pi(P,R)}\int |x-y|^p\,\pi(dx,dy) \right)^{1/p}.   
\end{equation*}
Define the ambiguity sets for an arbitrary center $R\in \mathcal M(\R)$, 
the choice of which will be specified in each instance:
\begin{equation}\label{eqn:ambiguity_sets_considered}
    \mathcal W_{p,\delta} = \{P:\mathcal D_W^{(p)}(P,R) \leq \delta\} \quad \text{ and }\quad \Dphi(R) = \{P : \mathcal D_\phi(P,R) \leq \delta\}. 
\end{equation}
Here $R$ will typically play the role of the nominal model (e.g., $R=P_0$ or a data-driven surrogate). Throughout, $\delta>0$ is the radius specifying the extent of perturbation.

\subsection{Assumptions on Risk Measure} 
For a distribution \(P\) and tail level \(\beta\in(0,1)\), the Value-at-Risk (VaR) is
\begin{equation}\label{eqn:VaR}
v_{1-\beta}(P) = \inf\{\,u: P(Z>u)\leq \beta\}.
\end{equation}
\begin{assumption}[\textbf{Tail-Weighted VaR}]\label{assume:tail_risk_measures}
There exist $\kappa>-1$ and  $\ell$ that is slowly varying at infinity, such that as $t\to0$, $
w(t) \sim t^{\kappa} \ell(1/t)$, 
$w:(0,1]\to[0,\infty)$ and $\int_0^1 w(t)\,dt=1$. For $P\in\mathcal M(\R)$ suppose that
\begin{equation}\label{eqn:drm_tails}
\rho_{1-\beta}(P) \;=\; \int_0^1 w(t)\, v_{1-\beta t}(P)\,dt ,
\qquad \beta\in(0,1).
\end{equation}
\end{assumption}
Assumption~\ref{assume:tail_risk_measures} is satisfied by popular measures of tail risk use as CVaR, Power distortion measures and the Wang transform; if $w(\cdot)$ is decreasing then $\rho_{1-\beta}(\cdot)$ is a coherent risk measure.
The regularity  of $w(\cdot)$ around $0$ allows us to characterize the asymptotic behavior of $\rho_{1-\beta}$ as $\beta\to 0$: indeed \cite{asimit2016extremes} derive asymptotics under a similar assumption on the weight function $w(\cdot)$, although in that work, they restrict themselves to coherent risk measures.
For $P\in \mathcal M(\R)$ with $Z\sim P$, define its $\beta$-tail restriction as the conditional distribution of $Z$ given that it exceeds its $(1-\beta)$th  quantile:
\begin{equation*}
    F_{P,\beta}(x)  =P(Z\leq x \mid Z> v_{1-\beta}(P))
\end{equation*}
We henceforth restrict attention to distributions $P$ such that 
$P\left(Z = v_{1-\beta}(P)\right)=0$, so that the upper tail beyond $v_{1-\beta}(P)$ 
carries exactly mass $\beta$, i.e. $P(Z>v_{1-\beta}(P))=\beta$.
In line with \cite{liu2021theory}, $\rho:\mathcal M(\mathbb R) \to [0,\infty]$ is said to be a $\beta$-tail risk measure if the following holds: $\rho(P) = \rho(R)$ for any $(P,R)$ such that $F_{P,\beta} = F_{R,\beta}$.  
The interpretation is that $\beta$-tail risk measures only depend on the behavior of the underlying measure beyond its $(1-\beta)$th quantile; the behavior of the bulk does not affect the evaluation.
\begin{lemma}\label{lem:tail_risk_measure}
    For any $\beta\in (0,1)$ a risk measure $\rho_{1-\beta}$ that has a representation as in \eqref{eqn:drm_tails} is a $\beta$-tail risk measure.
\end{lemma}
The intuition behind Lemma~\ref{lem:tail_risk_measure} is that $\rho_{1-\beta}(P)$ only depends on
$v_{1-s}(P)$ for $s\in(0,\beta]$. For each such $s$, $v_{1-s}(P)$ is the $(1-s/\beta)$ quantile of $F_{P,\beta}$, so if $F_{P,\beta}=F_{R,\beta}$ then $v_{1-s}(P)=v_{1-s}(R)$ for all $s\in(0,\beta]$.
Concrete examples of risk measures which satisfy Assumption~\ref{assume:tail_risk_measures} are presented in Table~\ref{tab:trm}. Verification that these risk measures satisfy Assumption~\ref{assume:tail_risk_measures} is presented in the appendix.

\begin{table}[htbp]
\centering
\setlength{\tabcolsep}{5pt}
\renewcommand{\arraystretch}{1.15}
\caption{Examples of weight functions $w(s)$ satisfying Assumption~\ref{assume:tail_risk_measures}
and the corresponding risk measures. We list the weight function $w(s)$ on $(0,1]$ and whether the resulting risk measure is coherent.}
\label{tab:trm}
\begin{tabular}{|p{2.8cm}|p{6.8cm}|>{\raggedright\arraybackslash}p{4.5cm}|}
\hline
\textbf{Name of Risk Measure} & \textbf{Weight function $w(s)$} & \textbf{Coherent?} \\
\hline
Expected Shortfall (CVaR)
& $w(s)=1$
& \makecell[l]{Yes} \\ \hline

Power distortion
& $w(s)=k\,s^{\,k-1}$ \quad($k>0$)
& \makecell[l]{Yes if $0<k\leq 1$;\\No otherwise} \\ \hline

Wang transform 
& $w(s)=\exp\!\big(-\lambda\,\Phi^{-1}(s)-\tfrac{\lambda^2}{2}\big)$
& \makecell[l]{Yes (for $\lambda\ge 0$)} \\ \hline

Log–power tail weight
& $w(s)=\dfrac{p^{\,q+1}}{\Gamma(q+1)}\,s^{\,p-1}\,(-\log s)^{q}$ \quad($p>0,\ q>-1$)
& \makecell[l]{Yes if $0<p\leq 1$ and $q\ge 0$;\\No otherwise} \\ \hline

Beta–tail weight
& $w(s)=\dfrac{1}{\mathrm{B}(p,q)}\,s^{\,p-1}(1-s)^{\,q-1}$ \quad($p>0,\ q>0$)
& \makecell[l]{Yes if $0<p<1$\\(any $q>1$)} \\ \hline

Slowly varying (polylog) tail weight
& $w(s)=\dfrac{1}{\Gamma(q+1)}\,(-\log s)^{q}$ \quad($q>-1$)
& \makecell[l]{Yes if $q\geq 0$} \\ \hline
\end{tabular}
\end{table}

\subsection{Distributionally Robust CVaR}
The special case $\rho_{1-\beta}=C_{1-\beta}$ (CVaR at tail level $\beta$) admits a reformulation (see \cite[Theorem 2]{rockafellar2002conditional}):
\[
C_{1-\beta}(P)=\inf_{u\in\R}\Big\{\,u+\beta^{-1}E_P[(Z-u)^+]\,\Big\}.
\]
Let $\mathcal P$ denote either $\mathcal W_{p,\delta}$ or $\Dphi(R)$. 
Both ambiguity sets are convex: 
\[
\mathcal W_{p,\delta}(R)=\{P:W_p(P,R)\le \delta\}=\{P:W_p^p(P,R)\le \delta^p\},
\]
and the map $P\mapsto W_p^p(P,R)$ is convex, hence its sub-level set is convex (see \cite[Proposition 7.9]{villani2008optimal}); 
likewise $P\mapsto D_{\phi}(P,R)$ is convex whenever $\phi$ is convex, so $\Dphi(R)$ is convex.
A min-max interchange (see \cite[Lemma~1]{natarajan2014probabilistic})  yields 
\begin{equation}\label{eqn:wc_cvar}
C_{1-\beta}(\mathcal P)
=\sup_{P\in\mathcal P} C_{1-\beta}(P)
=\inf_{u\in\R}\Big\{\,u+\beta^{-1}\,\sup_{P\in\mathcal P} E_P[(Z-u)^+]\,\Big\}.
\end{equation}
Thus worst–case CVaR reduces to worst–case {expectations} of $(Z-u)^+$, enabling dual reformulations and efficient computational evaluation under both Wasserstein and $\phi$–divergence balls. While our asymptotic results apply to the general tail-weighted VaR class in
Assumption~\ref{assume:tail_risk_measures}, we will return to the formulation
\eqref{eqn:wc_cvar} in Section~\ref{sec:simulation} as a concrete example on which we conduct numerical studies.


\subsection{Assumptions on Tail Behavior}
We impose assumptions $P_0$ which are standard in literature on modeling tails. 

\begin{assumption}[Heavy tails]\label{assume:heavy_tailed_data}
There exists a $\gamma>0$  such that $\bar F_{P_0} \in \RV(-\gamma)$ .
\end{assumption}
\begin{assumption}[Weibull-type tails]\label{assume:weibullian_tails}
There exists $\gamma>0$ such that the cumulative hazard $\Lambda_{P_0}(x)=-\log \bar F_{P_0}(x)$ is regularly varying with index $\gamma$; i.e., $\Lambda_{P_0}\in \RV(\gamma)$.
\end{assumption}

Assumption~\ref{assume:weibullian_tails} complements Assumption~\ref{assume:heavy_tailed_data} by covering light–tailed laws whose tail cdf decays faster than any power law.
Indeed, $\Lambda_{P_0}\in \RV(\gamma)$ implies
$\bar F_{P_0}(x)=\exp\{-x^\gamma \ell(x)\}$ with $\ell$ slowly varying, whereas under Assumption~\ref{assume:heavy_tailed_data} we have the power–law form $\bar F_{P_0}(x)=x^{-\gamma}\ell(x)$. Examples of distributions satisfying Assumption~\ref{assume:weibullian_tails} include exponential and gamma (effective index $\gamma=1$), Weibull ($\gamma$ equals the Weibull shape), and Gaussian ($\gamma=2$).
When Assumption~\ref{assume:weibullian_tails} holds, $P_0\in \MDA(0)$ (Gumbel domain of attraction); see \cite[Theorem 2]{de2016approximation}.

Interestingly, the regular variation of $w(\cdot)$ near the origin with the tail regularity of $P_0$ translate to regularity on the behavior of tail risk measures. 
\begin{lemma}\label{lem:CVaR_regularity} Suppose Assumption~\ref{assume:tail_risk_measures} holds. Then, 
\begin{enumerate} 
    \item[(i)] Suppose that a distribution $P$ satisfies Assumption~\ref{assume:heavy_tailed_data} with $1/\gamma<\kappa+1$. Then $\rho_{1-\beta}(P) = \beta^{-1/\gamma}\ell_1(1/\beta)$,  as $\beta\to 0$ where $\ell_1$ is slowly varying at $\infty$.
    \item[(ii) ]  Suppose that a distribution $P$ satisfies Assumption~\ref{assume:weibullian_tails}. Then $\rho_{1-\beta}(P) = \log^{1/\gamma}(1/\beta)\ell_2(\log(1/\beta))$ as $\beta\to 0$, where $\ell_2$ is slowly varying at $\infty$.
\end{enumerate}
\end{lemma}
The above two results imply the following:
for distributions with a power law type tail, the risk measure grows polynomially, while for those with Weibullian tails, it grows poly-logarithmically in $(1/\beta)$. We point out that the extra condition  $1/\gamma<\kappa+1$ is required in order for $\rho_{1-\beta}(P_0)$ to be finite (for CVaR, this reduces to the well known condition, $\gamma>1$). These observations serve as a starting point for quantifying the extent of conservativeness incurred upon the
use of DRO.

\section{Conservativeness  of Wasserstein DRO} \label{sec:wass_dro}
Recall the following \emph{exact} worst case risk evaluation over $\mathcal W_{p,\delta}(P_0)$ (e.g. \cite[Theorem 5]{wu2025generalization} and \cite[Proposition 6.18]{kuhn2025distributionally}): whenever $\rho_{1-\beta}$ satisfies Assumption~\ref{assume:tail_risk_measures}
\begin{equation}\label{eq:additive-wp}
\rho_{1-\beta}\left(\mathcal W_{p,\delta}(P_0)\right)
\;=\;
\rho_{1-\beta}(P_0)\;+\;C_0\,\left(\delta/\beta^{1/p}\right),
\qquad \beta\in(0,1),
\end{equation}
for an appropriate constant $C_0>0$ that depends on $p$ and the weight function $w(\cdot)$, but is uniform in $\beta$ and $\delta$.
Under Assumption~\ref{assume:heavy_tailed_data}, by Lemma~\ref{lem:CVaR_regularity}
$
\rho_{1-\beta}(P_0)=\beta^{-1/\gamma}\ell_1(1/\beta)$ 
and under Assumption~\ref{assume:weibullian_tails}, by Lemma~\ref{lem:CVaR_regularity}(ii)  
$\rho_{1-\beta}(P_0)=  (\log(1/\beta))^{1/\gamma}\ell_2(\log(1/\beta))$ for $(\ell_1,\ell_2)\in \RV(0)$. 
Consequently, so long as $\gamma>p$, $\rho_{1-\beta}(\mathcal W_{p,\delta}(P_0)) \sim C_0 (\delta/\beta^{1/p})$. 
The theorem below now follows:
\begin{theorem}\label{thm:Robust_distortion_measures}
Suppose Assumption~\ref{assume:tail_risk_measures} holds and let $\delta>0$ be fixed. Then, 
\begin{enumerate}
\item[(i)] \textbf{Heavy tails.}
If in addition $P_0$ satisfies Assumption~\ref{assume:heavy_tailed_data} with index $\gamma>p$\footnote{The $\gamma=p$ case can also be handled, but it requires imposition of second order regular variation. This is a direction we do not pursue in this paper.}, (hence $E_{P_0}[|Z|^p]<\infty$) and $1/\gamma<\kappa+1$. Then
\[
\lim_{\beta\downarrow 0}
\frac{\log \rho_{1-\beta}(\mathcal W_{p,\delta}(P_0))}{\log \rho_{1-\beta}(P_0)}
\;=\;\frac{\gamma}{p}.
\]
\item[(ii)] \textbf{Light tails.}
If instead $P_0$ satisfies Assumption~\ref{assume:weibullian_tails}, then
\[
\lim_{\beta\downarrow 0}
\frac{\log \rho_{1-\beta}(\mathcal W_{p,\delta}(P_0))}{\log \rho_{1-\beta}(P_0)}
\;=\;\infty .
\]
\end{enumerate}
\end{theorem}
The representation \eqref{eq:additive-wp} shows that Wasserstein robustification induces a penalty 
that dominates the nominal risk evaluation:
if $P_0$ has a power law type tail, then the worst case risk grows at a polynomially faster rate; in case the tail of $P_0$ decays faster than polynomial, the Wasserstein robustification induces a risk inflation that grows polynomially in $(1/\beta)$ (and therefore super-polynomially relative to the nominal risk).
\begin{remark}\em
In the setting of Theorem~\ref{thm:Robust_distortion_measures}, for any fixed $\delta>0$ the penalty term in \eqref{eq:additive-wp}
does not remain comparable to $\rho_{1-\beta}(P_0)$ as $\beta\downarrow0$.
One can let $\delta=\delta(\beta)\downarrow0$ and tune
its decay so that the penalty term in \eqref{eq:additive-wp} remains comparable to
$\rho_{1-\beta}(P_0)$. However this requires $\beta$ and tail-dependent tuning.
RPEV-DRO (Section~\ref{sec:calib_uncert}), on the other hand, is designed to match the tail growth rate of $\rho_{1-\beta}$ uniformly over $\delta$, thereby removing the need to tune $\delta$ with $\beta$ (formalized as \emph{rate preservation} in Section~\ref{sec:calib_uncert}).
\end{remark}
\subsection{Characterization of Worst Case Distribution}
We now specialize to the case where $\rho_{1-\beta} = C_{1-\beta}$ and derive the worst case distribution in order to illustrate the mechanism behind the excess pessimism from Theorem~\ref{thm:Robust_distortion_measures}.  The following duality result 
 (see \cite{blanchet2019quantifying,gao2023distributionally}) is well known: 
\begin{equation}\label{eqn:wass_duality}
     \sup_{P\in \mathcal W_{p,\delta}(P_0) } E_P[f(Z)]= \inf_{\lambda\geq 0}\left\{ \lambda \delta^p  + E_{P_0}\left[ \sup_{y} \left\{ f(y) - \lambda|y-Z|^p\right\} \right]\right\}.
\end{equation}
Set $f(y) = (y-u)^+$ and use \eqref{eqn:wc_cvar}
to get 
\[
C_{1-\beta}(\mathcal W_{p,\delta}(P_0)) = \inf_{(u,\lambda)} \left\{u+\lambda\beta^{-1} \delta^p +\beta^{-1} E_{P_0}\left[\sup_{y}\left\{(y-u)^+ - \lambda |y-Z|^p\right\}\right]\right\}.
\]
\noindent One can characterize the worst case distribution by solving the problem above.
\begin{theorem}[\textbf{Characterizing the Worst Case Distribution}]\label{thm:Dual_Mult}

If $p>1$ and  $F_{P_0}$ is continuous at $v_{1-\beta}(P_0)$. Then
 the 
 unique worst-case distribution in $\mathcal W_{p,\delta}(P_0)$ is
    \[
    P_{\beta, \tt wc} = P_0 \circ T_{\beta,\delta}^{-1}, \text{ where } T_{\beta,\delta}(x) = x\mv{1}\{x\leq v_{1-\beta}(P_0)\} + (x+ \delta/\beta^{1/p})\mv{1}\{x>v_{1-\beta}(P_0)\}
    \]    
   If $p=1$, any optimal coupling between $P_0$ and $P_{\beta, \tt wc}$ is supported on the region
    \begin{equation}\label{eqn:p=1}
         \bigcup_{c\geq 0}\left\{(x,y): y = x \mv{1}\{x\leq v_{1-\beta}(P_0)\} + (x+c)\mv{1}\{x>v_{1-\beta}(P_0)\}\right\}.
    \end{equation}
\end{theorem}

An important consequence of Theorem~\ref{thm:Dual_Mult} is that for each fixed $\beta$, both $P_0$ and $P_{\beta,\mathrm{wc}}$ share the same tail parameter $\gamma>0$: if $P_0$ satisfies Assumption~\ref{assume:heavy_tailed_data} (or \ref{assume:weibullian_tails}), then so does $P_{\beta,\mathrm{wc}}$.  For $p>1$ and fixed $\beta$, the transport map $T_{\beta,\delta}$ is invertible, hence
\[
\bar F_{P_{\beta,\mathrm{wc}}}(t)=\bar F_{P_0}\!\big(T_{\beta,\delta}^{\leftarrow}(t)\big),\qquad
\Lambda_{P_{\beta,\mathrm{wc}}}(t)=\Lambda_{P_0}\!\big(T_{\beta,\delta}^{\leftarrow}(t)\big).
\]
For \(t>v_{1-\beta}(P_0)\) one has \(T_{\beta,\delta}^{\leftarrow}(t)=t-\delta/\beta^{1/p}\), and therefore
\[
\bar F_{P_{\beta,\mathrm{wc}}}(t)=\bar F_{P_0}\!\left(t-\frac{\delta}{\beta^{1/p}}\right) ,\qquad \Lambda_{P_{\beta,\mathrm{wc}}}(t)=\Lambda_{P_0}\!\left(t-\frac{\delta}{\beta^{1/p}}\right).
\]
For each fixed $\beta$, the shift $\delta/\beta^{1/p}$ is constant; since regular
variation is stable under fixed translations, $P_{\beta,\mathrm{wc}}$ inherits the
tail index of $P_0$.
View the worst case as an adversary with budget $\delta^p$. 
Then fully reshaping the tail to decay like $x^{-p}$ is too costly under $p$-Wasserstein distance the adversary instead shifts only tail mass just enough that CVaR at level $\beta$ grows \emph{at the same rate} as it would under an $x^{-p}$-type tail.

\begin{remark}\label{rem:p=1}\em
    When $p=1$, the
    map $x\mapsto T_{\beta,\delta}(x)$ continues to produce an optimal coupling since it satisfies \eqref{eqn:p=1} and $E_{P_0}|T_{\beta,\delta}(Z) - Z | =  \delta$ so long as there is no  atom at $v_{1-\beta}(P_0)$.  We point out that this coupling need not be unique.
\end{remark}

\begin{illustration}\label{illustrate:wass_dro}\em 
   Suppose that $P_0$ has a distribution function 
   \begin{equation}\label{eqn:nominals}
         F(x) =\begin{cases} 1- \left(1+\frac{\alpha x}{\sigma}\right)^{-1/\alpha}\quad\quad \text{ when } P_0 \text{ satisfies Assumption~\ref{assume:heavy_tailed_data} and}\\
   1- \exp(-cx^{q}) \quad\quad \quad \text{ when }  P_0 \text{ satisfies Assumption~\ref{assume:weibullian_tails}}.
   \end{cases}
   \end{equation}
For this illustration, we fix $\alpha=1/3$, $q=1.5$, and $c=\sigma=1$ in \eqref{eqn:nominals}.
Figures~\ref{fig:wass_heavy}–\ref{fig:wass_light} plot $\log C_{1-\beta}(\mathcal W_{1,\delta})$ against $\beta$, alongside the baseline $\log C_{1-\beta}(P_0)$. The budget $\delta$ is indicated in each panel.
Consistent with theory, the worst case curve lies above the baseline across the range of $\beta$, indicating severe risk inflation; this is more pronounced for light-tailed $P_0$.

 \begin{figure}[htpb]
 \centering
 \begin{subfigure}{0.48\textwidth}
         \centering
         \includegraphics[width=\textwidth]{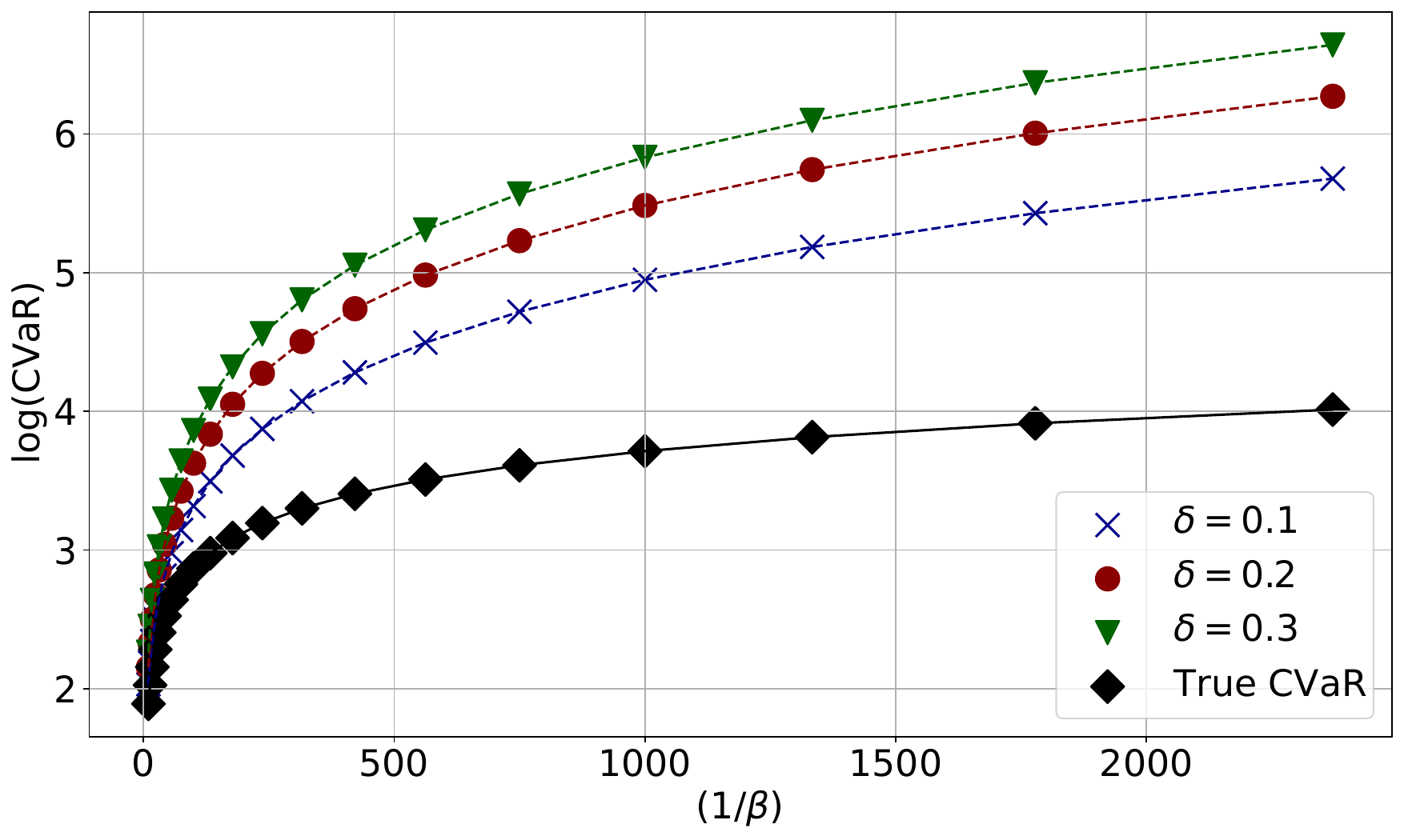}
         \caption{$P_0$ satisfies Assumption~\ref{assume:heavy_tailed_data}}
         \label{fig:wass_heavy}
     \end{subfigure}
     \hfill
     \begin{subfigure}{0.48\textwidth}
         \centering
         \includegraphics[width=\textwidth]{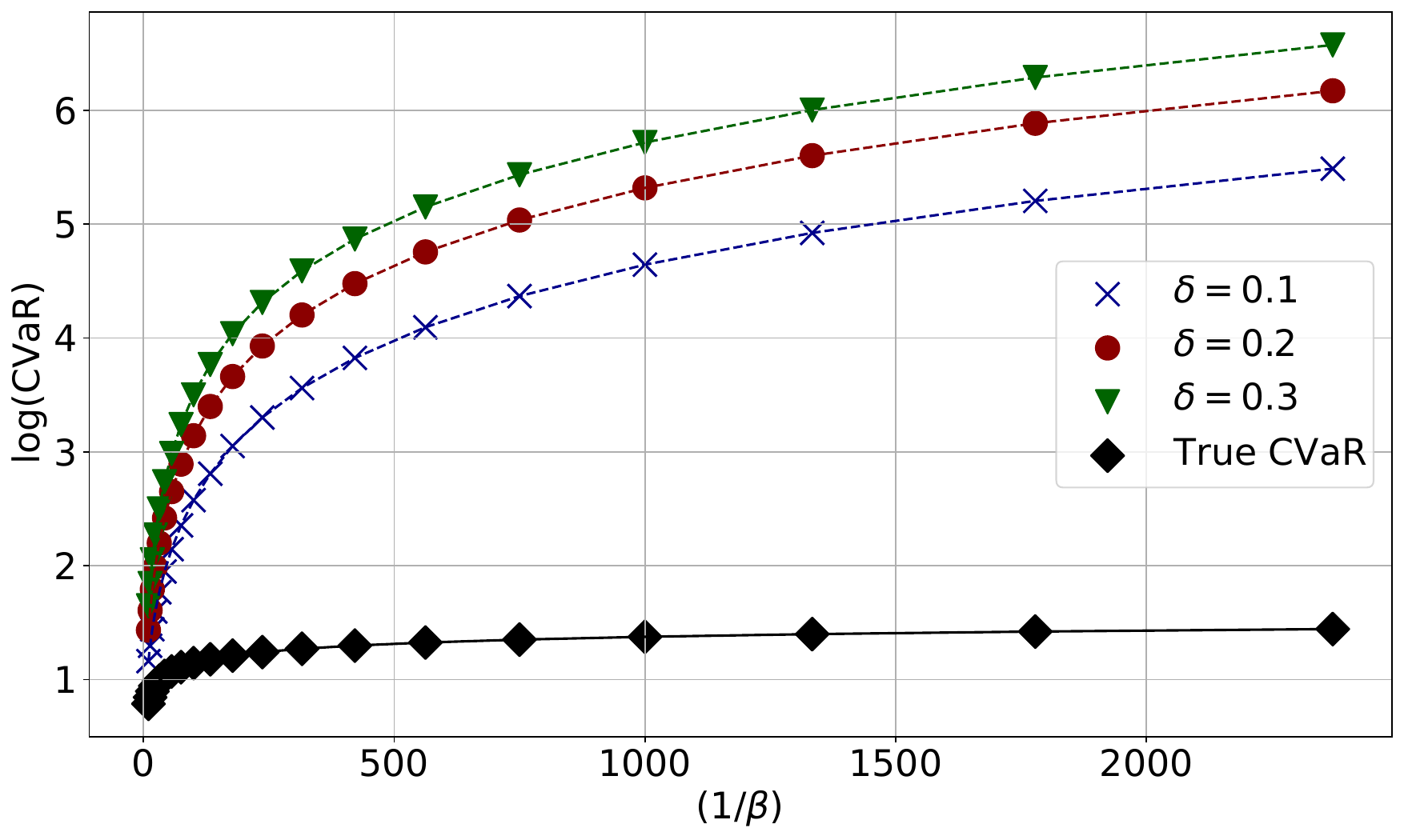}
         \caption{$P_0$  satisfies Assumption~\ref{assume:weibullian_tails}}
         \label{fig:wass_light}
     \end{subfigure}
     \caption{Worst-case CVaR evaluation under Wasserstein uncertainty}\label{fig:cvar_wc_wass}
\end{figure}


Figures~\ref{fig:wc_heavy}–\ref{fig:wc_light} show the worst-case distributions corresponding to Figures~\ref{fig:wass_heavy}–\ref{fig:wass_light} with $\beta=0.001$ and $\delta=0.1$; histogram frequencies are on a logarithmic scale.
There are two key takeaways: (i) The worst-case law is the nominal with its upper-$\beta$ tail shifted to the right by a constant, as in Theorem~\ref{thm:Dual_Mult}; accordingly, the nominal (red) and worst-case (blue) tails decay at the same rate and (ii) with the same transport budget, the absolute shift is identical across panels, but relative to the tail scale it moves mass farther from the bulk in the light-tailed case, explaining the larger conservativeness observed there.

 \begin{figure}[htpb]
 \centering
 \begin{subfigure}{0.48\textwidth}
         \centering
         \includegraphics[width=\textwidth]{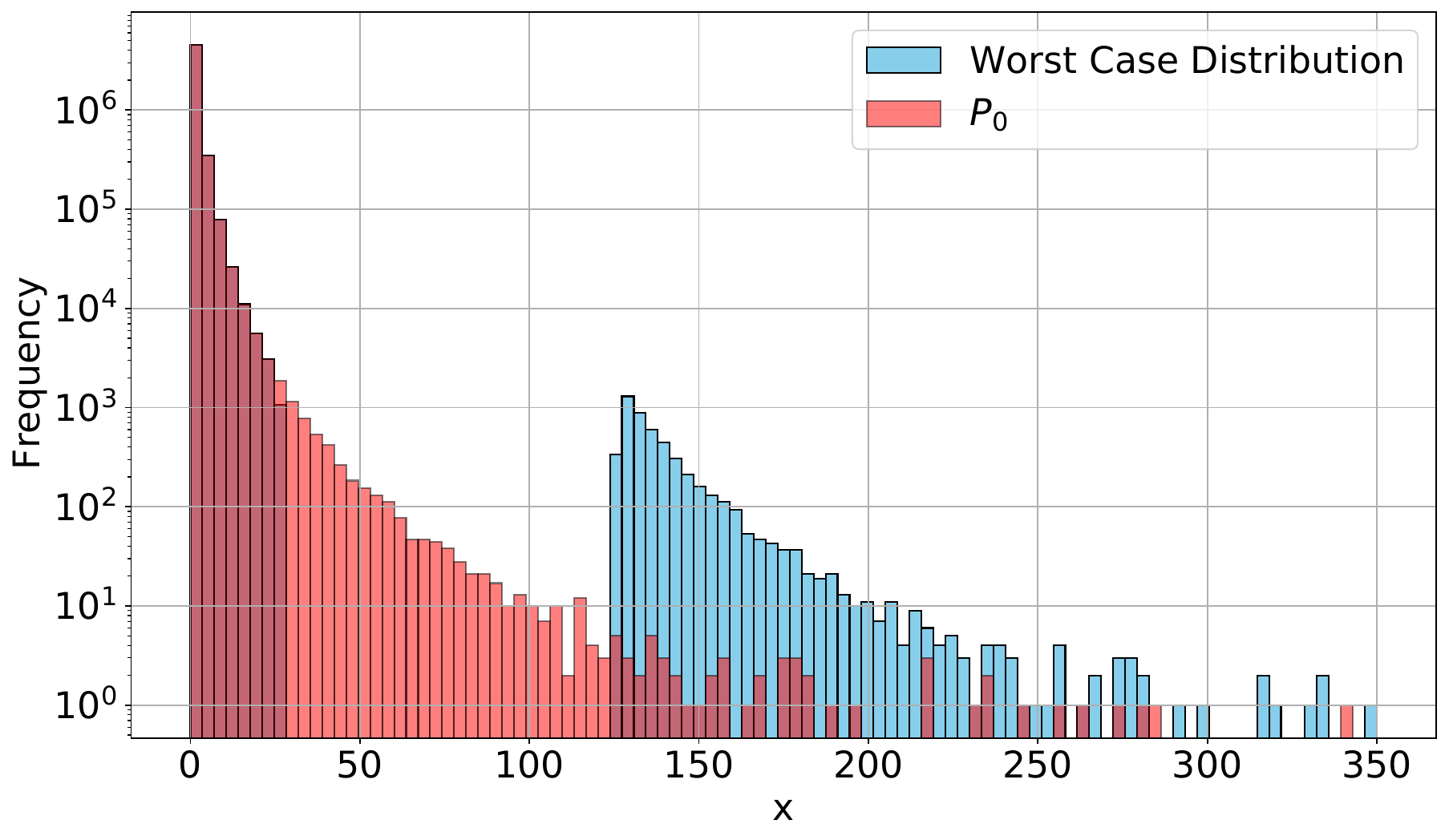}
         \caption{$P_0$ satisfies Assumption~\ref{assume:heavy_tailed_data}}
         \label{fig:wc_heavy}
     \end{subfigure}
     \hfill
     \begin{subfigure}{0.48\textwidth}
         \centering
         \includegraphics[width=\textwidth]{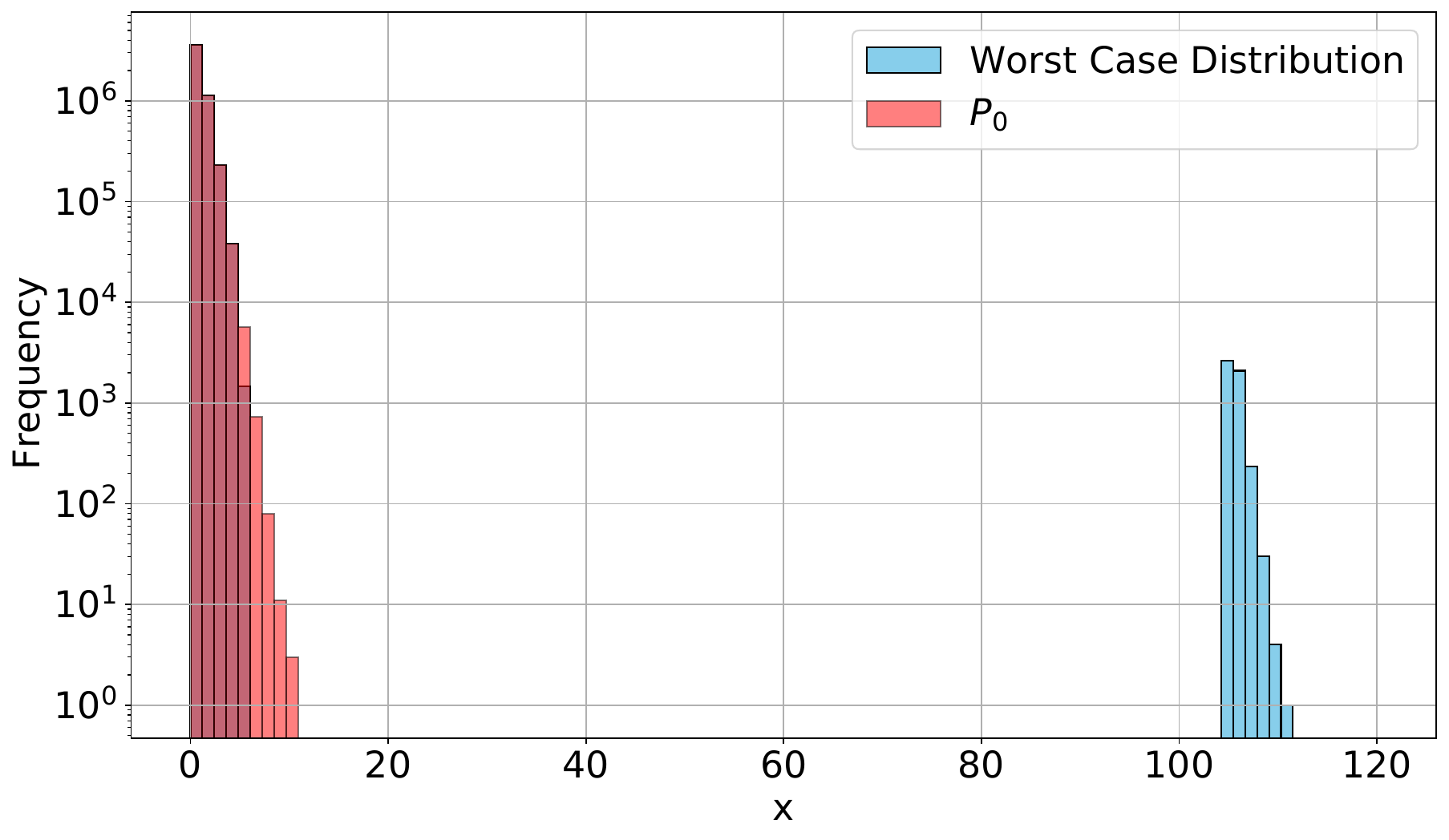}
         \caption{$P_0$ satisfies Assumption~\ref{assume:weibullian_tails}}
         \label{fig:wc_light}
     \end{subfigure}
\end{figure}
\end{illustration}
\section{Conservativeness of Divergence Based DRO}\label{sec:divergence_dro}
In this section we derive asymptotics for $\rho_{1-\beta}(\Dphi(P_0))$, where $\Dphi(P_0)$ is defined in \eqref{eqn:ambiguity_sets_considered}. We make the following standard assumptions on $\phi(\cdot)$:
\begin{assumption}[\textbf{Smoothness of $\phi$-function}]\label{assume:f-assume}
    Suppose $\phi$ is a twice differentiable, convex function such that $\phi^\prime(1) = \phi(1) =0$ and that $\phi(x)/x\to \infty$ as $x\to\infty$.
\end{assumption}
Assumption~\ref{assume:f-assume} captures popular cases such as KL divergence, $\chi^2$ divergence and the various Cressie-Read divergences (see \cite{lam2019recovering,duchi2021statistics}).
To build intuition for the conservativeness of $\phi$–divergence DRO, we characterize tail behaviors that can occur inside the ball
$\Dphi(P_0)=\{P:\mathcal D_\phi(P,P_0)\le\delta\}$. 
Similar results were shown in \cite[Proposition~4]{kruse2019joint}; however our statements hold under milder assumptions (they assume exactly Pareto/Weibull distributions, while we only regularly varying tails conditions). We also give constructive proofs: for light tails, exponential tilting thickens tails without changing the Weibull‐type exponent; for heavy tails, power tilts yield  distributions with strictly heavier power laws while keeping the divergence finite.

\begin{proposition}[\textbf{Contents of divergence ball}]
The conclusions below hold for any \textit{given} $\delta>0$:
\label{prop:f_div_contents}
    \begin{enumerate}
        \item[(i)] There exists $P_1\in \Dphi(P_0)$ with $E_{P_1}[Z^+] =\infty$, if $\phi(x) \sim x\log x$ and either Assumption~\ref{assume:heavy_tailed_data} holds with $\gamma> 1$\footnote{If $\gamma\leq 1$, then  $P_0\in \Dphi(P_0)$ and $E_{P_0}[Z^+]=\infty$.}, or Assumption~\ref{assume:weibullian_tails} holds with $\gamma< 1$. If instead $\phi\in \RV(p)$ and $P_0$ satisfies Assumption~\ref{assume:heavy_tailed_data} 
        with $1<\gamma< \frac{p}{p-1}$, there exists $P_1\in \Dphi(P_0)$ with $E_{P_1}[Z^+]=\infty$. 
        \item[(ii)] Suppose that for some $p>1$, $\phi\in \RV(p)$, and let $P_0$ satisfy Assumption~\ref{assume:heavy_tailed_data}. Then, for any $\gamma^\prime \in\left( \frac{p-1}{p}\gamma,\gamma\right)$ there exists $P_1\in \Dphi(P_0)$ for which $\bar F_{P_1}\in \RV(-\gamma^\prime)$.
    \item[(iii)] Suppose $\phi\in\RV(p)$ for some $p>1$ and $P_0$ satisfies Assumption~\ref{assume:weibullian_tails}. 
    Then, for any $\varsigma>0$ there exists $ P_1\in\Dphi(P_0)$ and $t_0 >0$ 
    such that for all $t>t_0$,
    \[
  \frac{\Lambda_{P_1}(t)}{\Lambda_{P_0}(t)}  \leq  \left(\frac{p-1}{p}\right)+\varsigma.
    \]
    \end{enumerate}
\end{proposition}
Unlike existing characterizations of worst-case tail probabilities within divergence balls, Proposition~\ref{prop:f_div_contents} is used here as a structural obstruction: for any fixed $\delta>0$ the ball necessarily contains heavier-tailed alternatives, implying unavoidable inflation of tail-risk functionals and ruling out rate preservation under standard $\phi$-divergence DRO: 
(i) if $\phi$ has KL-divergence type growth, then even when the baseline $P_0$ has finite mean, the $\phi$-divergence ball may contain alternatives with infinite mean; hence many tail risk functionals become infinite in the worst case.
(ii) In the heavy-tailed case $\bar F_{P_0}\in RV(-\gamma)$, the ball contains distributions whose survival function is $\RV(-\gamma^\prime)$ for some  $\gamma'<\gamma$. Since this corresponds to a \emph{heavier} power-law tail, the ball necessarily includes strictly heavier-tailed models than $P_0$.
(iii) In the Weibull-type case $\Lambda_{P_0}\in RV(\gamma)$, the ball contains a distribution $P_1$ with the \emph{same} Weibull exponent $\gamma$ but a smaller hazard scale (equivalently, $\Lambda_{P_1}(t)\le c\,\Lambda_{P_0}(t)$ with $c<1$), hence a \emph{more slowly} decaying (heavier) tail.
These containment results directly yield lower bounds on worst-case tail risk and foreshadow the sharp asymptotics in Lemma~\ref{lem:lbb_phi_div}.
\begin{lemma}[\textbf{Lower Bound for Robust Risk}]\label{lem:lbb_phi_div}
Suppose Assumption~\ref{assume:tail_risk_measures} holds. Then,
    \begin{enumerate}
        \item[(i)] Suppose the conditions of  Proposition~\ref{prop:f_div_contents}(i) hold there exists $m>0$  such that  $\inf_{t\in (0,m)}w(t) > 0$.  Then, $\rho_{1-\beta}(\Dphi(P_0)) = \infty$.
        \item[(ii)] Suppose the conditions of  Proposition~\ref{prop:f_div_contents}(ii) hold and that $\kappa > \frac{1}{\gamma}\frac{p}{p-1}-1$. Then, given any $\gamma^\prime \in\left(\frac{p-1}{p}\gamma,\gamma\right)$ there exists $P_1\in\Dphi(P_0)$ with $\rho_{1-\beta}(P_1)  = \beta^{-1/\gamma^\prime } \ell(1/\beta)$ as $\beta\downarrow 0 $, where $\ell$ is slowly varying at $\infty$. Consequently, for any $\varsigma>0$,  $\log \rho_{1-\beta}(\Dphi(P_0)) > \left(\frac{p}{p-1} - \varsigma\right) \log \rho_{1-\beta}(P_0)$ for all sufficiently small $\beta$.
        \item[(iii)]Suppose the conditions of Proposition~\ref{prop:f_div_contents}(iii) hold. Then for any $\varsigma> 0$, there exists a distribution $P_1\in \Dphi(P_0)$ such that for all sufficiently small $\beta$, $\rho_{1-\beta}(P_1) > ((\frac{p}{p-1})^\frac{1}{\gamma}-\varsigma)\rho_{1-\beta}(P_0)$
    \end{enumerate}
\end{lemma}

As a consequence of Lemma~\ref{lem:lbb_phi_div},  for polynomially growing divergences, the (log of) worst-case tail risk is asymptotically bounded \emph{below} by a fixed multiplicative factor strictly greater than $1$, i.e., there is a significant risk inflation.
Theorem~\ref{thm:poly_divergence} below derives upper bounds that match those derived above.
\begin{theorem}[\textbf{Worst-Case Risk under $\phi-$Divergence Ambiguity}]\label{thm:poly_divergence}
    Suppose Assumption~\ref{assume:f-assume} hold, and further $\phi\in \RV(p)$ for some $p>1$. 
    \begin{enumerate}
        \item[(i)]  If in addition Assumptions~\ref{assume:tail_risk_measures} and \ref{assume:heavy_tailed_data} hold with $\kappa > \frac{1}{\gamma} \frac{p}{p-1} -1$, then 
        \[
        \log \rho_{1-\beta}(\Dphi(P_0)) \sim \frac{p}{p-1} \log \rho_{1-\beta}(P_0) \text{ as $\beta\to 0$}.
        \]
        \item[(ii)] If Assumptions \ref{assume:tail_risk_measures} and \ref{assume:weibullian_tails} holds, then
        \[
        \rho_{1-\beta}(\Dphi(P_0)) \sim \left(\frac{p}{p-1}\right)^{1/\gamma} \rho_{1-\beta}(P_0) \text{ as $\beta\to 0$}
        \]
    \end{enumerate}
\end{theorem}
The implication of Theorem~\ref{thm:poly_divergence} is that, for heavy tails the inflation is polynomial with exponent \(p/(p-1)\), whereas for Weibullian tails it is a constant factor \((p/(p-1))^{1/\gamma}\) (contrast Wasserstein, whose log-ratio limit is \(\gamma/p\)).
\begin{illustration}\label{illustrate:poly_div}\em
    In order to show the implications of the asymptotics from Theorem~\ref{thm:poly_divergence}, we numerically compute $C_{1-\beta}(\Dphi(P_0))$ and compare it with $C_{1-\beta}(P_0)$. The worst-case CVaR is obtained by applying the duality from \cite{shapiro2017distributionally}, Section 3.2 to \eqref{eqn:wc_cvar},   
    \begin{equation}\label{eqn:phi_div_dual}
        C_{1-\beta}(\Dphi(P_0)) = \inf_{u,\lambda > 0,\eta\in \R } \left\{u+ \beta^{-1}\left(\eta+ \delta \lambda + \lambda E_{P_{0}}\left[\phi^*\left(\frac{(Z-u)^+-\eta}{\lambda}\right)\right]\right)\right\}.
    \end{equation}
    where $\phi^*(s) = \sup_{t}\{st-\phi(t)\}$ is the convex dual of $\phi$.
All expectations in \eqref{eqn:phi_div_dual} are evaluated by Monte Carlo (with $N=10^6$ samples) under $P_0$.
In Figure~\ref{fig:WC_evals}, we compare the worst case evaluation against CVaR evaluated at the nominal model. We set $\phi(t) = \frac{1}{2}(t-1)^2$, which corresponds to the $\chi^2-$divergence. The nominal distribution $P_0$ and budget parameter $\delta$ are as indicated in the respective sub-figure; the parameters of each distribution are as in Numerical Illustration~\ref{illustrate:wass_dro}.
 When $P_0$ is heavy tailed, observe that $C_{1-\beta}(\Dphi(P_0))$ grows as \(\log C_{1-\beta}(\Dphi(P_0)) \sim \tfrac{p}{p-1}\log C_{1-\beta}(P_0)\), hence strictly faster on a log scale. Conversely, if $P_0$ is of a Weibullian type, then \textit{both} $C_{1-\beta}(\Dphi(P_0))$ and $C_{1-\beta}(P_0)$ appear to grow at the same rate, differing instead by a constant. In both the figures below, we keep the vertical axes linear so as to emphasize the extent of risk inflation that occurs due to robustification.

 \begin{figure}[htpb]
 \centering
 \begin{subfigure}{0.48\textwidth}
         \centering
         \includegraphics[width=\textwidth]{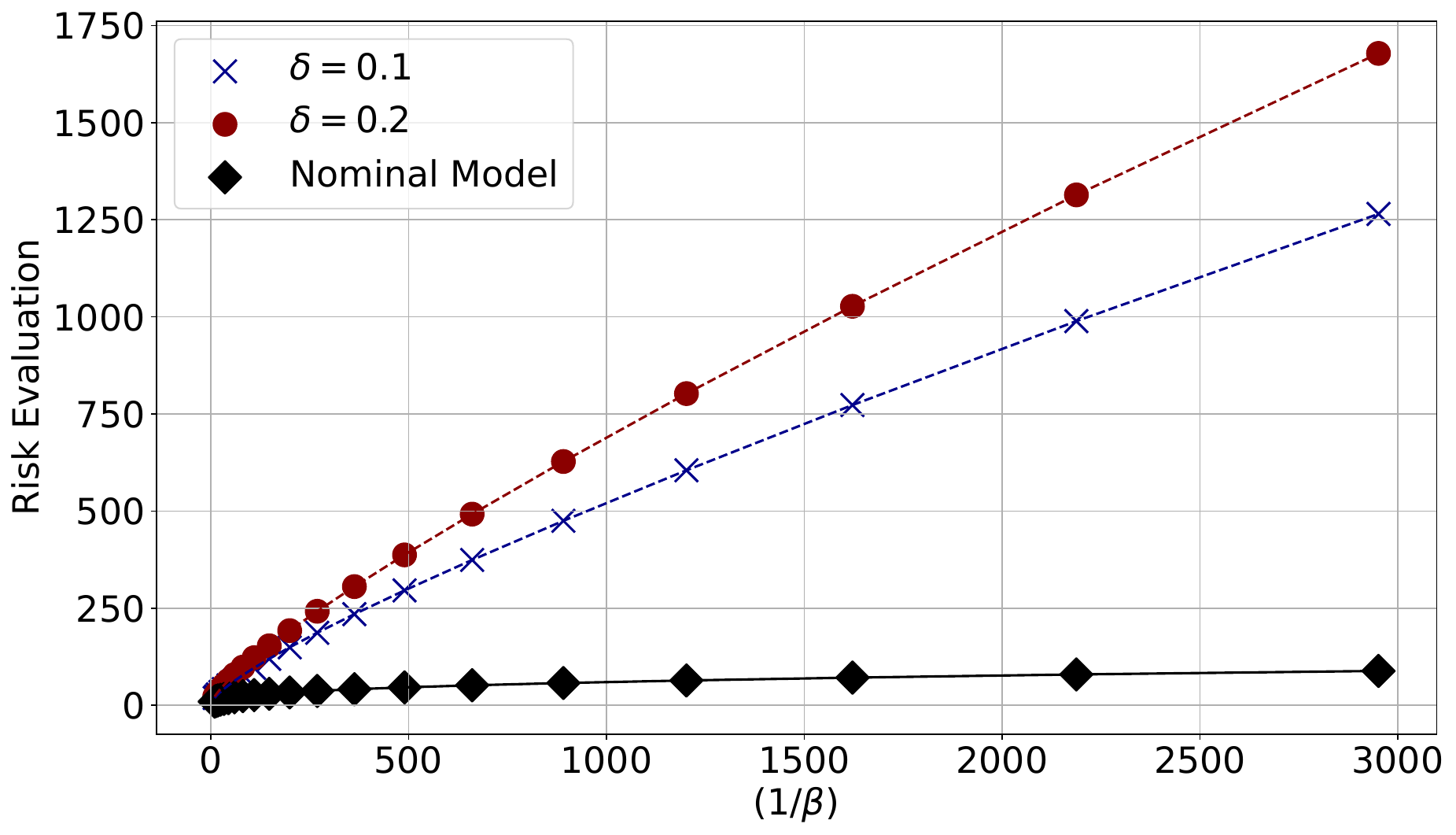}
         \caption{$P_0$ satisfies Assumption~\ref{assume:heavy_tailed_data}}
         \label{fig:pareto_wc}
     \end{subfigure}
     \hfill
     \begin{subfigure}{0.48\textwidth}
         \centering
         \includegraphics[width=\textwidth]{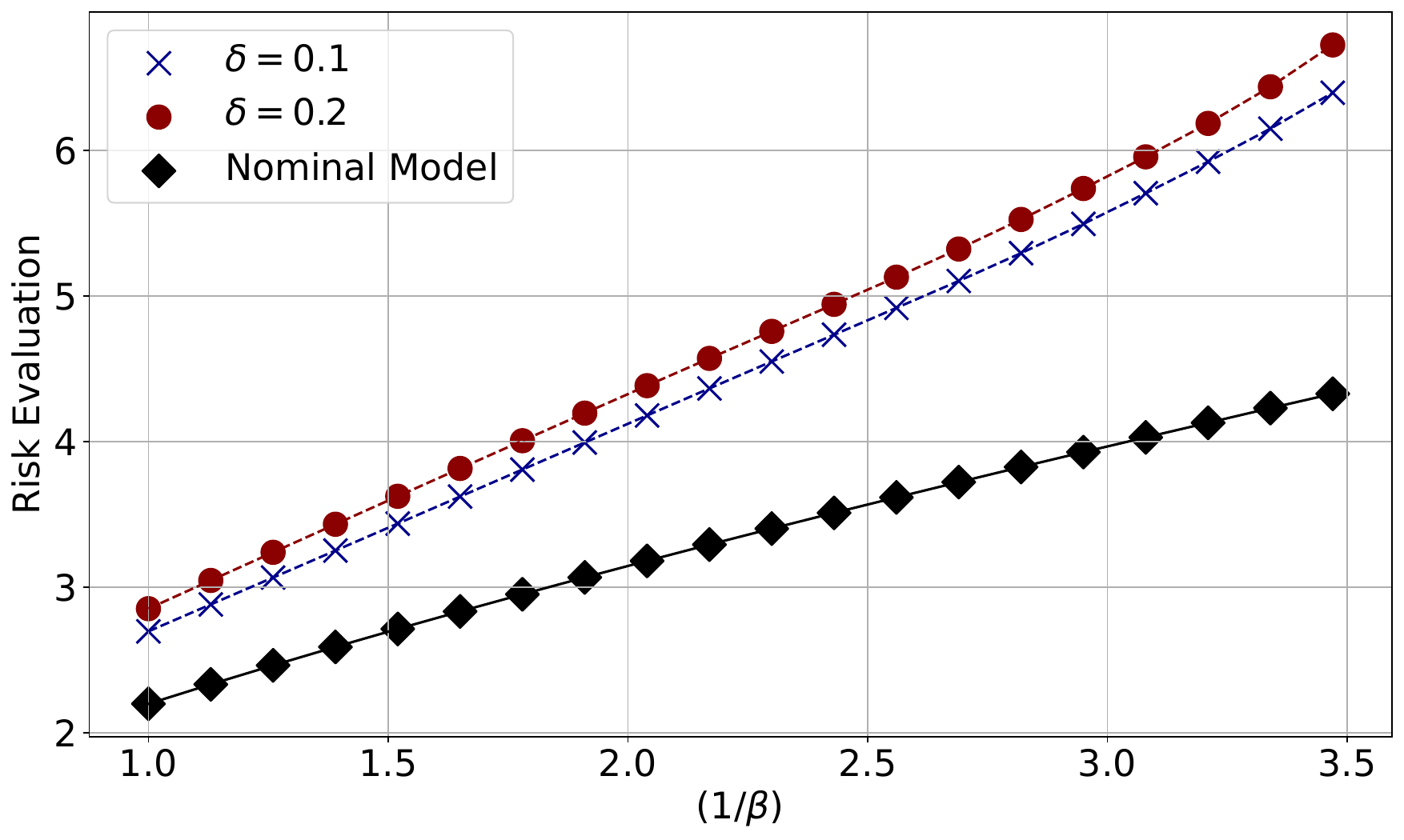}
         \caption{$P_0$ satisfies Assumption~\ref{assume:weibullian_tails}}
         \label{fig:weibull_wc}
     \end{subfigure}
     \caption{Worst-case CVaR evaluation with $\chi^2$-divergence ambiguity set}
\end{figure}\label{fig:WC_evals}
\end{illustration}

\section{Design of Representative DRO Formulations}\label{sec:calib_uncert}
The previous two sections show that DRO based on common discrepancy measures can be overly conservative. 
To identify formulations whose worst-case value tracks the true tail risk, we introduce the notion of rate-preserving ambiguity families. Let  $\Delta \subset [0,\infty)$ be compact.
\begin{definition}[Rate-Preserving ambiguity family]\label{def:Scale_preserving} 
   Given a discrepancy measure $\mathcal D$\footnote{that is, $\mathcal D(P,Q) = 0$ only if $P=Q$. In this paper, we  consider either $\mathcal D = \mathcal D_\phi$ or $\mathcal D = \mathcal D_W^{(p)}$}, nominal distributions  $Q_{\beta}^{(\theta)}$  parametrized by $\theta$ taking values in a compact set $\Theta$, for $(\theta,\delta) \in \Theta\times \Delta$, define $\mathcal Q_{\beta,\delta}^{(\theta)} = \{P: \mathcal D(P,Q_\beta^{(\theta)})\leq \delta\}$, and let $Q\in \mathcal M(\mathbb R)$ be a fixed distribution. 
   We say the family $\{\mathcal Q_{\beta,\delta}^{(\theta)}:(\theta,\delta)\in\Theta\times\Delta\}$ is
\emph{$Q$-weakly rate-preserving} if
\[
\lim_{\beta\downarrow 0}\sup_{\substack{\theta\in \Theta,\\\delta\in \Delta}}\left\vert\frac{\log \rho_{1-\beta}(\mathcal Q_{\beta,\delta}^{(\theta)})}{\log \rho_{1-\beta}(Q)} -1\right\vert = 0, 
\]
and \emph{$Q$-strongly rate–preserving} if
\[
\lim_{\beta\downarrow 0}\sup_{\substack{\theta\in \Theta,\\\delta\in \Delta}} \left\vert\frac{\rho_{1-\beta}(\mathcal Q_{\beta,\delta}^{(\theta)})}{\rho_{1-\beta}(Q)} - 1\right\vert =0.
\]
\end{definition}
Intuitively, rate preservation means the worst–case tail risk grows at the same asymptotic rate as the true tail risk $\rho_{1-\beta}(Q)$ as $\beta\downarrow 0$ (in the weak version, on the log scale).
The supremum enforces uniformity over tuning choices $(\theta,\delta)$ so that hyperparameter selection does not significantly impact worst-case tail risk estimates.\footnote{Our definition allows $\delta=\delta_\beta$ to depend on $\beta$; for example, $\delta_\beta\to 0$ as $\beta\to 0$. Under additional second–order tail assumptions, such choices can yield convergence rates for $\rho_{1-\beta}(\mathcal Q_{\beta,\delta_\beta}^{(\theta)})\to\rho_{1-\beta}(Q)$, but we focus here on first–order tail asymptotics.}

Observe that strong rate preservation implies weak rate preservation; the converse can fail due to slowly varying terms.
The results from the previous sections imply the following for $Q_\beta=Q$ and fixed $\delta$: (i) $\mathcal W_{p,\delta}(Q)$ ($\gamma>p$ if $\bar F_Q\in \RV(-\gamma)$) is not even weakly rate-preserving under either heavy or Weibull/Gumbel–type $Q$ for the class of risk measures that satisfy Assumption~\ref{assume:tail_risk_measures} (the additive $(\delta/\beta)^{1/p}$ term dominates for small $\beta$); 
(ii) for $\phi$–divergences with $\phi\in\RV(p)$, the worst-case log-slope changes to $p/(p-1)$ if $Q$ has heavy tails and the strong rate-preserving property fails, while for Weibull-type $Q$ a worst-case inflation factor $(p/(p-1))^{1/\gamma}$ arises so strong rate preservation fails but weak rate preservation holds.
Our notion of rate preservation is intentionally first-order (log or level scale); we do not pursue second-order refinements here.

A sufficient condition for ambiguity sets $\mathcal Q_{\beta,\delta}^{(\theta)}$ to be $Q-$weakly (strongly) rate preserving can be derived by comparing the tail of the worst case distribution with that of $Q$. This is the content of Proposition~\ref{prop:rate_preserving_suffcient} below.  
\begin{proposition}[\textbf{Sufficient Condition for Rate Preserving Ambiguity sets}]\label{prop:rate_preserving_suffcient}
    Let the family $\{\mathcal Q_{\beta,\delta}^{(\theta)}:(\theta,\delta) \in \Theta\times \Delta\}$ be  as in Definition~\ref{def:Scale_preserving} and let $\rho_{1-\beta}$ satisfy Assumption~\ref{assume:tail_risk_measures}. Then if $\bar F_{Q}\in \RV(-\gamma)$ with $\kappa>\frac{1}{\gamma}- 1$, $\mathcal Q_{\beta,\delta}^{(\theta)}$ is $Q-$weakly rate preserving so long as 
    \begin{equation}\label{eqn:suff_cond}
        \liminf_{\beta\to 0} \inf_{\substack{s\in(0,\beta], \\\theta\in\Theta}} \frac{\log v_{1-s}(Q_\beta^{(\theta)})}{\log v_{1-s}(Q)} \geq 1\quad \text{ and }\quad \limsup_{\beta\to 0} \sup_{\substack{s\in(0,\beta], \\\theta\in \Theta,\delta\in \Delta}} \frac{\log v_{1-s}(\mathcal Q_{\beta,\delta}^{(\theta)})}{\log v_{1-s}(Q)} \leq 1.
    \end{equation}
 If instead $\Lambda_Q\in \RV(\gamma)$, then $\mathcal Q_{\beta,\delta}^{(\theta)}$ is $Q$-strongly rate preserving if 
 for every $q>1$, 
 \begin{equation}\label{eqn:suff_cond_lt}
        \liminf_{\beta\to 0} \inf_{\substack{s\in(\beta^{q},\beta]\\ \theta\in \Theta}}  \frac{v_{1-s}(Q_{\beta}^{(\theta)})}{v_{1-s}(Q)} \geq 1 \quad \text{ and }\quad \limsup_{\beta\to 0} \sup_{\substack{s\in(\beta^{q},\beta]\\ \theta\in \Theta,\delta\in \Delta}} \frac{ v_{1-s}(\mathcal Q_{\beta,\delta}^{(\theta)})}{v_{1-s}(Q)} \leq 1, 
    \end{equation}
    and in addition the upper bound in \eqref{eqn:suff_cond} holds uniformly over $(\theta,\delta)$ on $s\in (0,\beta^q]$.
\end{proposition}
The conditions in Proposition~\ref{prop:rate_preserving_suffcient} formalize our design principles for  constructing tail-aligning DRO formulations: The lower bounds in
\eqref{eqn:suff_cond}-\eqref{eqn:suff_cond_lt} involve only the nominal distributions
$Q_\beta^{(\theta)}$ and ensure these track the true tail. Therefore, these are independent of  $\delta$. The upper
bounds, taken \emph{uniformly} in $(\theta,\delta)\in\Theta\times\Delta$, control
 inflation due to the worst-case evaluation. Together these conditions imply that, uniformly over tuning choices,
the family $\{\mathcal Q_{\beta,\delta}^{(\theta)}\}$ is $Q$-weakly (resp.\ strongly)
rate-preserving: the worst-case tail risk grows at the same asymptotic
rate as under $Q$.

When $\bar F_Q\in \RV(-\gamma)$, the conditions in \eqref{eqn:suff_cond} together with Lemma~\ref{lem:CVaR_regularity}, imply that for small $\beta$, $\log v_{1-\beta}(\mathcal Q_{\beta,\delta}^{(\theta)})$ is uniformly within a $(1+o(1))$ of $v_{1-\beta}(Q)$ over $t\in(0,\beta]$, which is  the weak rate-preservation property. When $\Lambda_Q\in \RV(\gamma)$, 
the main contribution to $\rho_{1-\beta}(\cdot)$ comes from quantiles with $t\in(\beta^q,\beta]$, where $q>1$; the lower bound in \eqref{eqn:suff_cond_lt} controls this leading term, while the upper bounds (on $(\beta^q,\beta]$ and on $(0,\beta^q]$) ensure that the residual probability mass of order $\beta^q$ remains asymptotically negligible. 

Since the condition in
Proposition~\ref{prop:rate_preserving_suffcient} is in terms of worst case quantiles, it is amenable to verification (as we do next) using several well-studied characterizations. 
This provides a clean pathway to designing DRO formulations that lead to representative tail-risk evaluations and are readily implementable using available data samples.

\subsection{Rate-Preserving EVT Design (RPEV–DRO)} We now combine Proposition~\ref{prop:rate_preserving_suffcient}
with EVT asymptotics to obtain a rate-preserving construction.
We refer to this as Rate-Preserving Extreme-Value DRO (RPEV-DRO).

\noindent\textbf{Choice of nominal distribution: }If 
$\bar F_Q\in \RV(-\gamma)$,  $\bar F_Q(tx) \sim x^{-\gamma}\bar F_Q(t)$. If $t_0$ is chosen to be large, but such that there is enough data to accurately infer $\bar F_Q(t_0)$, then conceptually, one may approximate $\bar F_Q(t)$ by $(t/t_0)^{-\gamma}\bar F_Q(t_0)$. 
If  instead $\Lambda_Q \in \RV(\gamma)$, the above observations can be repeated with $\bar F_Q$ replaced by $\Lambda_Q$. Choose $\beta_0=\beta^\theta$ with $\theta\in(0,1)$ and set the nominal distribution
\begin{align}\label{eqn:nominal_dist}
     dQ_{\beta}^{(\theta)}(t) &=  dQ(t) \mv{1}\{t\leq v_{1-\beta_0}(Q)\} + dG_{\beta}(t) \mv{1}\{t>v_{1-\beta_0}(Q)\}, \text{ where }\nonumber  \\
G_{\beta}(t) &= \begin{cases}
    1-\beta_0 \left(\frac{t}{v_{1-\beta_0}(Q)}\right)^{-\gamma} \quad \quad\quad\quad\  \quad\text{ if $\bar F_Q\in \RV(-\gamma)$}\\
1-\exp\left(\log(\beta_0)\left(\frac{t}{v_{1-\beta_0}(Q)}\right)^\gamma\right)      \quad\text{ if } \Lambda_Q\in \RV(\gamma).
    \end{cases} 
\end{align}
Thus $Q_{\beta}^{(\theta)}$ coincides with $Q$ on the data-rich bulk $t< v_{1-\beta_0}(Q)$ and extrapolates the tail beyond $v_{1-\beta_0}(Q)$ using regular variation of $\bar F_Q$ or $\Lambda_Q$\footnote{Let $u = v_{1-\beta_0}(Q)$, and set $\beta_0^+ = \bar F_Q(u).$ If there is a point mass at $u$, then redefine $G_\beta$ such that $G_\beta(u) = 1-\beta_0^+$ (that is change $\beta_0$ to $\beta_0^+$  everywhere) in \eqref{eqn:nominal_dist}.}. Observe that
\begin{align}\label{eqn:nominal_tail}
   \bar F_{Q_\beta^{(\theta)}}(x) &= \int_{x}^{v_{1-\beta_0}(Q)} dQ(t) + \int_{x\lor v_{1-\beta_0}(Q)}^\infty  dG_{\beta}(t)\nonumber\\
   &=\begin{cases}  Q((x,v_{1-\beta_0}(Q)]) + \beta_0\left(\frac{x\lor v_{1-\beta_0}(Q)}{v_{1-\beta_0}(Q)}\right)^{-\gamma} \quad\quad  \text{ if } \bar F_Q\in \RV(-\gamma)\\
   Q((x,v_{1-\beta_0}(Q)]) + \beta_0^{\left(\frac{x\lor v_{1-\beta_0}(Q)}{v_{1-\beta_0}(Q)}\right)^{\gamma}} \quad\quad \quad\quad \text{ if } 
   \Lambda_Q\in \RV(\gamma).
   \end{cases}
\end{align}
We use \eqref{eqn:nominal_tail} to derive Lemma~\ref{lem:lb_suff} which shows that $Q_\beta^{(\theta)}$ satisfies the lower bound in \eqref{eqn:suff_cond}. For $\varepsilon<1/2$ let $\Theta = [\varepsilon, 1-\varepsilon]$.
\begin{lemma}\label{lem:lb_suff}
    Let the tail cdf of $Q_\beta^{(\theta)} $ be given by \eqref{eqn:nominal_tail}
    with $\beta_0  = \beta^\theta$. Then, the family $\{ Q_\beta^{(\theta)} : \theta\in \Theta\}$ satisfies the lower bounds in \eqref{eqn:suff_cond}-\eqref{eqn:suff_cond_lt}.
\end{lemma}

\noindent\textbf{Choice of Divergence Function:} 
Recall that from Proposition~\ref{prop:f_div_contents}, the conservativeness of $\phi-$divergence DROs with polynomially growing $\phi$ occurs because the growth of $\phi(\cdot)$ is
insufficient to exclude distributions with tails heavier than $Q$ from the ambiguity set.
Motivated by this discussion, let us assume that
\begin{equation}\label{eqn:phi_growth}
    \log \circ \ \phi  \in \RV(p) \text{ for some } p>0.
\end{equation}
Setting $\mathcal Q_{\beta,\delta}^{(\theta)} = \Dphi(Q_\beta^{(\theta)})$ with $\phi(\cdot)$ satisfying
\eqref{eqn:phi_growth} 
is sufficient to ensure the quantile upper bound in \eqref{eqn:suff_cond}-\eqref{eqn:suff_cond_lt}. This hence leads to a rate-preserving formulation.

\begin{theorem}[\textbf{Rate-Preserving Behaviour of RPEV-DRO}]\label{thm:scale_preserving_dro} Let the conditions of Proposition~\ref{prop:rate_preserving_suffcient} hold.
  Suppose $\phi(\cdot)$ satisfies Assumption~\ref{assume:f-assume} and the growth condition in \eqref{eqn:phi_growth}. Let  the tail cdf of $Q_\beta^{(\theta)}$ be given by \eqref{eqn:nominal_dist}  where  $\beta_0 =   \beta^\theta$. 
   If $\bar F_Q\in \RV(-\gamma)$ ,  then $\{\Dphi(Q_{\beta}^{(\theta)}) : (\theta,\delta) \in \Theta\times \Delta\}$ is a $Q$-weakly rate-preserving family of ambiguity sets for the risk functions $\rho_{1-\beta}$. 
   If  $\Lambda_Q\in \RV(\gamma)$ then $\{\Dphi(Q_{\beta}^{(\theta)}) : (\theta,\delta) \in \Theta\times \Delta\}$ is $Q$-strongly rate-preserving.
\end{theorem}
We henceforth refer to the DRO formulation resulting from the construction in Theorem~\ref{thm:scale_preserving_dro} as RPEV-DRO.
A key feature RPEV-DRO is the \emph{uniform} control of the worst–case evaluation over both the radius $\delta$ and the calibration exponent $\theta$ (and hence on the intermediate quantile $\beta_0$): the bounds hold simultaneously for all $\delta\in\Delta$ and $\theta\in[\varepsilon,1-\varepsilon]$. Consequently, the formulation is insensitive to fine-tuning and retains rate preservation across a 
range of parameter choices, rather than only at a single calibrated value.

\begin{remark}\label{rem:wc_cvar}\em
    While selecting a divergence function $\phi(\cdot)$ which satisfies \eqref{eqn:phi_growth}, it is important that its convex conjugate $\phi^*$ be analytically tractable (for example, so as to ensure that the worst case CVaR in \eqref{eqn:wc_cvar} can be computed efficiently). This may be achieved for instance
 by setting $\phi^*(s) = (1+s)\ln(1+s)$. An application of \cite{jin2024constructing}, Proposition 2 shows that $\phi^*(s)$ is the convex conjugate of $\phi(x) = e^{x-1} - x$, which satisfies \eqref{eqn:phi_growth}. 
\end{remark}

\begin{illustration}\label{illustrate:tailored_dro}\em
  We compute the worst case CVaR evaluation under three specifications: (i) $\phi(t) = 0.5(t-1)^2$, and nominal distribution $Q_\beta^{(\theta)}$ as in \eqref{eqn:nominal_dist} (dashed blue line), (ii) $\phi(t) = e^{t-1}-t$ and  $Q_\beta^{(\theta)}$ as in \eqref{eqn:nominal_dist} (dashed red line), corresponding to RPEV-DRO and (iii)  $\phi(t) = 0.5(t-1)^2$ and the nominal distribution is normal with mean and variance identical to that of $Q$ (dashed green line).   For reference, we also plot $C_{1-\beta}(Q)$ (solid black line). 
  The data distribution $Q$ in Figures~\ref{fig:tailoed_ht}-\ref{fig:tailored_lt} are identical to those used for Figures~\ref{fig:wass_heavy}-\ref{fig:wass_light} respectively. The parameters $\beta_0$ and $\delta$ are as indicated in the figure.

 Observe the following: first, when the nominal distribution accurately captures the tails of $Q$ (i.e, we use $Q_\beta^{(\theta)}$ as the nominal), the $\chi^2$-divergence DRO significantly overestimates the actual tail-risk. 
 Second,  if the tailored divergence is used to capture distributional ambiguity but the nominal distribution is mis-representative (i.e., we use a Gaussian nominal), the resulting DRO severely underestimates the tail-risk. 
 Contrast this to RPEV-DRO, where the choice of ambiguity set and the nominal distribution ensures that the worst case tail risk is robust to tails, while not being too conservative.

\begin{figure}[htpb]
 \centering
 \begin{subfigure}{0.48\textwidth}
         \centering
         \includegraphics[width=\textwidth]{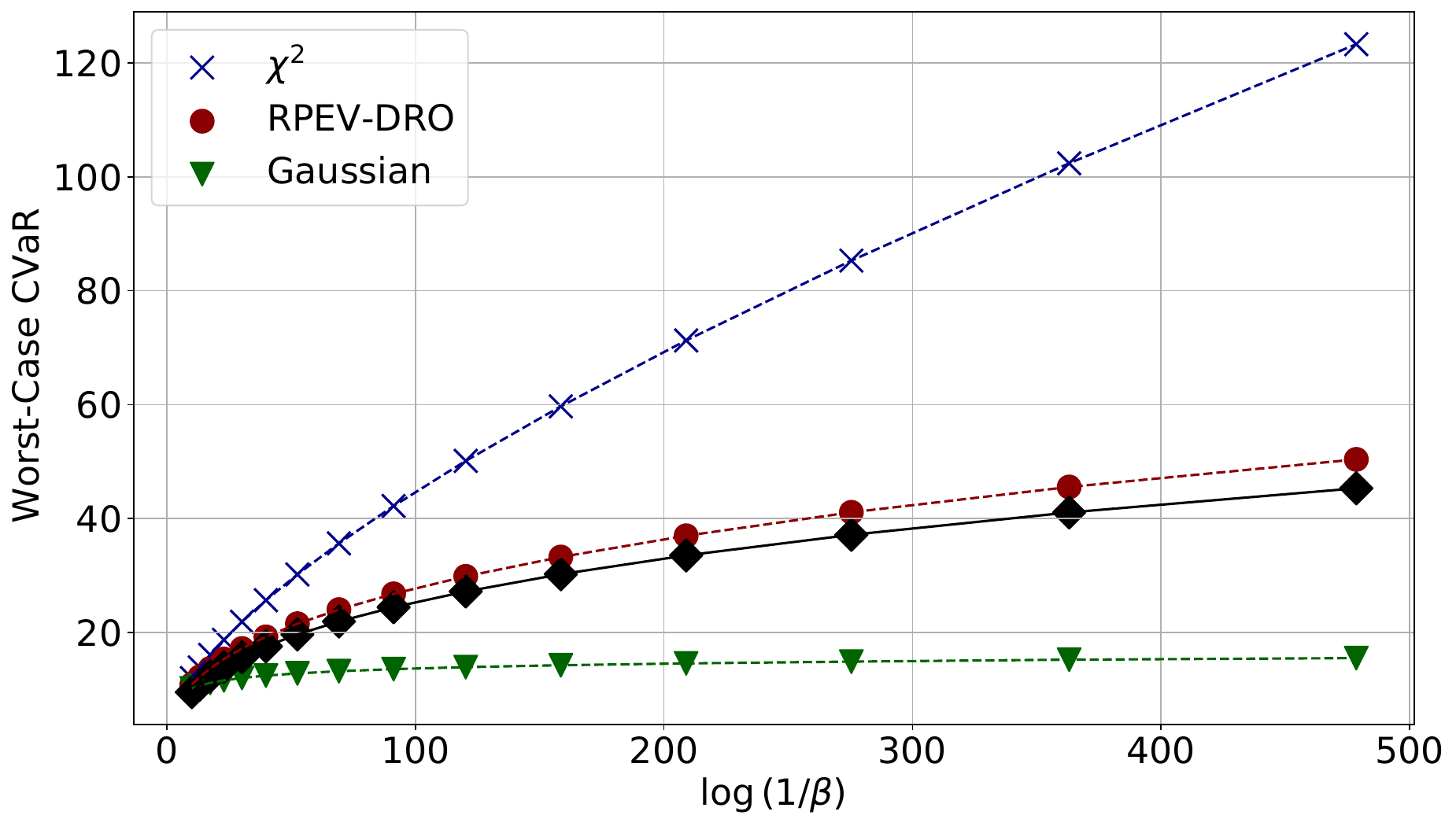}
         \caption{$\bar F_Q\in \RV(-\gamma)$}
         \label{fig:tailoed_ht}
     \end{subfigure}
     \hfill
     \begin{subfigure}{0.48\textwidth}
         \centering
         \includegraphics[width=\textwidth]{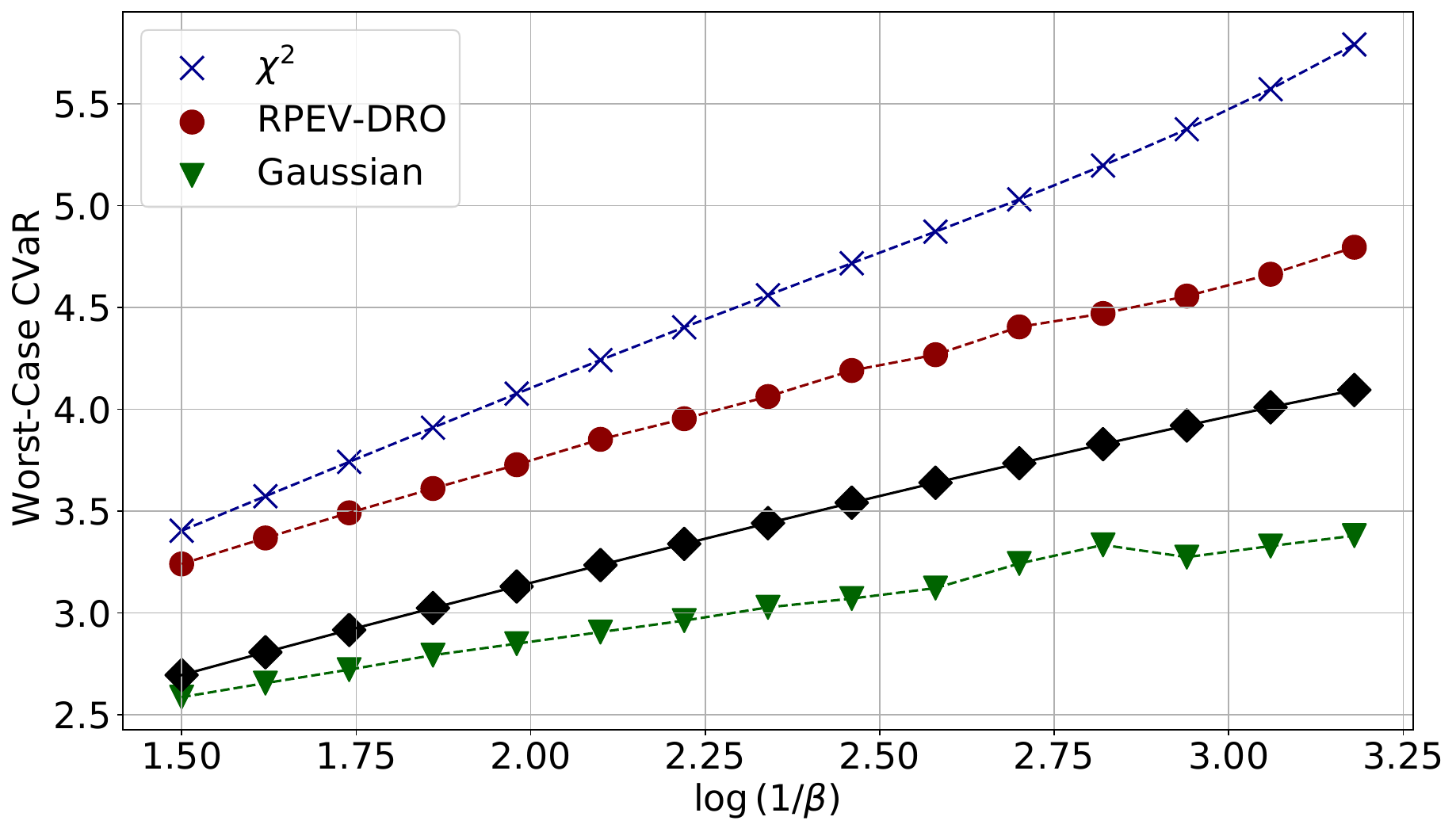}
         \caption{$\Lambda_Q\in \RV(\gamma)$}
         \label{fig:tailored_lt}
     \end{subfigure}
     \caption{ Common parameters: $\delta=0.1$, $\beta_0 =\min\{0.1, \beta^{0.5}\}$. Solid black line is the true CVaR. At $\beta=10^{-2}$, in Figure~\ref{fig:tailoed_ht}, RPEV-DRO is $1.1\, C_{1-\beta}(Q)$, $\chi^2$-DRO with EVT nominal is $1.9\,C_{1-\beta}(Q)$ and the Gaussian nominal $\chi^2$-DRO is $0.3\, C_{1-\beta}(Q)$. Similarly, in Figure~\ref{fig:tailored_lt} RPEV-DRO is $1.2 \,C_{1-\beta}(Q)$, $\chi^2$-DRO with EVT nominal is $1.4\,C_{1-\beta}(Q)$ and the Gaussian nominal $\chi^2$-DRO is $0.7\, C_{1-\beta}(Q)$.}
     \label{fig:WC_evals_proposed_method}
\end{figure}
\end{illustration}

\subsection{Data Driven Implementation} While the nominal distribution in \eqref{eqn:nominal_dist} leads to formulations that are rate-preserving relative to $Q$ (in the sense of Definition~\ref{def:Scale_preserving}), the quantities
$v_{1-\beta_0}(Q)$ and $\gamma$ still need to be estimated from data. 
Suppose $(Z_1,\ldots,Z_n)\sim Q$ are available
data samples 
from which worst-case risk has to be estimated and let $\hat Q^{(n)}$ denote the empirical distribution of these samples. Let $ v_{1-\beta_0}(\hat Q^{(n)}) = Z_{(\lfloor n\beta_0\rfloor)}$, where $Z_{(i)}$ is used henceforth to denote the $i$th largest element of the collection $\{Z_1,\ldots,Z_n\}$. In all the results that follow, we let $k_{n,\theta} = \lfloor n^{1-\theta}\rfloor$.
\begin{lemma}\label{lem:intermidiate_var}
    Suppose that either $\bar F_Q\in \RV(-\gamma)$ or 
    $\Lambda_Q\in \RV(\gamma)$ and let $\beta_0 := \beta_0(n,\theta) = n^{-\theta}$. Then, with $\Theta = [\varepsilon,1-\varepsilon]$ for $\varepsilon<1/2$,
    \[
    \lim_{n\to\infty}\sup_{\theta\in \Theta} \left\vert\frac{ v_{1-\beta_0}(\hat Q^{(n)})}{v_{1-\beta_0}(Q)} - 1 \right \vert =0\quad \text{ almost surely}.
    \]
\end{lemma}
The intuition behind Lemma~\ref{lem:intermidiate_var} is that since $n\beta_0 \to \infty$, there are sufficient samples from which $v_{1-\beta_0}(Q)$ is estimated, even as $\beta_0\to 0$. Therefore the statistical errors associated in estimation of $v_{1-\beta_0}(Q)$  are small. 
Since $\sup_{\theta\in \Theta}\beta_0\to 0$,  irrespective of the choice of hyper-parameter $\theta,$ samples in excess of $v_{1-\beta_0}(\hat Q^{(n)})$ are representative of the distribution tail. We hence 
consider estimates of  $\gamma$ that are obtained from the $k_{n ,\theta}$\footnote{For brevity we denote $k_n:=k_{n,\theta}$. Note that
$k_{n,\theta}=\lfloor n\beta_0\rfloor$ with $\beta_0=n^{-\theta}$.} samples in excess of $v_{1-\beta_0}(\hat Q^{(n)})$. We denote such an estimate generically by $\hat \gamma(k_{n,\theta})$. Define the sampled nominal distribution
\begin{equation}\label{eqn:sampled_measure}
d\hat Q_\beta^{(n,\theta)}(t)  = d\hat Q^{(n)}(t) \mv{1}\left\{t\leq  v_{1-\beta_0}(\hat Q^{(n)})\right\} +d\hat G_\beta^{(n)}(t)  \mv{1}\left\{t> v_{1-\beta_0}(\hat Q^{(n)})\right\},
\end{equation}
where $\hat G_\beta^{(n)}$\footnote{in the definition of $\hat G_\beta^{(n)}$, instead of $\beta_0$, we use $\hat \beta_0 = \frac{k_n-1}{n}$. This ensures that $dQ_\beta^{(\theta)}$ defines a probability measure.} is as in \eqref{eqn:nominal_dist}, but with $\gamma$ and $v_{1-\beta_0}(Q)$ replaced by $\hat\gamma(k_{n,\theta})$ and $v_{1-\beta_0}(\hat Q^{(n)})$, respectively. Under consistency assumptions on $\hat \gamma(k_n)$, the sampled version of RPEV-DRO is almost surely rate-preserving. For $0<c_- <c_+<\infty $ and $M >1 $ define the set $\mathcal B_n = \{\beta(n) : \beta(n) = c_0 n^{-q}: q\in [1,M], c_0\in [c_{-},c_+]\}$.

\begin{theorem}{\textbf{(Uniform Strong Consistency of Data-Driven DRO):}}\label{thm:consistency_of_dro} If $\phi(\cdot)$ satisfies \eqref{eqn:phi_growth} and the Assumptions of Proposition~\ref{prop:rate_preserving_suffcient} hold. Let $\hat\gamma(k_{n,\theta})$ satisfy the following uniform consistency over $\theta:$
\begin{equation}\label{eqn:uc_gamma}
    \lim_{n\to\infty} \sup_{\theta\in \Theta} |\hat \gamma(k_{n,\theta}) -\gamma| = 0 \text{ almost surely}.
\end{equation}
        Suppose the data generating distribution $Q$ satisfies $\bar F_Q\in \RV(-\gamma)$. Then, 
        \[
        \lim_{n\to\infty}\sup_{\substack{\delta\in \Delta,\theta\in \Theta\\ \beta(n)\in\mathcal B_n}}\left\vert\frac{\log\rho_{1-\beta}(\Dphi (\hat Q_{\beta}^{(n,\theta)}))}{\log \rho_{1-\beta}(Q)} -1 \right\vert  = 0 \quad \text{ almost surely.}  
        \]
        If instead $\Lambda_Q \in \RV(\gamma)$, then 
        \[
       \lim_{n\to\infty}\sup_{\substack{\delta\in \Delta, \theta\in \Theta,\\  \beta(n)\in\mathcal B_n}}\left\vert\frac{\rho_{1-\beta}(\Dphi (\hat Q_{\beta}^{(n,\theta)}))}{ \rho_{1-\beta}(Q)} -1 \right\vert  =0  \quad \text{ almost surely.}  
        \]
\end{theorem}
Theorem~\ref{thm:consistency_of_dro} shows that, as long as the tail parameter is estimated consistently (see below), the resulting DRO evaluation remains asymptotically valid uniformly over the tuning set. Consequently, in implementation one may choose the operational parameters by a grid search over this set, without requiring additional optimization, which simplifies deployment.

\noindent \textbf{Estimation of $\gamma$:} When $\bar F_Q\in \RV(-\gamma)$,  the reciprocal of the Hill estimator (\cite{deHaan}, Chapter 3) can be used to estimate $\gamma$:
\begin{equation}\label{eqn:hill_est}
    \hat \gamma(k_{n})  = \left(\frac{1}{k_{n}} \sum_{i=1}^{k_n}\log \left(\frac{ Z_{(i)}}{ Z_{(k_n+1)}}\right)\right)^{-1}. 
\end{equation}
 In case $\Lambda_Q\in \RV(\gamma)$, we estimate $\gamma $ as follows: set $k_{1,n} = \lfloor n\beta_1(n) \rfloor$ where $\beta_1(n)= [\beta_0(n)]^{\kappa_1}$ for a fixed $\kappa_1\in(0,1)$, and define
\begin{equation}\label{eqn:lambda_lt}
     \hat \gamma(k_n) = \frac{\log (1/\kappa_1)}{\log \left(\frac{Z_{(k_n)}}{Z_{(k_{1,n})}}\right)}.
\end{equation}
The lemma below demonstrates that these tail parameter estimates lead to the strong consistency in Theorem~\ref{thm:consistency_of_dro}. 
\begin{lemma}\label{lem:consistency_of_our_estimates}
    The estimators in \eqref{eqn:hill_est} and  \eqref{eqn:lambda_lt} satisfy the uniform consistency in \eqref{eqn:uc_gamma}. 
\end{lemma}

While sample average approximation (SAA) does not rely on any parametric assumptions to define the nominal distribution, in order to 
estimate $\rho_{1-\beta}(Q)$ to within a relative error of $\varepsilon$ with at least $(1-p) \times 100\%$ confidence, order $\beta^{-1}\varepsilon^{-2}p^{-1} $ samples are required. Therefore  
SAA has a vanishing relative error only when $\beta(n) \in \omega (1/n)$. 
On the other hand, RPEV-DRO is able to  produce an approximation which is representative of the true risk even for $\beta(n) \in O(1/n)$, while only requiring under standard assumptions on tail regularity (Assumptions~\ref{assume:heavy_tailed_data}-\ref{assume:weibullian_tails}).

\noindent \textbf{Monte-Carlo Implementation:} From the duality in \eqref{eqn:phi_div_dual},
\[
C_{1-\beta}(\Dphi(\hat Q_{\beta}^{(n,\theta)})) = \inf_{u,\lambda\geq 0,\eta} \left\{u+ \beta^{-1} \left( \eta+ \delta\lambda+  \lambda E_{\hat Q_{\beta}^{(n,\theta)}}\left[\phi^*\left(\frac{(Z- u)^+-\eta}{\lambda}\right)\right)\right]\right\}.
\]
Algorithm~\ref{alg:robust_cvar_estimation} provides a sample-average solution to the above to provide a distributionally robust CVaR estimate using RPEV-DRO.
Two implementation facts: (i) the objective is \emph{jointly convex} in $(u,\eta,\lambda)$ with $\lambda\ge 0$ and 
(ii) the convex conjugate $\phi^*$ enforces its \emph{own domain} taking a value of $+\infty$ outside it  (in our case  $\phi(t)=e^{t-1}-t$, $\phi^*(y)=(1+y)\log(1+y)$ for $y>-1$ and $+\infty$ otherwise). 
Consequently, the minimization can be solved reliably with an off-the-shelf convex solver. 
\begin{algorithm}[htbp] 
  \caption{Distributionally Robust CVaR  using RPEV-DRO}\label{alg:robust_cvar_estimation}
  \ \vspace{-2pt}\\
  \textbf{Input: } Intermediate risk level $\beta_0 =n^{-\theta},$ target level $\beta$, data samples
    $Z_1,\ldots,Z_n$
    
      \noindent \textbf{1. Estimate tail index and baseline VaR: }
      
      \noindent Estimate $\hat \gamma_n$ using \eqref{eqn:hill_est} if $\bar F_{Q}\in \RV(-\gamma)$, and \eqref{eqn:lambda_lt} if  $\Lambda_Q\in \RV(\gamma) $. Set $ v_{1-\beta_0}(\hat Q^{(n)}) = Z_{(\lfloor n\beta_0\rfloor)}$.\\

       \noindent \textbf{2. Generate tail samples from nominal distribution: } 

        \noindent Draw samples $U_{1},\ldots U_{N} \sim \text{Unif}\ [1-\beta_0,1]$, i.i.d. For $i\in[N]$ set 
        \begin{equation}\label{eqn:inverse_samples}
            \tilde Z_i = \begin{cases}
         v_{1-\beta_0}(\hat Q^{(n)}) \left(\frac{1-U_i}{\beta_0}\right)^{-1/\hat \gamma_n} \quad\quad\quad\quad \ \ \text{ if $\bar F_{Q}\in \RV(-\gamma)$}\\
         v_{1-\beta_0}(\hat Q^{(n)}) \left(\frac{\log(1-U_i)}{\log(\beta_0)}\right)^{1/\hat \gamma_n} \quad \quad\quad\quad\text{ if $\Lambda_{Q}\in \RV(\gamma)$}.
        \end{cases}
        \end{equation}
\ \\ 
\noindent \textbf{3. Evaluate Robust CVaR: } Return:
       \begin{align}\label{eqn:dro_estimate}
 \hat C_{1-\beta}^{(N)}(\Dphi(\hat{Q}_{\beta}^{(n,\theta)})) = \inf_{u,\lambda \geq 0,\eta } \Bigg\{u+ \beta^{-1}\Bigg(\eta+ \delta \lambda &+ \frac{\lambda}{n}\sum_{i=1}^{n} \phi^*\left(\frac{(Z_{i}-u)^+-\eta}{\lambda}\right)\mv{1}\left\{Z_i\leq v_{1-\beta_0}(\hat Q^{(n)})\right\} \nonumber\\
 &+  \frac{\beta_0\lambda }{N} \sum_{i=1}^N \phi^*\left(\frac{(\tilde Z_{i}-u)^+-\eta}{\lambda}\right)\Bigg)\Bigg\}.          
       \end{align}
    \end{algorithm}
Note also that Algorithm~\ref{alg:robust_cvar_estimation} uses a Monte-Carlo simulation to evaluate the worst case expectation.
\begin{proposition}\label{prop:correctness_of_algo}
For any given problem instance ($n$, $\beta$ and $\beta_0$ fixed),  the output of Algorithm~\ref{alg:robust_cvar_estimation} is consistent, i.e. $\hat C_{1-\beta}^{(N)}(\Dphi(\hat Q_{\beta}^{(n,\theta)})) \to C_{1-\beta}(\Dphi(\hat Q_{\beta}^{(n,\theta)}))$ as $N\to\infty$.
\end{proposition}
Proposition~\ref{prop:correctness_of_algo} demonstrates
that it is possible to approximate $C_{1-\beta}(\Dphi(\hat Q_\beta^{(n,\theta)}))$ to any degree of accuracy. 
To reduce the computational burden, one can use variance reduction techniques such as importance sampling  to solve the stochastic program \eqref{eqn:dro_estimate} (see \cite{deo2021efficient} for one potential implementation).

\section{\textbf{Multivariate Extension}}\label{sec:application}
In this section, we extend the DRO of Section \ref{sec:calib_uncert} to the case where the $\XX$ is vector-valued and the task is to evaluate tail risk of the scalar loss functional 
$Z=L(\XX)$. 
Let $\XX$ be a $\mathbb R_+^d$–valued random vector with law $\tilde Q\in\mathcal M(\mathbb R_+^d)$, and let $L:\mathbb R_+^d\to\mathbb R_+$ be a loss functional. 
Write $Q=\tilde Q\circ L^{-1}$ for the law of $Z:=L(\XX)$. Our objective is to evaluate $\rho_{1-\beta}(Q)$. We make the following assumptions on $L(\cdot)$ and the distribution of $\XX$:

\begin{assumption}[Asymptotic homogeneity of $L$]\label{assume:loss_functional}
There exists $\vartheta>0$ and a continuous function $L^*:\mathbb S_+^{d-1}\to(0,\infty)$ such that as $t\to\infty$,
\[
\sup_{\zz\in K}\left|\frac{L(t\zz)}{t^{\vartheta}}-L^*(\zz)\right|\to0
\]
for every compact $K\subset \mathbb S_+^{d-1}$, where $\mathbb S_+^{d-1}=\{\zz\in\mathbb R_+^d:\|\zz\|=1\}$.
\end{assumption}
Let $\tilde Q$ have marginals $(Q_1,\ldots, Q_d)$ and let $\Lambda_{Q_i}$ denote the hazard function of the $i$th marginal of $\tilde Q$. Further for any $\xx\in \R^d_+$, define $\mv\Lambda(\xx) = (\Lambda_{Q_1}(x_1),\ldots \Lambda_{Q_d}(x_d))$.
\begin{assumption}\label{assume:mrv}
One of the following holds: for some $\alpha>0$,
\begin{itemize}
\item[(i)] \textbf{Multivariate Regular Variation (MRV):} There exists $a_n\in \RV(1/\alpha)$ and a limiting measure $\mu(\cdot)$ such that 
\[
n\tilde Q(a_n^{-1}\XX \in \cdot) \to \mu(\cdot) \quad \ \ 
\text{vaguely on $\R^d_+\setminus \{\mv 0\}$ }\footnote{A sequence of measures $m_n$ converges vaguely to $m$ if for all Borel sets $A\subset \R_+^d\setminus\{\mv 0\}$ with $m(\partial A)=0$, one has $m_n(A)\to m(A)$}.
\]
\item[(ii)] \textbf{Log–Regular Variation (LRV):} Suppose that for each $i\in[d]$, $\Lambda_{Q_i}\in \RV(\alpha)$. There exists a function $I:\mathbb R_+^d\to[0,\infty]$ such that for every Borel $A\subset\mathbb R_+^d$ bounded away from $0$,
\[
\lim_{t\to\infty}t^{-1}\log \mathbb P\!\left(\boldsymbol\Lambda(\XX)\in t\,A\right)
=-\inf_{\zz\in A} I(\zz).
\]
\end{itemize}
\end{assumption}
%

Assumption~\ref{assume:loss_functional} holds for standard scale–type losses, including norms and weighted norms $L(z)=\|Wz\|_p$ ($\vartheta=1$), linear forms $L(z)=c^\top z$ ($\vartheta=1$), and power gauges $L(z)=\|z\|_p^{\,\theta}$ ($\vartheta=\theta$); in each case $L^*$ is $L$ restricted to $\mathbb S^{d-1}_+$. We note that exact homogeneity is not required: for instance, if $L(\zz) = \Phi(\zz)$, where $\Phi(\cdot)$ is a function approximation based on a feature map, then Assumption~\ref{assume:loss_functional} holds. We point the reader to \cite{deo2023achieving} for additional examples.

Assumption~\ref{assume:mrv} provides a natural extension of univariate extremes to the multivariate risk factors (see \cite{resnick2008multivariate,deValk}).  Indeed,
Assumption~\ref{assume:mrv}(i) (MRV) is met by
(i) elliptical distributions with heavy–tailed radial part (e.g., multivariate-$t$), (ii) Marshall-Olkin type multivariate Pareto distributions and (iii) heavy tailed distributions with dependence structure encoded by a graph (see \cite{engelke2020graphical}) among others.
Assumption~\ref{assume:mrv}(ii) (LRV) directly re-states  \cite[Assumption 2]{deo2023achieving} and is satisfied by
(i) elliptical distributions with Weibull-type radial tail (Gaussian: $\alpha=2$; multivariate Laplace: $\alpha=1$), (ii) Multivariate double–exponential ($\alpha=1$) and  (iii) Gaussian/t–copula models with Weibull–type margins.
A common non-example is the lognormal: it lies in the Gumbel MDA but its hazard function is not regularly-varying. Hence it is excluded by Assumption~\ref{assume:mrv}(ii).
See \cite{deo2023achieving} for additional examples. The assumptions above ensure that the scalar loss $Z=L(\XX)$ inherits the same type of tail behavior (regular variation vs log–regular variation), with index rescaled by the homogeneity degree $\vartheta$.

\begin{proposition}\label{prop:mv_cvar}
Under Assumptions~\ref{assume:loss_functional} and \ref{assume:mrv}(i), the tail cdf of $Z=L(\XX)$ satisfies
\[
\bar F_Q(t)=Q\{Z>t\}\ \in\ \RV\!\left(-\frac{\alpha}{\vartheta}\right).
\]
Under Assumptions~\ref{assume:loss_functional} and \ref{assume:mrv}(ii),
\[
\Lambda_Q(t):=-\log Q\{Z>t\}\ \in\ \RV\!\left(\frac{\alpha}{\vartheta}\right).
\]
\end{proposition}
By Proposition~\ref{prop:mv_cvar}, the loss distribution $Q$ satisfies the tail assumptions of 
Section~\ref{sec:calib_uncert} (regular variation or log regular variation with index
$\alpha/\vartheta$), so Theorems~\ref{thm:scale_preserving_dro}–\ref{thm:consistency_of_dro} 
apply directly to give the corollary below.
\begin{corollary}\label{cor:mv_extension}
Let Assumptions~\ref{assume:loss_functional}–\ref{assume:mrv} hold. 
Let $Q$ be the law of $Z=L(\XX)$, and define $Q_\beta$ as in \eqref{eqn:nominal_dist}. 
Then the conclusions of Theorems~\ref{thm:scale_preserving_dro}–\ref{thm:consistency_of_dro} continue to hold in the multivariate setting.
\end{corollary}
Proposition~\ref{prop:mv_cvar} and Corollary~\ref{cor:mv_extension} show that our one-dimensional
results extend to multivariate risk factors $\XX$ by applying them to the loss
$Z=L(\XX)$: all tail assumptions and rate-preserving guarantees are stated in terms of
the loss law $Q=\text{Law}(Z)$, and do not require a nominal model for $\XX$ itself.

\begin{lemma}[\textbf{Lifting the DRO}]\label{lem:lift_to_univariate}
    Let $\rho$ be any law invariant risk measure on $\mathcal M(\R)$ and $\phi(\cdot)$ be convex. Then with $P_Z = P_0\circ L^{-1}$,
    \[
    \sup_{\tilde P \in \Dphi(P_0)}\rho(\tilde P\circ L^{-1}) = \sup_{P\in \Dphi(P_Z)}\rho(P).
    \]
\end{lemma}

\begin{lemma}[\textbf{Existence of a lift}]\label{lem:loss_to_factor_lift}
Let $L:\mathbb R_+^d\to\mathbb R_+$ be Borel and let $Q_\beta\in\mathcal M(\mathbb R_+)$ satisfy
$Q_\beta\!\big(L(\mathbb R_+^d)\big)=1$.
Then there exists $\tilde Q_\beta\in\mathcal M(\mathbb R_+^d)$ with $Q_\beta=\tilde Q_\beta\circ L^{-1}$.
\end{lemma}


The two lemmas above clarify how this loss-based formulation relates to more traditional
factor level $\phi$–divergence DRO. Note that since $\rho$ is law invariant, its evaluation depends only on the distribution of $L(\XX).$ Lemma~\ref{lem:lift_to_univariate} shows that if one
starts from a nominal $P_0$ on $\XX$ and considers a $\phi$-divergence ball
around $P_0$, then evaluating the worst–case loss is equivalent to working with the
corresponding $\phi$–divergence ball around $P_Z := P_0\circ L^{-1}$ on the loss space: both
formulations yield the same worst-case value of any law-invariant risk measure. \footnote{These results  extend  \cite[Proposition 5]{kruse2021toolkit} which applies to the special case where $\rho(Q) = E_Q(Z)$. }.
Lemma~\ref{lem:loss_to_factor_lift} shows conversely that any loss–level nominal $Q_\beta$
supported on $L(\R_+^d)$ is consistent with at least one multivariate nominal
$\tilde Q_\beta$ satisfying $\tilde Q_\beta\circ L^{-1}=Q_\beta$.
Thus our ambiguity sets may be formulated entirely at the loss level without loss of
generality for law-invariant risks, while remaining consistent with factor level
interpretations.\\


\begin{example}[\textbf{Application: Evaluating Risk of Contagion in a Financial Network}]\em \label{eg:distribution_nw}
    Consider a financial system with $K$ firms (agents) and $d$ risky objects (assets) (see \cite{elliott2014financial}). 
When object $j$ suffers a loss $\xi_j$, the portion borne by firm $i$ is
\[
f_i(\xi_j)\;=\;W_{ij}\,\xi_j\,\mv 1\{i\leftrightarrow j\},
\]
where $W_{ij}\in[0,1]$ is the contractual weight and $\mv 1\{i\leftrightarrow j\}$ indicates that firm $i$ is exposed to object $j$. 
Define the direct exposure matrix $A\in\R^{K\times d}$ by
\[
A_{ij}\;:=\;W_{ij}\,\mv 1\{i\leftrightarrow j\},
\qquad
\XX:=(\xi_1,\ldots,\xi_d)^\top.
\]
Thus $A\XX$ is the vector of \emph{direct} portfolio losses. Firms may also hold equity stakes in each other. Let $C\in[0,1]^{K\times K}$ be the cross-holdings matrix with 
\emph{column} sums $\sum_{i=1}^K C_{ij}\le 1$, where $C_{ij}$ is the fraction of firm $j$ owned by firm $i$. 
Let $\hat C:=\mathrm{Diag}(1-\sum_{i=1}^K C_{ij})_{j=1}^K$ be the diagonal matrix of external ownership shares - these are shares owned outside the financial system. 
Assuming that $(I-C)$ is invertible\footnote{This is the case, for example when the columns of $\mv C$ sum up to less than $1,$ that is, there are non-zero external holdings}, the vector of market-value losses is
\[
\mv F(\XX)\;=\;\hat C\,(I-C)^{-1}\,A\,\XX,
\]
A regulator may summarize systemic losses via the $p$-norm
\[
L_p(\zz)\;=\;\|\mv F(\zz)\|_p,\qquad p\in[1,\infty],
\]
capturing different stress notions: $p=1$ aggregates \emph{total} system loss (overall contagion), while $p=\infty$ reports the \emph{largest} single-firm loss. Observe that for any $p$, $L_p(\zz)$ satisfies Assumption~\ref{assume:loss_functional}. 
Let $\tilde Q$ denote the law of $\XX$; then the induced loss distribution is $Q = \tilde Q\circ L_p^{-1}$. 
A regulator who evaluates a tail risk $\rho_{1-\beta}$ for $Z:=L_p(\XX)$ can work entirely at the loss level: we center the ambiguity set at a loss-level nominal $Q_\beta$ calibrated to the tail of $Z$. By Lemma~\ref{lem:lift_to_univariate}, this yields the same worst–case value as any factor-level formulation after push-forward; if desired, a factor nominal exists via a lift $\tilde Q_\beta$ with $\tilde Q_\beta\circ L_p^{-1}=Q_\beta$ (Lemma~\ref{lem:loss_to_factor_lift}). Hence, the rate-preserving guarantees and algorithms from the uni-variate setting apply to $Z=L_p(\XX)$ without further modification.
\end{example}

\section{Simulation Experiments}\label{sec:simulation}
The following procedure is used to compare the performance of RPEV-DRO with other ambiguity sets commonly used in literature: given a target level $\beta$, intermediate level $\beta_0$ (indicated in the respective figures) and number of samples $n$:

\begin{enumerate}
\item[(i)] \textbf{Draw Samples:} Generate i.i.d.\ samples $\{Z_1,\ldots,Z_n\}$ from the data distribution $Q$. 
Compute the worst–case tail risk $\widehat{\rho}_{1-\beta}$ for each ambiguity set considered.
\item[(ii)] \textbf{Replications:} Repeat (i) ${\tt reps}$ times to obtain $\widehat{\rho}_{1-\beta}^{(1)},\ldots,\widehat{\rho}_{1-\beta}^{({\tt reps})}$.
\item[(iii)] \textbf{Compute Ground Truth.} Approximate $\rho_{1-\beta}(Q)$ by a large Monte Carlo run with $N=5\times 10^6$ samples. 

\item[(iv)] \textbf{Compute performance summaries and coverage:} Report the median and interquartile range (IQR) of $\widehat{\rho}_{1-\beta}^{(r)}$ across replications and coverage:
\[
\widehat{\text{cov}}(\beta) = \frac{1}{\tt reps}\sum_{r=1}^{\tt reps} \mv 1\{\rho_{1-\beta}(Q)\le \widehat{\rho}_{1-\beta}^{(r)}\}.
\]
This is the empirical probability that the worst–case risk exceeds the true risk.
\end{enumerate}

For the multivariate experiments, step (i) is modified as follows: we construct $Z_i=L(\XX_i)$, where $\{\XX_1\ldots ,\XX_n\}$ are generated from the distribution of risk factors. 
Our asymptotic theory assumes either a regularly varying tail $\bar F_Q\in\RV(-\gamma)$  or a log-weibull tail with $\Lambda_Q\in \RV(\gamma)$ with finite tail index $\gamma>0$. 
Under standard conditions (see \cite{deHaan}, Theorem 3.2.5), if $\bar F_Q\in \RV(-\gamma)$, for an intermediate sequence $k:=k_n$, $\sqrt{k}(\hat \gamma_n-\gamma) \to N(0,\gamma^2)$ as $n\to\infty $. We use this CLT heuristically to build a diagnostic that distinguishes regularly varying from log–Weibull tails. 
Let $H_0: \gamma \geq M$ and $H_a:\gamma<M $. Given $n$ input samples of data, a confidence level $(1-\alpha)$, and Hill estimator $\hat \gamma_n$ computed using $k$ tail samples, ${\tt REJECT}$ $H_0$ if $\hat \gamma_n < M (1-z_{1-\alpha}/\sqrt{k})\footnote{we plug in $\gamma=M$, the least favorable boundary value $H_0$, which yields a conservative rejection rule.}$ and treat this as evidence consistent with a regularly varying, heavy-tailed loss distribution with tail index $\gamma<M$. If $H_0$ is not rejected, there is  insufficient evidence to suggest a regularly varying tail,
and we therefore use the log-Weibull tail model as our nominal. Parameter choices $(\delta,\beta_0,M,\alpha,{\tt reps})$ are made explicit in the respective experiments.
 
\subsection{Univariate Example}\label{sec:univariate}
Consider first a simple univariate example where $Z$ represents the loss due to an insurance claim. We consider two data-generating models for $Z$ whose tails satisfy (i) $\bar F_{Q}(x) \sim 0.2 x^{-3.4}\log(x)$ and (ii) $\Lambda_Q(x) \sim x^{0.9}\log^{1.8}(1+x)$. Note that the first example has $\bar F_Q\in \RV$ and the second has $\Lambda_Q\in \RV$. 
For each of the above distributions, we first evaluate using $n$ samples of data, the distributionally robust CVaR under the following $\phi-$divergence based ambiguity sets: (i) Gaussian nominal distribution whose mean and variance are evaluated from data with $\phi(t) = 0.5(t-1)^2$, (ii) nominal distribution $\hat Q_{\beta}^{(n)}$ with $\phi(t) = 0.5(t-1)^2$, and (iii)  nominal distribution $\hat Q_{\beta}^{(n)}$ with $\phi(t) = e^{t-1}-t$. The latter corresponds to RPEV-DRO. For fair comparison, we use common random numbers to draw samples for each of the above evaluations.  
We plot  only the band from the median down to the lower quartile (shaded region), to emphasize the extent of potential underestimation of risk; overestimation is less critical in our application  of risk evaluation.
For reference, we also plot the true CVaR (solid black line). Results of the experiment are given in Figure~\ref{fig:data_univaruiate} below.

\begin{figure}[htpb]
 \centering
 \begin{subfigure}{0.48\textwidth}
         \centering
         \includegraphics[width=\textwidth]{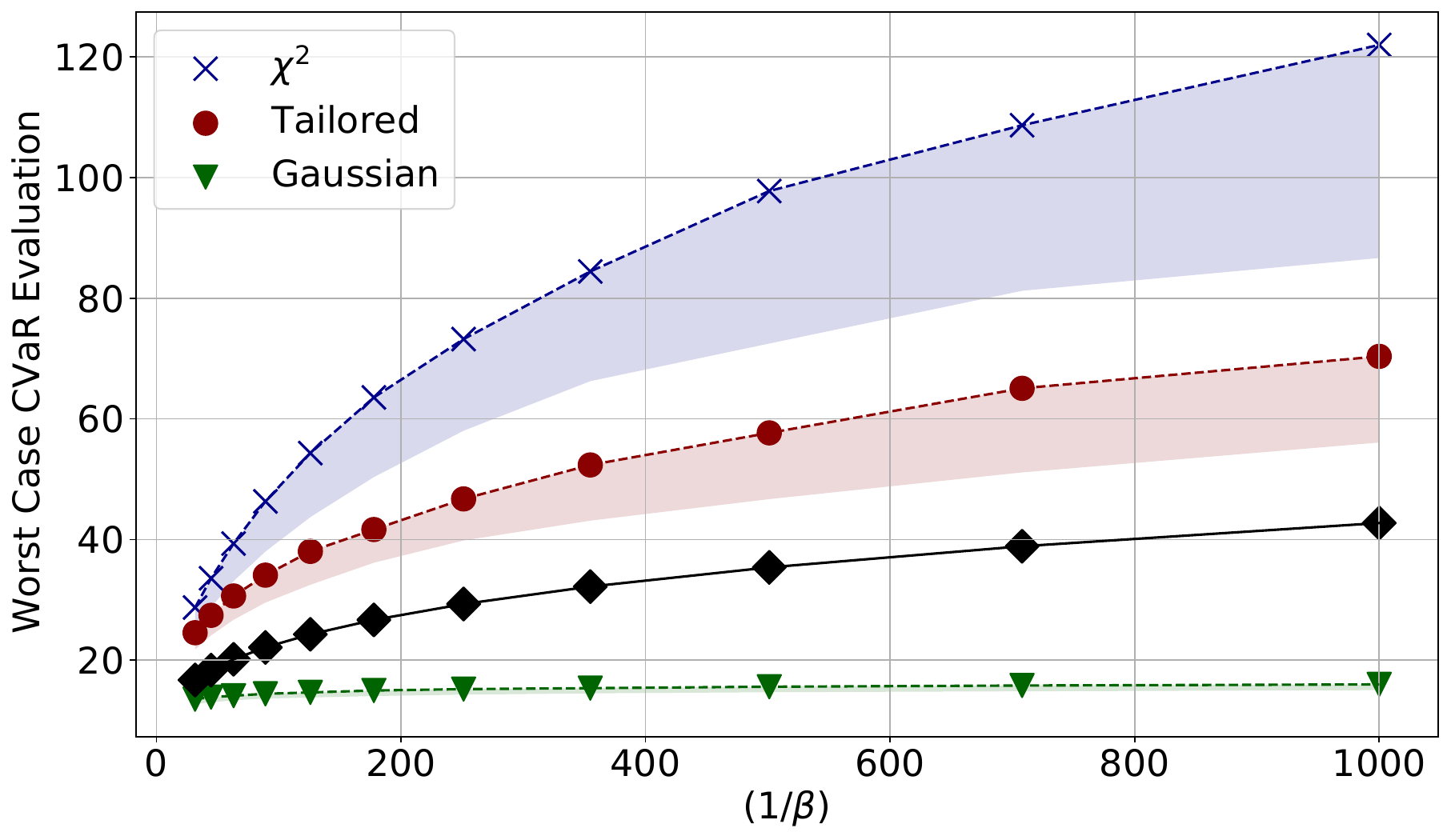}
         \caption{Heavy-Tailed data: $\bar F_Q\in \RV$}
         \label{fig:data_drive_ht}
     \end{subfigure}
     \hfill
     \begin{subfigure}{0.48\textwidth}
         \centering
         \includegraphics[width=\textwidth]{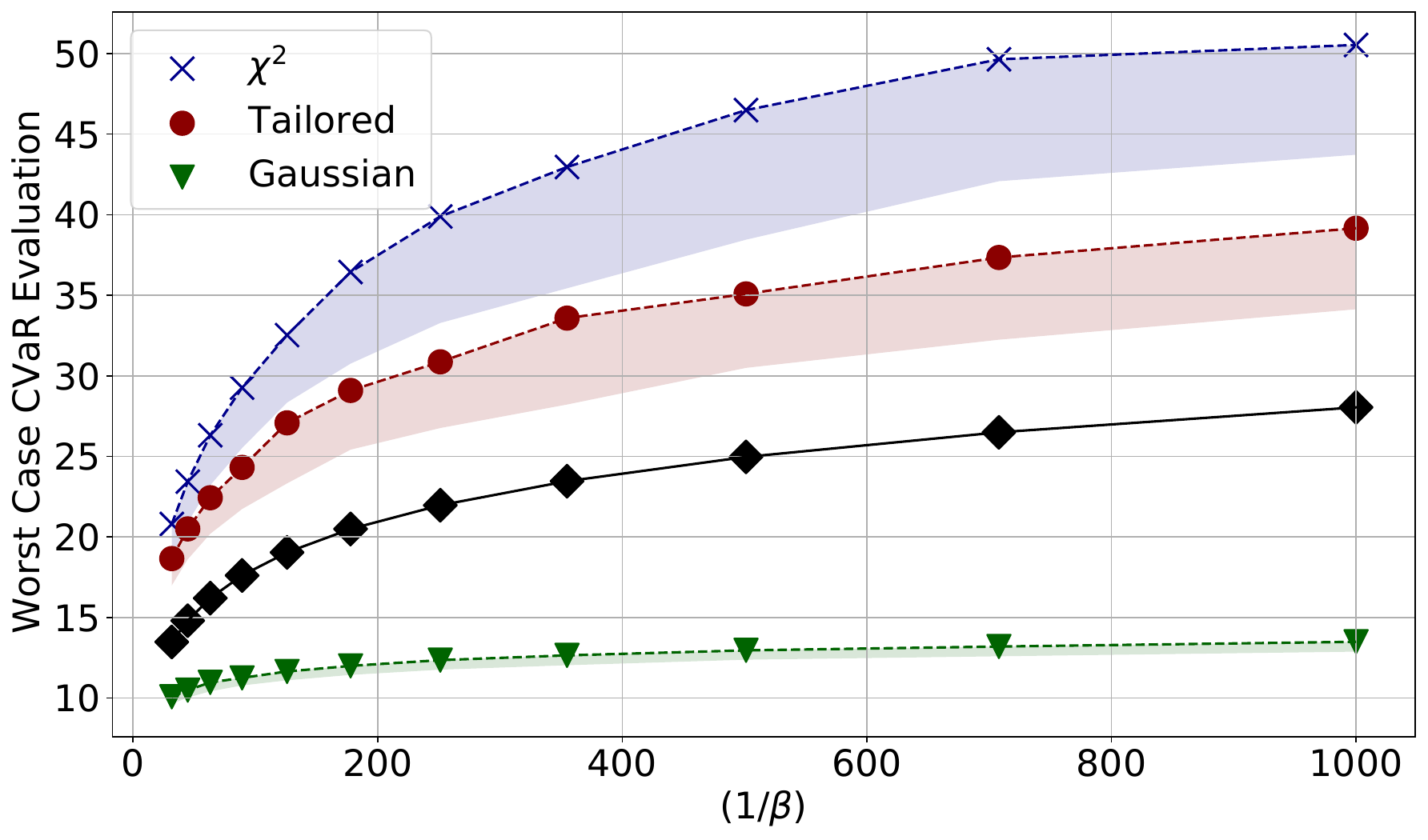}
         \caption{Light Tailed data: $\Lambda_Q\in \RV$}
         \label{fig:data_driven_lt}
     \end{subfigure}
     \caption{Common parameters: $n=500$, $\beta_0 = \min\{0.1, \ \beta^{0.5}\}$, $\delta = 0.1$, ${\tt reps}=100$.  Markers on each graph denote actual risk evaluations.}\label{fig:data_univaruiate}
\end{figure}

We point out a few important observations: in Figure~\ref{fig:data_drive_ht} where data is heavy tailed, the $\chi^2-$divergence DRO ends up overestimating the underlying risk by a significant amount, even when tails of the nominal distribution are representative of those of $Q$. Conversely using a Gaussian distribution leads to severe underestimation of risk even when a fairly generous choice of ambiguity set is used.
In Figure~\ref{fig:data_driven_lt} where data is
light tailed, the $\chi^2$-divergence based DRO is not too conservative; however, using the proposed DRO still ends up cutting the conservativeness by about 50\% as compared to $\chi^2$-DRO. Crucially, irrespective of the tails of $Q$ the coverage for RPEV-DRO exceeds $95\%$ over the entire range of $\beta$ considered indicating under-estimation of risk does not happen often. We point out that the extent of under-estimation of risk is small in case RPEV-DRO is used: indeed, the smallest evaluation of robust risk over $100$ replications only underestimates risk by $25\%$ (versus over $60\%$ for the Gaussian nominal).

\smallskip

\noindent \textbf{Robustness checks:} To study the robustness of our method to the underlying assumptions, we consider the following regime:
\begin{align*}
\text{\textbf{HT base + Lognormal contamination:}}\quad & Z_{1}
:= (1-B)\,H + B\,C \\
\text{\textbf{LT base + Lognormal contamination:}}\quad & Z_{2}
:= (1-B)\,L + B\,C.
\end{align*}
 where $L$ is a random variable with $\Lambda_L(x)  \sim x^{0.9}\log^{0.5}(1+x)$, and $H$ is a random variable with $\bar F_H(x) \sim 0.2 x^{-3.4}\log (x)$, $C$ is a log-normal contamination ($C\sim \exp(N(0,1))$), and $B$ is Bernoulli with success probability $\varepsilon = 0.1$ that is independent of $L$, $H$ and $C$. In this setup, we repeat the procedure in steps (i)-(iv) above to compare the robust risk evaluation with the true risk. 
 Since the Gaussian–$\chi^2$ baseline in Figure~\ref{fig:data_univaruiate}
significantly under-estimates risk, we compare our method to Wasserstein-1 DRO with SAA nominal (green dashed line).
We find that in the setup with HT base + Lognormal contamination, RPEV-DRO
maintains ${\widehat{\text{cov}}}(\beta)$ between $87\%$ and $93\%$ uniformly across $\beta$. For Wasserstein-1 DRO despite
smaller worst-case values,  $\widehat{\text{cov}}(\beta)$ drops to between $60$–$70\%$.
This demonstrates a favorable safety-efficiency trade-off: our method preserves tail safety
(coverage) while avoiding undue risk inflation. 

\begin{figure}[htpb]
 \centering
         \centering
          \begin{subfigure}{0.48\textwidth}
         \centering
         \includegraphics[width=\textwidth]{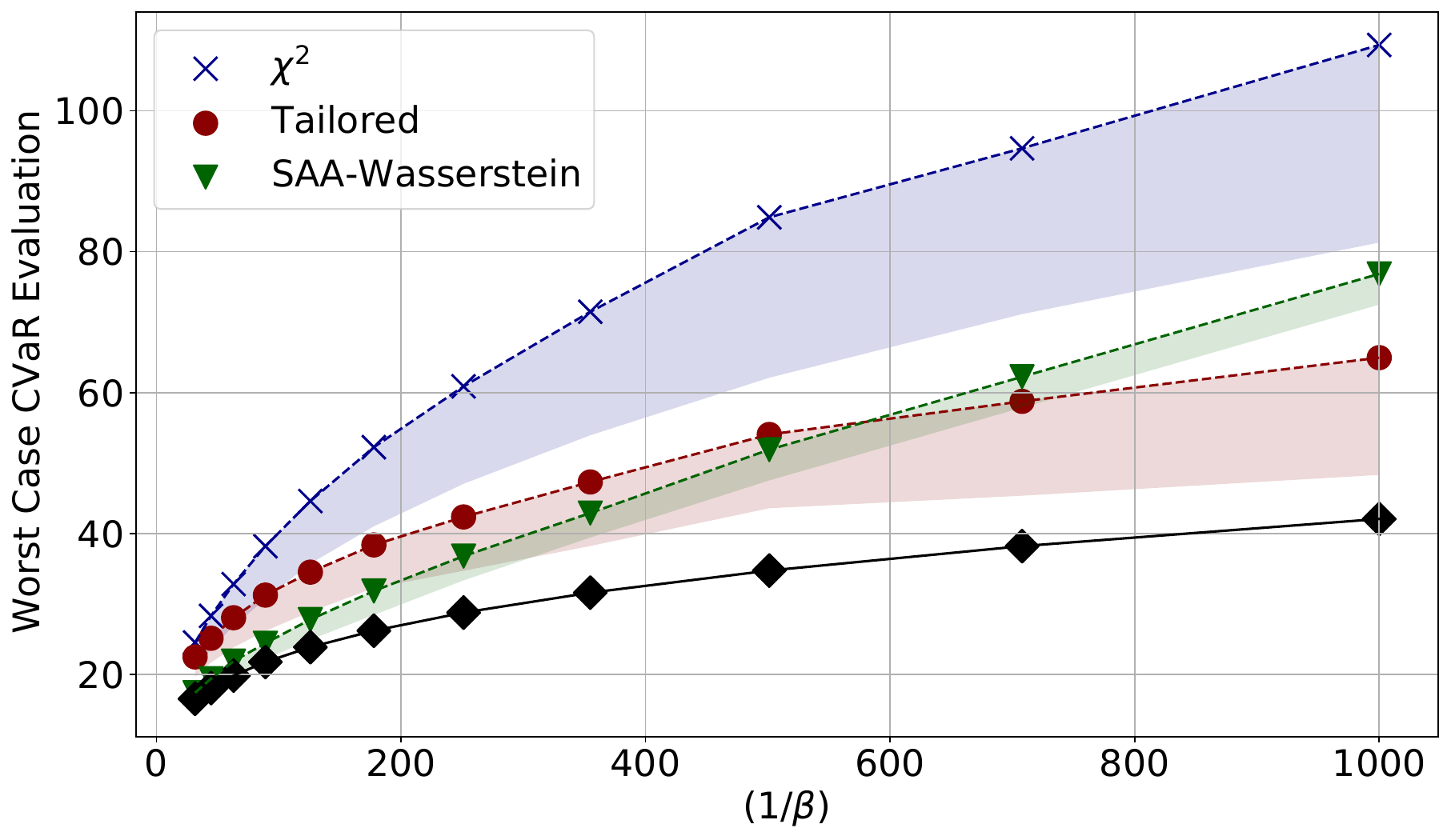}
         \caption{HT base+Lognormal contamination}
         \label{fig:ht_corrupt}
     \end{subfigure}
     \hfill
     \begin{subfigure}{0.48\textwidth}
        \includegraphics[width=\textwidth]{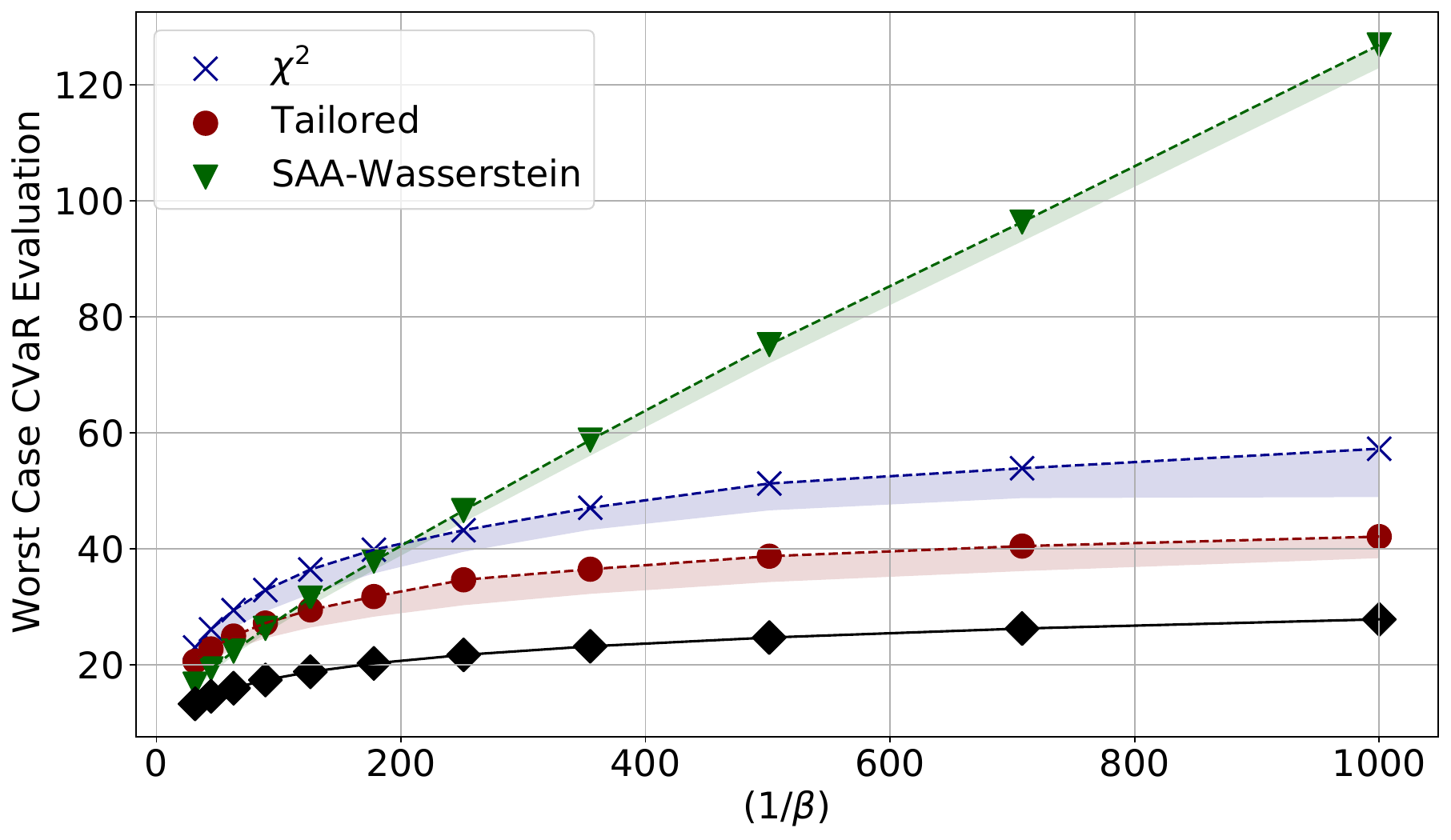}
         \label{fig:lt_corrupt}
          \caption{LT base+Lognormal contamination}
     \end{subfigure}
     \caption{Parameters: For illustration, we choose $\delta=0.05$, $\varepsilon=0.1$, ${\tt reps}=100$. Markers on each graph denote actual risk evaluations. Here $n=500$.}\label{fig:robustness_check}
\end{figure}

\begin{figure}[htpb]
 \centering
         \centering
          \begin{subfigure}{0.96\textwidth}
         \centering
         \includegraphics[width=\textwidth]{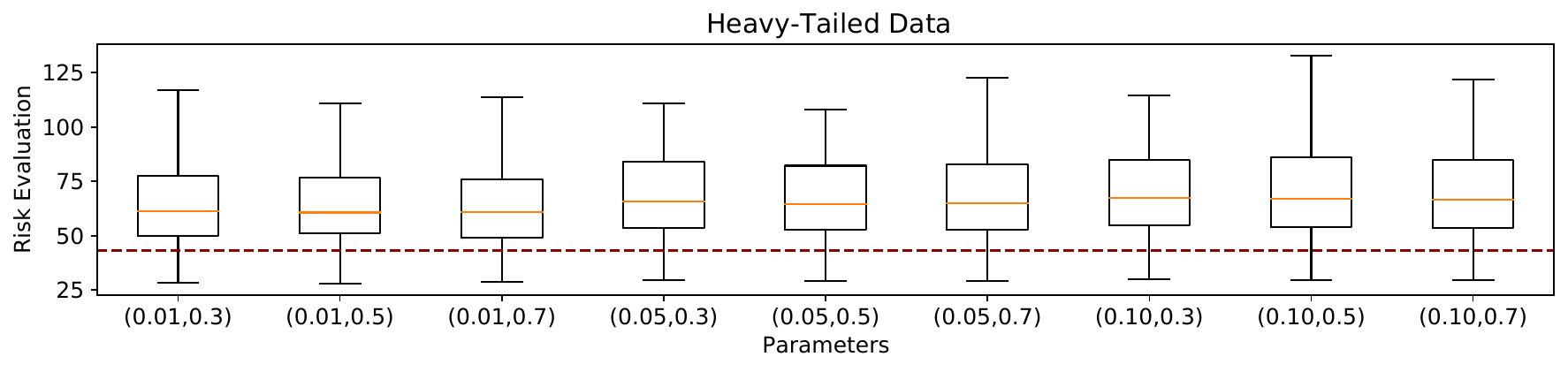}
         \caption{ }
         \label{fig:ht_param_sweep}
     \end{subfigure}
     \hfill
     \begin{subfigure}{0.96\textwidth}
        \includegraphics[width=\textwidth]{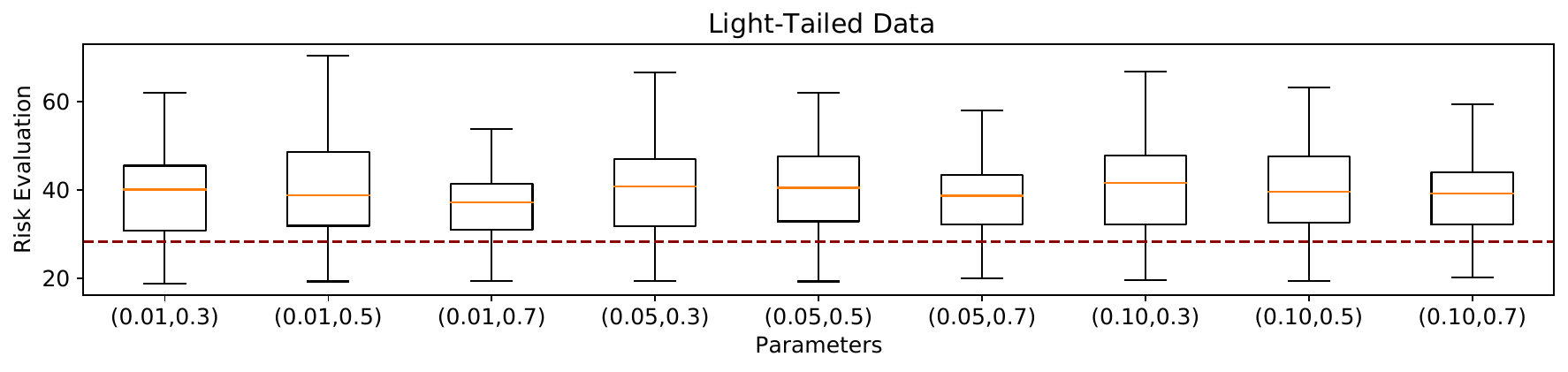}
         \label{fig:lt_para_sweep}
          \caption{ }
     \end{subfigure}
     \caption{${\tt reps}=100$, $\beta=0.001$ and $n=500$. Worst–case evaluations are stable over the grid
  $\delta\in\{0.01,0.05,0.10\}$ and $\theta\in\{0.3,0.5,0.7\}$; minimum empirical coverage over all $(\delta,\theta)$ is $\approx90\%$.}\label{fig:param_sweep}
\end{figure}

We next illustrate numerically that the asymptotic uniformity over $(\delta,\theta)$ is reflected in finite samples.
Figure~\ref{fig:param_sweep} shows box-plots of the worst–case evaluations as we vary
$\delta\in\{0.01,0.05,0.10\}$ and $\theta\in\{0.3,0.5,0.7\}$, where recall that $\beta_0 = n^{-\theta}$ was the nominal level. 
Each box-plot summarizes 100 replications of the RPEV–DRO worst-case CVaR estimate at the indicated $(\delta,\theta)$.
These plots 
are essentially insensitive to parameter choices:
both the medians and IQRs remain stable, and empirical coverage is unaffected (in excess of $90\%$ across all $(\delta,\theta)$ choices).

\subsection{Insurance Risk}\label{sec:Sys_risk}
\begin{figure}[htpb]
 \centering
 \begin{subfigure}{0.32\textwidth}
         \centering
         \includegraphics[width=\textwidth]{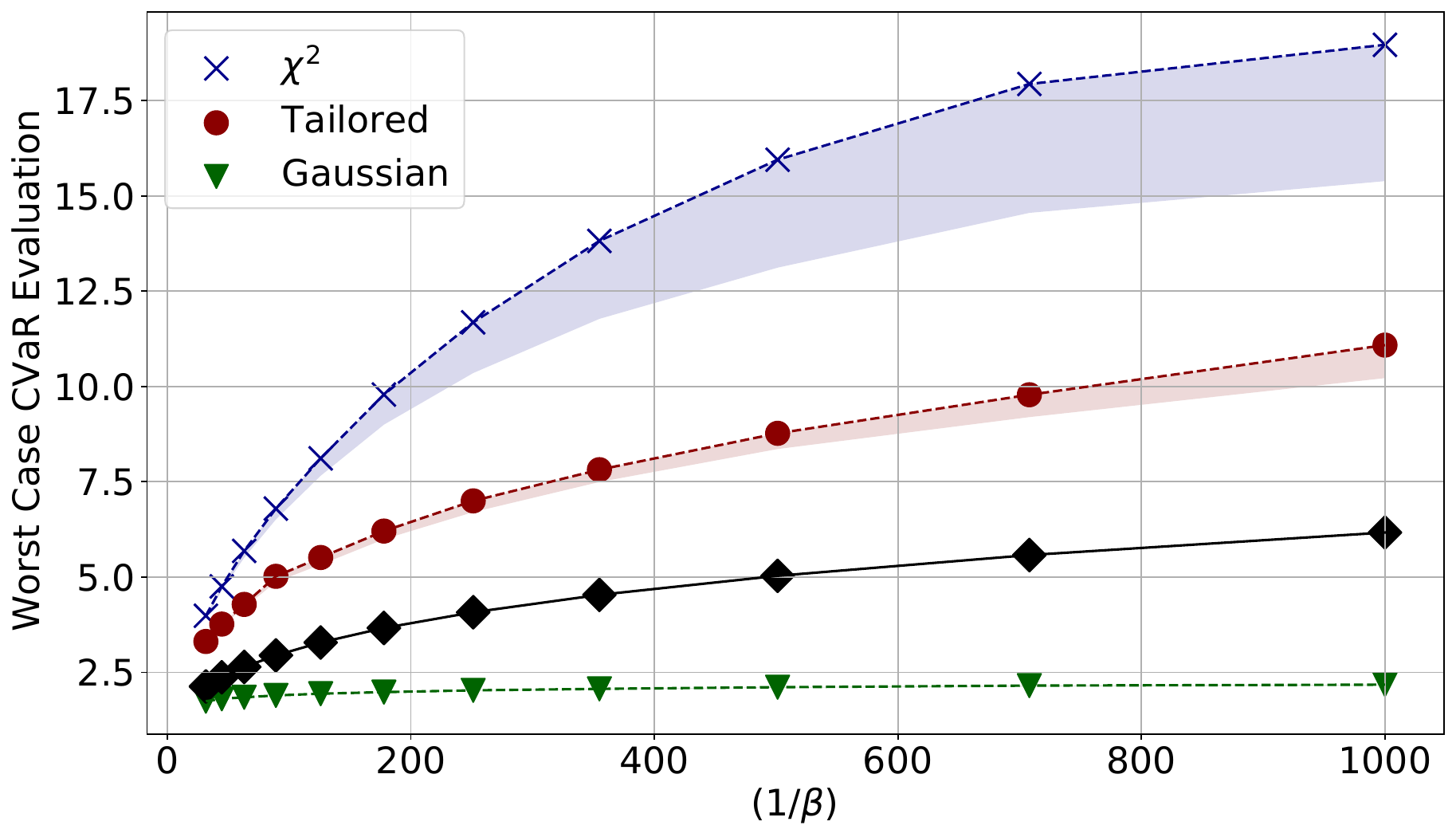}
         \caption{$\lambda=0$, $K=24$}
         \label{fig:lambda_1}
     \end{subfigure}
     \hfill
     \begin{subfigure}{0.32\textwidth}
         \centering
         \includegraphics[width=\textwidth]{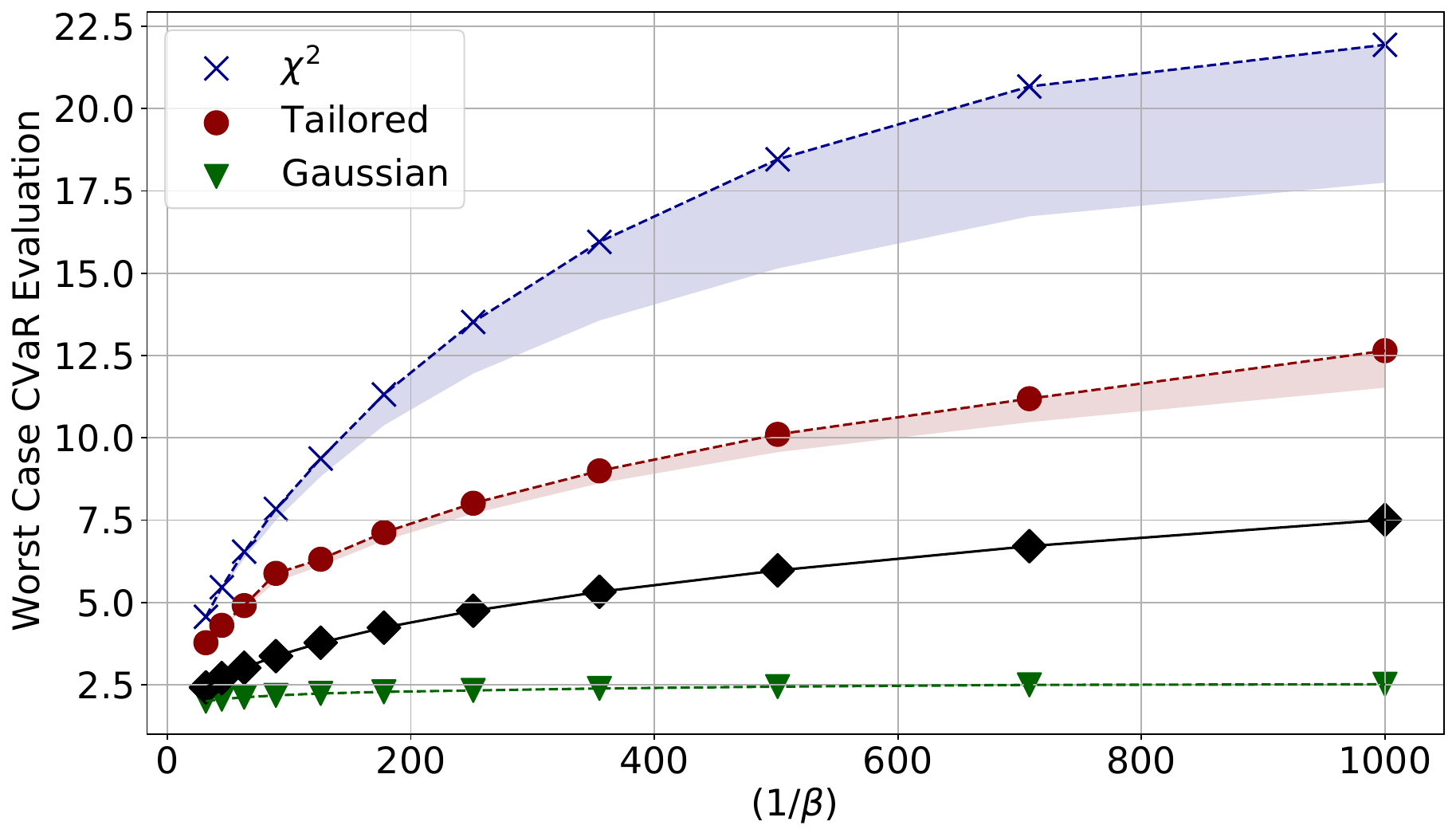}
         \caption{$\lambda=0.5$, $K=6$}
         \label{fig:lambda_2}
     \end{subfigure}
     \begin{subfigure}{0.32\textwidth}
         \centering
         \includegraphics[width=\textwidth]{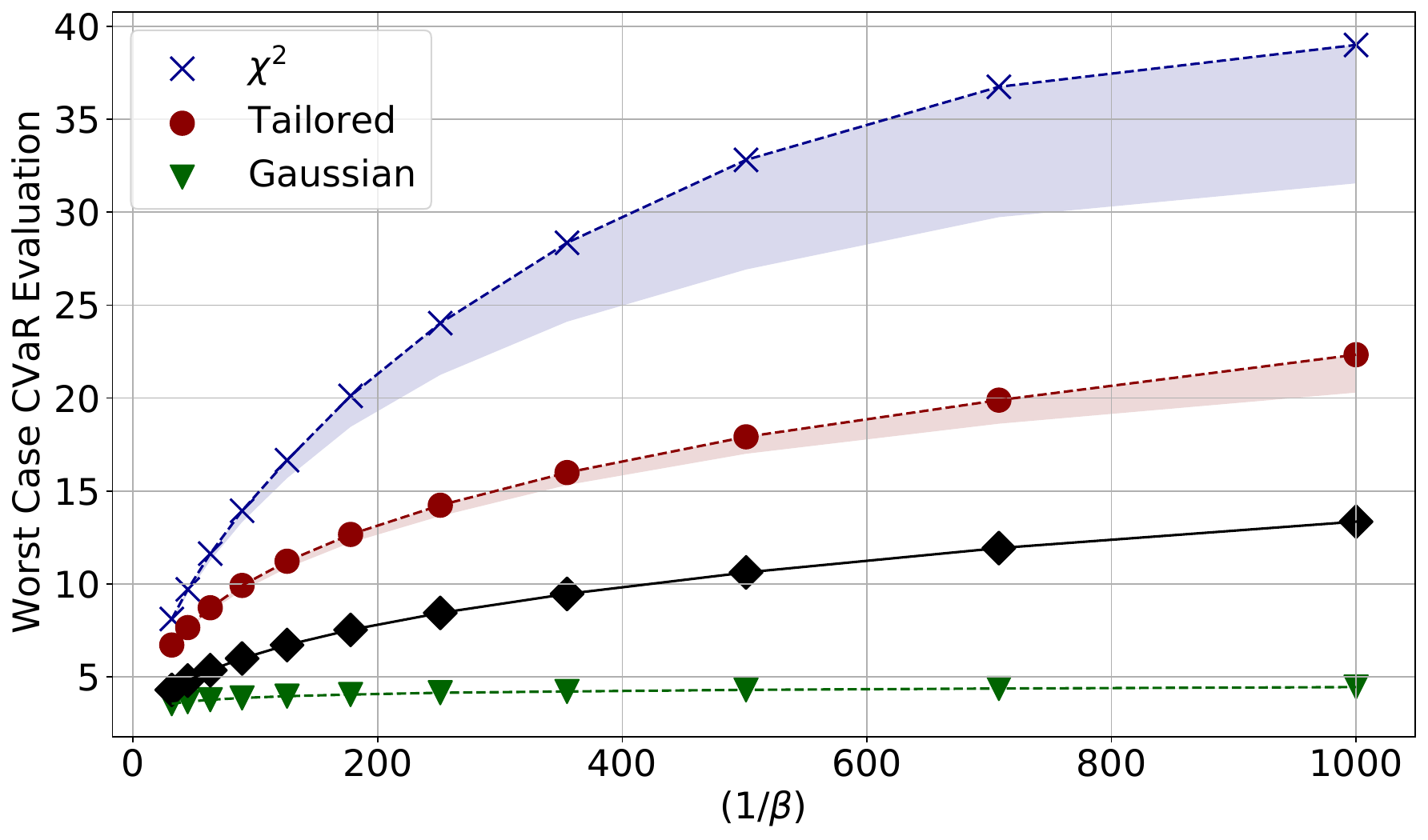}
         \caption{$\lambda=1$, $K=16$}
         \label{fig:lambda_3}
     \end{subfigure}
     \caption{Common parameters: $n=500$, $\delta=0.1$, $p=1$, $\beta_0 = \min\{0.1,\beta^{0.5}\}$ and ${\tt reps}=100$. Markers on each graph denote actual risk evaluations. True CVaR is shown by the solid black line.}\label{fig:systemic_risk}
\end{figure}

We revisit the network from Section~\ref{sec:application} with $d=48$ assets and $K$ firms ($q:=d/K\in\mathbb{N}$). 
The direct exposure matrix interpolates between disjoint and fully shared holdings:
\[
A_\lambda \;=\; (1-\lambda)A_0+\lambda A_1,\qquad \lambda\in[0,1],
\]
where 
\[
(A_0)_{ij}=\mv 1\!\{\,j\in\{(i-1)q+1,\ldots,iq\}\,\},\qquad (A_1)_{ij}=\frac{1}{K}.
\]
We set cross-holdings to zero ($C=0$, hence $\hat C=I$). Asset losses have generalized Pareto marginals and a multivariate $t$ copula with $\nu=4$ degrees of freedom and correlation $I_d$  (note that this \textit{does not} imply independence of asset losses). The systemic loss is
\[
L(\zz)=\frac{\|A_\lambda \zz\|_p}{d},
\]
and we evaluate worst–case $\mathrm{CVaR}_{1-\beta}$ across methods while varying $(\lambda,K)$, so as the capture various extents of network connectivity. As in the uni-variate example, the proposed formulation is simultaneously less conservative than $\chi^2-$divergence, while not underestimating risk like when Gaussian nominal distribution is used; see Figure~\ref{fig:systemic_risk} for results. We point out that RPEV-DRO gives $\widehat{\text{cov}}(\beta)$ of between $90\%$ and $95\%$ over every value of $\beta$ considered in this experiment.

\subsection{Discrete Delta Hedging} Consider a single stock $\{S_t: t\in[0,1]\}$ whose price follows a geometric Brownian motion with mean $\mu$ and volatility $\sigma^2$, and  a call option whose terminal
payoff  is $C(S_1) = (S_1-K)^+$. 
Suppose a portfolio aims to replicate the option payoff. We divide the overall time window into $m$ parts and re-balance the portfolio at the start of each time period.
At time step $t_i = i/m$, the total value of the portfolio is the sum of cash investments, ${\tt cash}(t_i)$ plus the stock value, ${\tt stock}(t_i)$. The number of shares of $S$ held between $[t_i,t_{i+1})$ is denoted by $\delta_{t_i}$, and is computed using the Black-Scholes formula. The stock and cash values are given  by the recursion 
\[
{\tt stock}(t_i)=\delta_{t_i}S_{t_i},\qquad 
{\tt cash}(t_i)={\tt cash}(t_{i-1})e^{r/m}-S_{t_i}(\delta_{t_i}-\delta_{t_{i-1}})-k_1\,S_{t_i}\,|\delta_{t_i}-\delta_{t_{i-1}}|,
\]
where $r$ is the risk-free interest rate, $k_1$ is the cost to transact per share.
The initial stock and cash, call them ${\tt stock}(0)$ and ${\tt cash}(0)$ are set to that portfolio which determines the option price when transaction costs are set to $0$.
The random variable of interest for us is the hedging error, or the difference between the option payoff and portfolio value at time $t=1$,
\[
X_{\tt err} = \left\vert(S_1-K)^+ - {\tt cash}(1) - {\tt stock}(1)\right\vert.
\]
We test the performance of the proposed DRO in hedging against model uncertainty in $X_{\tt err}$. To this end, for each $m$ we compute the CVaR of $X_{\tt err}$ at $\beta=0.01$ using a Monte-Carlo simulation with $10^6$ samples, and compare it with worst-case evaluation under the ambiguity sets from Section~\ref{sec:univariate}. In this example we treat the hedging error $X_{\tt err}$ purely as a black–box loss: the data–driven procedure from Section~\ref{sec:univariate} (Hill diagnostic and EVT-based nominal choice) is applied without analytically characterizing its domain of attraction.

\begin{figure}[htpb]
 \centering
 \begin{subfigure}{0.48\textwidth}
         \centering
         \includegraphics[width=\textwidth]{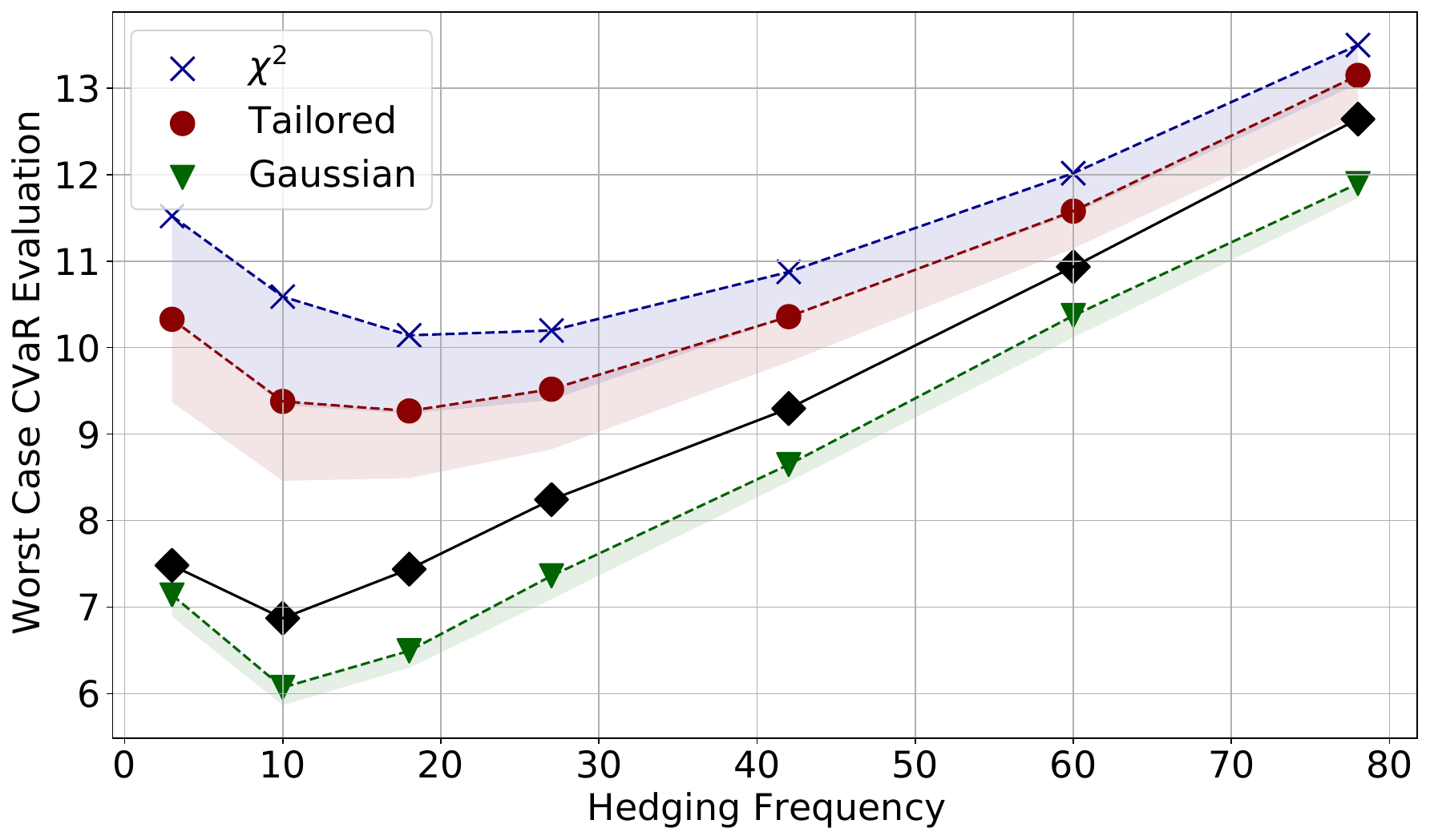}
         \caption{$\sigma^2=0.075$ $k_1=0.0025$}
         \label{fig:delta_1}
     \end{subfigure}
     \hfill
     \begin{subfigure}{0.48\textwidth}
         \centering
         \includegraphics[width=\textwidth]{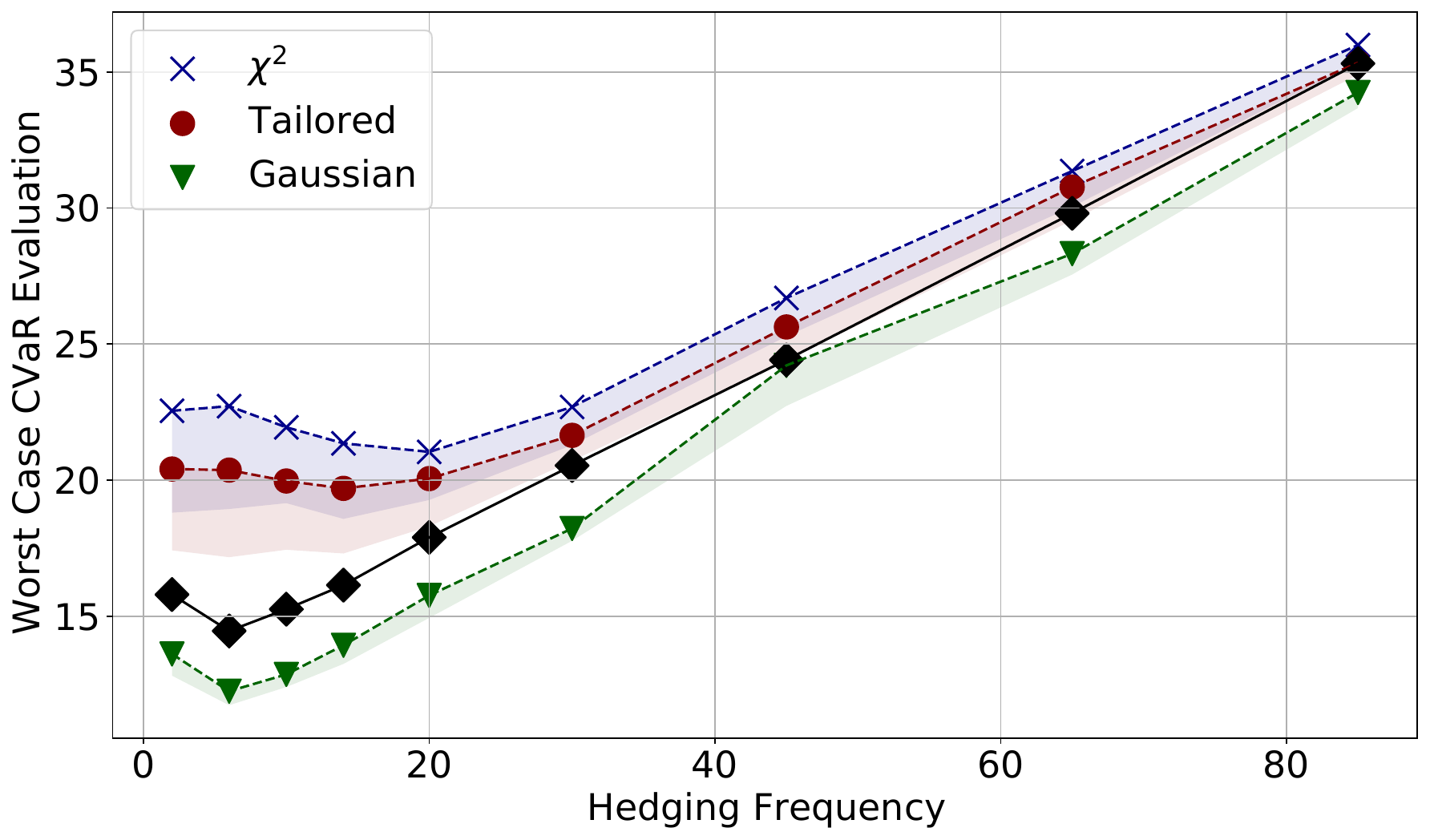}
         \caption{$\sigma^2=0.15$,$k_1=0.001$}
         \label{fig:delta_2}
     \end{subfigure}
     \caption{Common Parameters: $n=200$, $\beta_0=0.1$, $S_0=K=25$, $\delta=0.05$, $r=10\%$. Markers on the graph denote the actual value of risk evaluated using the respective DRO. The solid black line shows the true CVaR or $X_{\tt err}$. }\label{fig:delta_heedge}
\end{figure}

Figure~\ref{fig:delta_heedge} shows the characteristic U-shape of $\mathrm{CVaR}_{1-\beta}(X_{\tt err})$ versus $m$:
discretization error decreases with more frequent re-balancing, while transaction costs increase. 
Our DRO formulation closely tracks the truth across $m$ and selects a near-optimal frequency. 
In both parameter configurations in Figure~\ref{fig:delta_heedge}, the minimizer $\hat m$ of the RPEV–DRO worst–case evaluation satisfies the empirical guarantee 
$\mathrm{CVaR}_{1-\beta}(X_{\tt err};\hat m)\le 1.10\times \min_m \mathrm{CVaR}_{1-\beta}(X_{\tt err};m)$. This suggests that RPEV-DRO gives better estimates of both the CVaR, as well as a frequency of hedging that leads to smaller tail risk due to discretisation.


\section{Studies on Real Datasets}\label{sec:real_data}
We conclude by demonstrating the efficacy of the proposed methods in evaluation of distributionally robust CVaR for real datasets. 
Here it is not possible to simulate the data distribution. 
Therefore, obtaining replications of the worst-case risk evaluations as in step (ii) in Section~\ref{sec:simulation} is challenging. 
To circumvent this difficulty somewhat, we instead construct "rolling" windows of data from available samples which serve to act as proxies for data replications. Note that these windows need not be independent (for e.g. in case of overlap), and we use them as a stability diagnostic rather than a formal sampling distribution. Let $N$ be the total data samples available and let $\tt grid$ denote a set of values of $\beta$ over which risk is computed. Then the procedure below is used to compare performance of various DRO formulations.  
\begin{enumerate}
    \item[(i)]  For each $k=1,2,\ldots, {\tt reps}$, take as input data $Z_{sk+1},\ldots, Z_{sk+n}$. 
    Solve the robust CVaR problem for each $\beta\in \tt grid$ and number of samples fixed. 
    \item [(ii)] Compute the full-sample CVaR of the data for each $\beta\in {\tt grid}$ using $N$ samples of data. This serves as our benchmark for comparison.
    \item[(iii)] Compare the median and lower quartile of the risk evaluations computed over the ${\tt reps}$ input datasets constructed in step (i). 
\end{enumerate}
\subsection{Fire-Insurance Dataset} \label{sec:fire_insurance}
\begin{figure}[htpb]
 \centering
     \begin{subfigure}{0.48\textwidth}
         \centering
         \includegraphics[width=\textwidth]{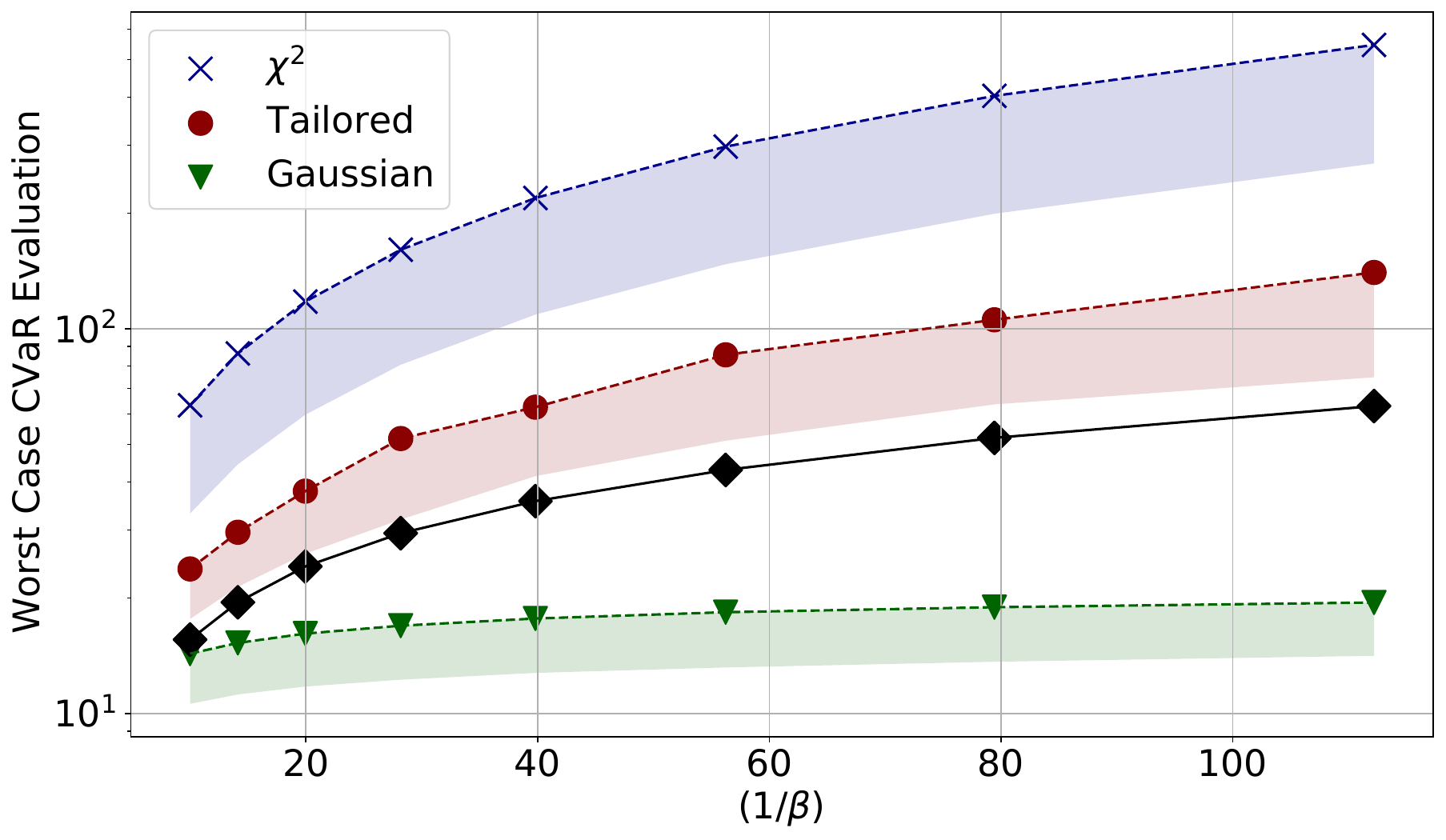}
         \caption{Comparison of Worst Case Evaluations}
         \label{fig:fire_main}
     \end{subfigure}
     \begin{subfigure}{0.48\textwidth}
         \centering
         \includegraphics[width=\textwidth]{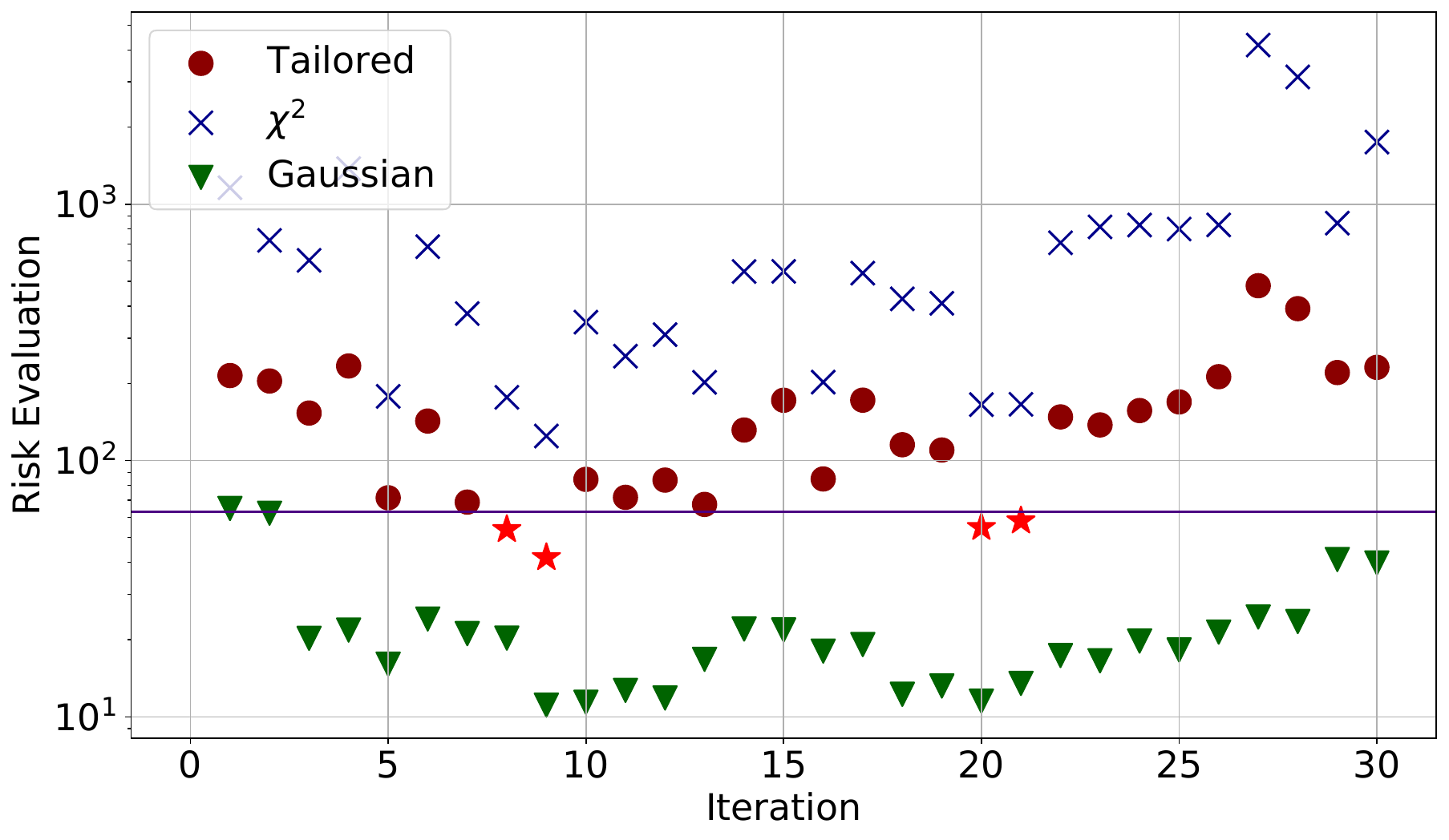}
         \caption{Scatter Plots for CVaR Evaluations}
         \label{fig:fire_box}
     \end{subfigure}
     \caption{Common parameters: $\delta=0.05$. With $M=8$, over ${\tt reps}=30$ runs $H_0$ was rejected  $R_0=30$ times at $\alpha=95\%$ confidence}\label{fig:fire}
\end{figure}

We consider a dataset of fire insurance claims in Denmark between 1980 and 1990 (in millions of Danish Krone).  The loss is a cumulative figure for the event of interest, and includes damage to buildings, damage to furniture and personal property as well as loss of profits. For these analyses the data have been adjusted for inflation to reflect 1985 values. 
Figure~\ref{fig:fire_main} compares the worst case CVaR evaluation using different methods with the benchmark CVaR. For this experiment, ${\tt grid} =\{10^{-0.1+0.15i}: i\in\{0,1,\ldots,8\}\}$, $n=200$ and  $s=60$.
These parameters are selected so as to ensure that data used in successive replications is as distinct (has minimal overlap) as is possible.
Figure~\ref{fig:fire_box} shows the scatter plots of the $30$ CVaR evaluations computed in step (i) above with $\beta=0.03$.
Note that for $26$ out of $30$ replications, the proposed method is both less conservative than $\chi^2$-DRO, and does not underestimate the true risk. For the remaining $4$ replications (indicated by the red star marks), the extent of underestimation far less than that of the DRO with Gaussian nominal distribution. 

\subsection{Fama-French Data} 
\begin{figure}[htpb]
 \centering
     \begin{subfigure}{0.32\textwidth}
         \centering
         \includegraphics[width=\textwidth]{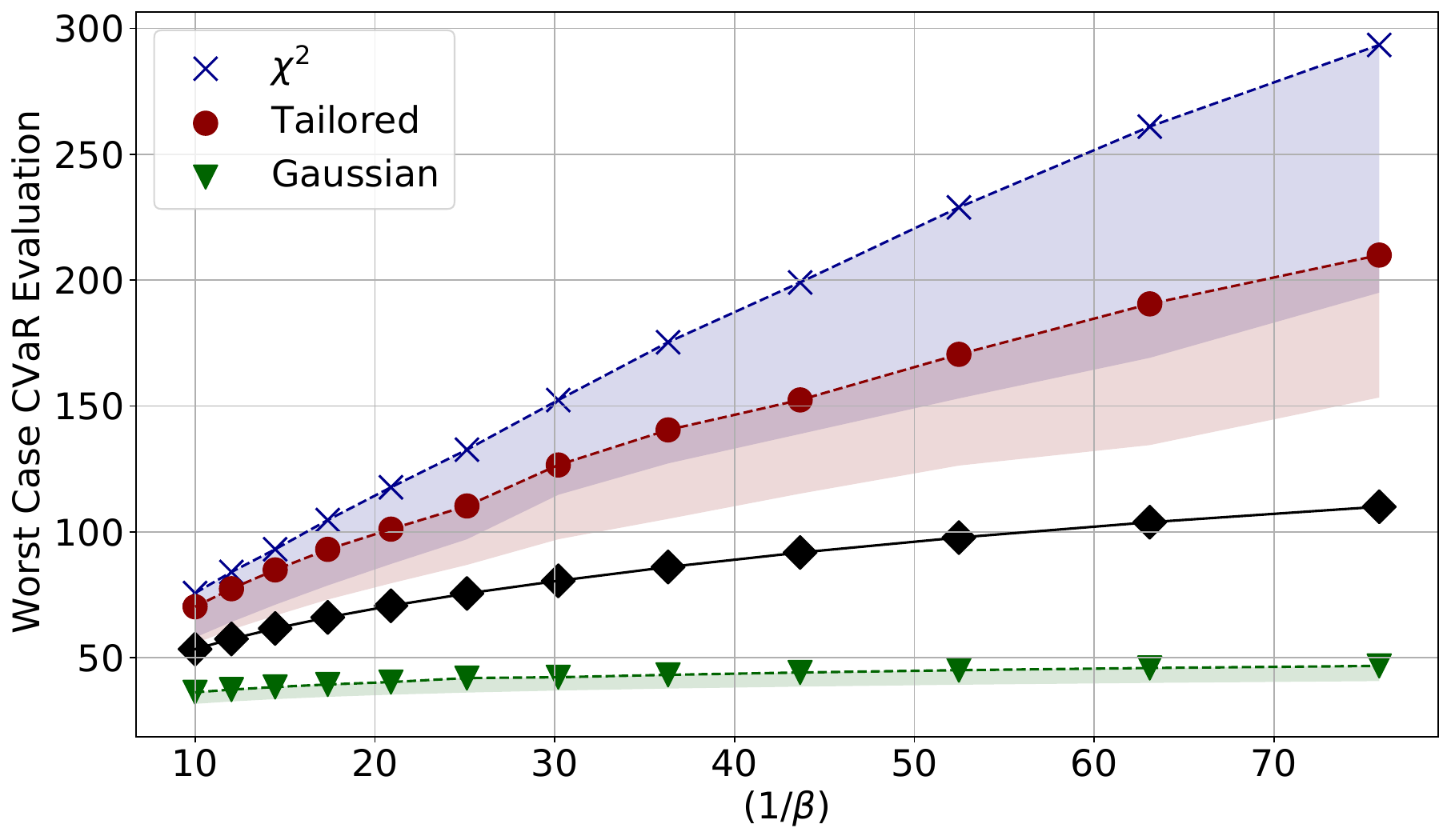}
             \caption{$\lambda=0$, $K=24$}
         \label{fig:ff_main_1}
     \end{subfigure}
     \begin{subfigure}{0.32\textwidth}
         \centering
         \includegraphics[width=\textwidth]{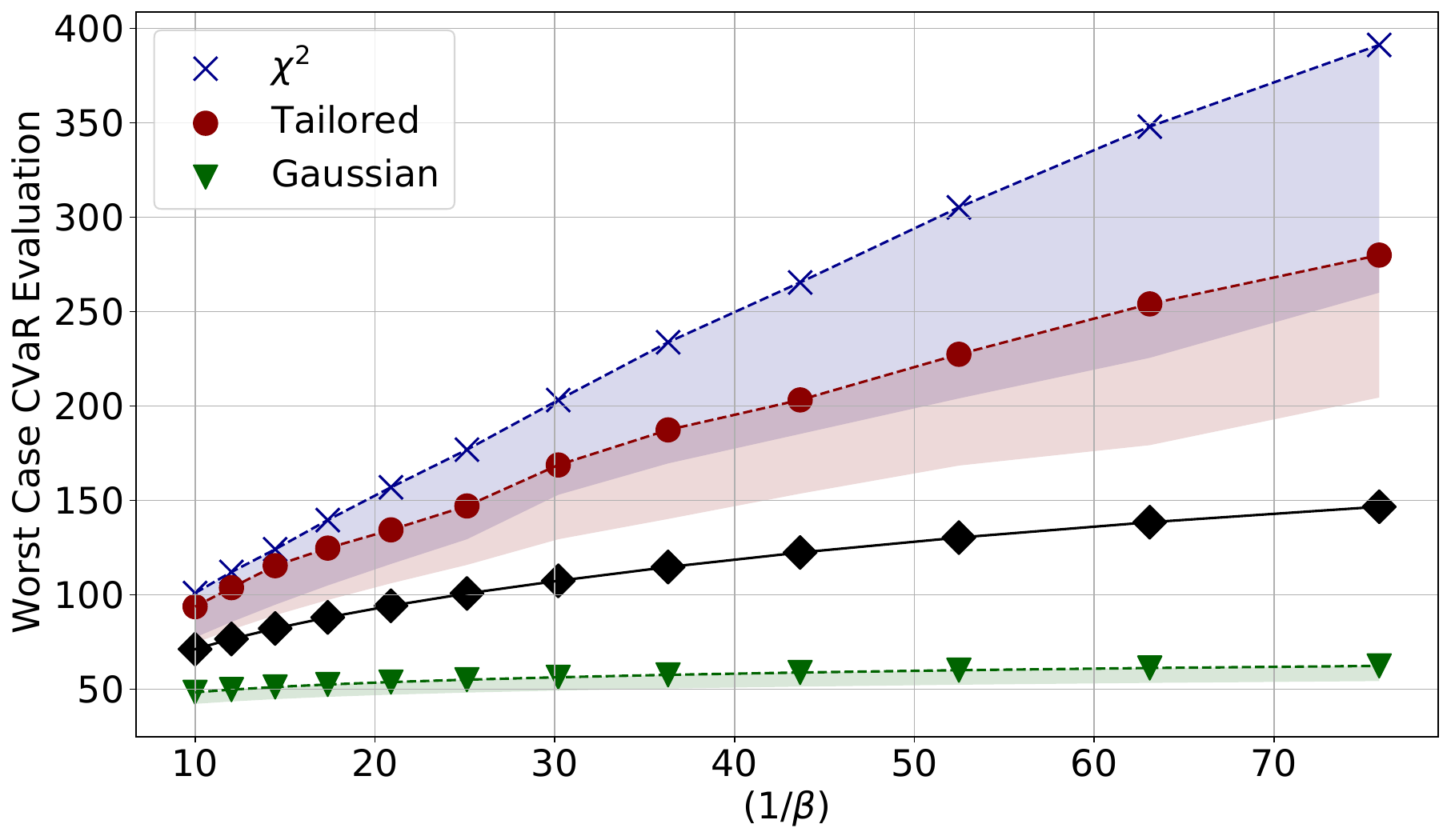}
 \caption{$\lambda=0.5$, $K=16$}
         \label{fig:ff_main_2}
     \end{subfigure}
     \begin{subfigure}{0.32\textwidth}
         \centering
         \includegraphics[width=\textwidth]{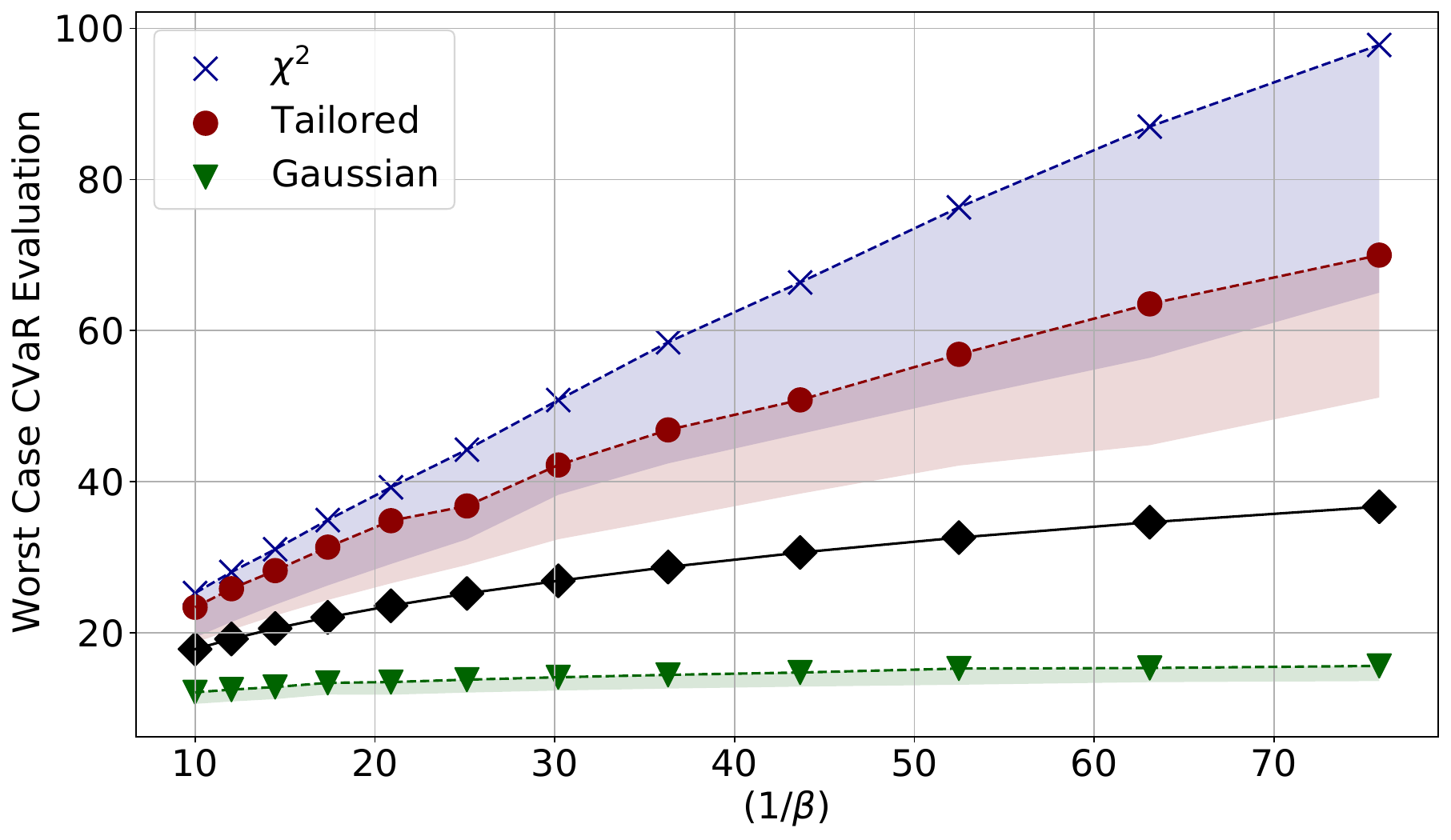}
         \caption{$\lambda=1$, $K=6$}
         \label{fig:ff_main_3}
     \end{subfigure}
     
     \caption{Parameters: $\delta=0.05,$ $p=2$, $d=48$ $n=150$, $s=165$ and $\tt reps =30$. With $M=6$, $H_0$ was rejected all $30$ times with $95\%$ confidence}\label{fig:ff}
\end{figure}

We consider the daily 48 industry portfolio data from the \href{https://mba.tuck.dartmouth.edu/pages/faculty/ken.french/Data\_Library/det\_5\_ind\_port.html} {FAMA French} database for a period of $5982$ days, spanning roughly 20 years starting July 2004 to December 2024. We suppose $K$ firms have investments in these portfolios, and there are no cross holdings. 
The objective is to compute the CVaR of the total loss incurred by the an agent insuring these firms against losses, given by $L(\zz) = \|A_\lambda (\zz\lor \mv 0)\|_p$. The connectivity matrix $A_\lambda$ is as given in Section~\ref{sec:Sys_risk}.
We repeat the experiment
from Section~\ref{sec:fire_insurance} 
with ${\tt grid} = \{10^{-0.1+0.08i}: i\in\{0,1,\ldots 15\}\}$
and plot the results in Figure~\ref{fig:ff}.  For $27$ out of $30$ replications, the benchmark CVaR was not underestimated. In each of these cases, the extent of conservativeness was significantly (between 30\%-50\%) less than that incurred by $\chi^2$-DRO. Over the remaining $3$ replications, the extent of underestimation is substantially less than if Gaussian nominal distribution is used. 
\bibliographystyle{acm}
\bibliography{ref}
\newpage

\begin{appendices}

\section{Proofs of Main Results}\label{appendix:proofs}

\subsection{Proofs from Section~\ref{sec:model}: }
\noindent \textbf{Proof of Lemma~\ref{lem:tail_risk_measure}: } 
Fix distributions $(P,Q)\in \mathcal M(\R)$ which are atom-less at their respective $(1-\beta)$ quantile such that $F_{P,\beta}=F_{Q,\beta}$. Note that then $P(Z>v_{1-\beta}(P)) = Q(Z>v_{1-\beta}(Q)) =\beta$. With $t\in(0,1]$,
\begin{align*}
    v_{1-t\beta}(P) &=\inf_{u} \left\{\frac{P(Z>u)}{\beta} \leq t \right\} = \inf_{u}\left\{{P_\beta(Z>u)} \leq t \right\}\\& = v_{1-t}(P_\beta) =v_{1-t}(Q_\beta) = v_{1-t\beta}(Q),
\end{align*}
where $P_\beta$ and $Q_\beta$ are the distributions whose cdfs are $F_{P,\beta}$ and $F_{Q,\beta}$ respectively. The first equality follows, since $\bar F_{P}(v_{1-\beta}(P)) = \beta$. For the second line note that since  $F_{P,\beta}=F_{Q,\beta}$, the conditional tail laws $P_\beta$ and $Q_\beta$ coincide, so 
$v_{1-t}(P_\beta)=v_{1-t}(Q_\beta)$. Finally, repeating the argument for $Q$,
$v_{1-t}(Q_\beta)=v_{1-\beta t}(Q)$.
Substituting into \eqref{eqn:drm_tails} yields
\[
\rho_{1-\beta}(P)
=\int_0^1 w(t)\,v_{\,1-t\beta}(P)\,dt
=\int_0^1 w(t)\,v_{\,1-t\beta}(Q)\,dt
=\rho_{1-\beta}(Q).
\]
Hence $\rho_{1-\beta}$ depends only on  $F_{P,\beta}$ and is therefore a $\beta$-tail risk measure.\qed

\noindent \textbf{Proof of Lemma \ref{lem:CVaR_regularity}: } The part of the  lemma is already established in \cite{zhu2012tail}, Example 2.6. For the second, suppose that $\Lambda_P\in \RV$. Then, for any $\varepsilon>0$, with $u_\beta = \log(1/\beta)$,
\begin{align}
    \rho_{1-\beta}(P) &= \int_0^1 w(t) v_{1-t\beta}(P)dt = \int_0^1 w(t) \Lambda^{\leftarrow}_P(\log(1/\beta) + \log(1/t)) \nonumber\\
    & = \underbrace{\int_0^\varepsilon w(t) \Lambda^\leftarrow_P(u_\beta + \log(1/t))  dt }_{(a)} + \underbrace{\int_\varepsilon^1 w(t) \Lambda^\leftarrow_P(u_\beta + \log(1/t))  dt }_{(b)} ). \label{eqn:rho_integral}
\end{align}
Recall that for a regularly varying function, for an compact set $K$,
\[
\sup_{y\in K}\left\vert\frac{g(x+y)}{g(x)} - 1\right\vert \to 0 \quad \text{ as } \quad x \to \infty.
\]
In \eqref{eqn:rho_integral}, setting $y= \log(1/t)$ and $x = u_\beta$ note that  
whenever $t\in [\varepsilon,1]$, $y\in[1,\log(1/\varepsilon)]$ and $x\to \infty$ as $\beta\to 0$. Therefore,  as $\beta\to 0$
uniformly over $t\in [\varepsilon,1]$, $\Lambda_P^{\leftarrow}(\log(1/\beta) + \log(1/u)) \sim \Lambda_P^{\leftarrow}(\log(1/\beta))$. Substitute this in $(b)$ to get 
\[
\int_\varepsilon^1 w(t) \Lambda^\leftarrow_P(u_\beta + \log(1/t))  dt  \sim  \Lambda_P^\leftarrow(\log(1/\beta))\int_\varepsilon^1 w(t) dt. 
\]

To handle the integral in $(a)$ we use Potter's bounds (see \cite{deHaan}, Proposition B.1.9): if $g\in RV(p)$ then for any $\delta>0$, there exists a $t_0$ such that for all $t>t_0$ and $x>1$
\begin{equation}\label{eqn:potter_bounds}
    \frac{g(tx)}{g(t)} \leq (1+\delta) x^{p+\delta}.
\end{equation}
Pick any $\beta_0$ such that $\Lambda_{P}^\leftarrow(\log(1/\beta_0)) > t_0$.  With $\Lambda_P^\leftarrow\in \RV(1/\gamma)$ and $\log(1/t)>1$ for $t\in (0,\varepsilon)$, \eqref{eqn:potter_bounds} implies 
\[
\Lambda_P^\leftarrow(\log(1/t\beta)) \leq  \Lambda_P^\leftarrow(\log(1/\beta)) (1+\delta)\left(\frac{\log(1/t\beta)}{\log(1/\beta)}\right)^{1/\gamma+\delta} \leq \Lambda_P^\leftarrow(\log(1/\beta))(1+\delta)\left(1+\frac{\log(1/t)}{\log(1/\beta_0)}\right)^{1/\gamma+\delta}
\]
for all $\beta<\beta_0$. Now, for all such $\beta$, bound the integral $(a)$  in \eqref{eqn:rho_integral} as 
\begin{equation}\label{eqn:int_bdd1}
\Lambda_P^\leftarrow(\log(1/\beta)(1+\delta)
\int_0^\varepsilon t^\kappa \ell(1/t) \left(1+\frac{\log(1/t)}{\log(1/\beta_0)}\right)^{1/\gamma+\delta} dt.    
\end{equation}
Since $\kappa>-1$ and $\ell (\cdot)$ is regularly varying at $\infty$ 
\[
t^\kappa \ell(1/t) \left(1+\frac{\log(1/t)}{\log(1/\beta_0)}\right)^{1/\gamma+ \delta} \leq t^{-1+\varsigma} \text{ for some }\varsigma>0.
\]
Therefore, the integral in \eqref{eqn:int_bdd1} can be made arbitrarily small. Hence  $(a)$ is $o(\Lambda_P^\leftarrow(\log(1/\beta))$. Since $\varepsilon>0$ was arbitrary, substituting this observation into \eqref{eqn:rho_integral}
\[
\rho_{1-\beta}(P)  \sim \Lambda_{P}^\leftarrow(\log(1/\beta)) \int_0^1 w(t)dt.  
\]
Since $\Lambda_P^\leftarrow \in \RV(1/\gamma)$, the conclusion of the lemma follows.
\qed

\noindent\textbf{Verification of Table~\ref{tab:trm}:} 

\noindent \textbf{i) Expected Shortfall:} It is evident that these satisfy the conditions of Assumption~\ref{assume:tail_risk_measures} with $\kappa = 0$ and $\ell(s) =1$. Further, coherence requires that $w(\cdot)$ be a non-increasing function, which is true for CVaR.

\noindent \textbf{ii) Power distortion: }For the power distortion, set $\ell(s) = k$ and $\kappa=k-1$. These also satisfy conditions of Assumption~\ref{assume:tail_risk_measures}. For coherence of power distortion, note that $w(\cdot)$ is decreasing only if $k\in (0,1]$.

\noindent\textbf{iii) Wang Transform: }Note the following identity: $\Phi^{-1}(s) \sim -\sqrt {2\log(1/s)}$ as $s\to 0$. Then, the weight defining the Wang transform satisfies 
\[
\log w(s) \sim \lambda\sqrt{2\log(1/s)} - \frac{\lambda^2}{2} 
\]
Consequently, $w(s)$ satisfies Assumption~\ref{assume:tail_risk_measures} with $\kappa=0$  and $\ell(s)= \exp(\lambda\sqrt{2\log(1/s)}-\lambda^2/2)$. To see the integrability condition, observe that using a change of variables $\Phi(p) = s$, we get
\[
\int_0^1 w(s) ds = e^{-\lambda^2/2}\int_{-\infty}^\infty e^{-\lambda p} \phi(p) dp  = 1
\]
where the last integral follows upon evaluating the moment generating function of  a standard Gaussian. Finally, so long as $\lambda\geq 0$, note that  $\log w$ (and therefore $w$) is a decreasing function.

\noindent \textbf{iv) Beta tail weight:} Clearly, for any $(p,q)>0$, $\int_0^1 s^{p-1}(1-s)^{q-1}ds = \text{B}(p,q)$, and hence $w(t)$ integrates to $1$. Observe that for $q>1$
\[
w^\prime(s) = \frac{p-1}{\mathrm{B}(p,q)} s^{p-1}(1-s)^{q-1} - \frac{q-1}{\textrm{B}(p,q)} s^{p-1}(1-s)^{q-2} <0 \implies \rho_{1-\beta} \text{ is coherent}.
\]

\noindent \textbf{v) Log-power tail weight: }Set $\ell(s) = \log^{q}(1/s) \times p^{q+1}/\Gamma(q+1)$ and $\kappa= p-1$ to see the representation in the form $w(s) \sim s^{\kappa} \ell(1/s)$ For coherence of log-power weighted distortion, note that $w(\cdot)$ is decreasing only if $p\in (0,1]$ and $ q\geq 0$. A standard change of variables $\log(1/s)= t$ demonstrates the condition $\int_0^1 w(s)ds=1$. Further, the function is decreasing for $p\in (0,1)$ so long as $q\geq 0$ \qed   



\subsection{Proofs from Section~\ref{sec:wass_dro}}
Define 
 \[
\vartheta(u,\lambda) = u - \frac{1-1/p}{(p\lambda)^{1/(p-1)}}.
\] 
Note that as $p\downarrow 1$, $\vartheta(u,\lambda) \to -\infty$ if $\lambda< 1$, $\vartheta(u,\lambda) \to u$ if $\lambda\geq 1$. The following lemmas are required for proving the results in this section

\begin{lemma}\label{lem:Dual_Mult}
    For any $(u,\lambda)$, the inner expectation in \eqref{eqn:wass_duality} evaluates to
\begin{equation}\label{eqn:cvar_WC}
    E_{P_0}\left[\sup_{y} \left\{ (y  - u )^+- \lambda |y -Z|^p\right\}\right] = E_{P_0}\left[(Z-\vartheta(u,\lambda))^+\right],
\end{equation}
where if $p=1$, $\nu(u,\lambda)$ is to be understood as its limit as $p\downarrow 1$. 
The worst case CVaR can be written as the value of following joint optimization problem in $(u,\lambda)$:
\begin{equation}\label{eqn:wc_variational_problem}
C_{1-\beta}(\mathcal W_{p,\delta}) = \inf_{u,\lambda\geq0}\left\{u+ \beta^{-1}\lambda \delta^p  + \beta^{-1}E_{P_0}[Z-\vartheta(u,\lambda)]^+\right\} = \inf_{u,\lambda\geq0}\varphi_\beta(u,\lambda)   . 
\end{equation}
Further, for a given $(\lambda,u)$ pair,  the point-wise supremum in on the left hand side of \eqref{eqn:cvar_WC} is attained at $y^* =  z\mv{1}\{z<\vartheta(u,\lambda)\}  + (z+(\lambda p)^{-1/(1-p)}) \mv{1}\{z\geq \vartheta(u,\lambda)\}$ if $p>1$ and at $y^* = z \mv{1}\{z<\vartheta(\lambda,u)\} + (z+c) \mv{1}\{y\geq \vartheta(\lambda,u)\} $, where $c>0$ is arbitrary, if $p=1$.
\end{lemma}

\begin{lemma}\label{lem:set_of_optimisers}
   Let $(\lambda^*,u^*)$ be optimisers to \eqref{eqn:wc_variational_problem}. Then, if $p>1$, $\vartheta(u^*,\lambda^*) = v_{1-\beta}(P_0)$ and $(\lambda^* p)^{-1/(p-1)} = \delta/\beta^{1/p}$. If $p=1$, then $u^*=v_{1-\beta}(P_0)$ and $\lambda^* = 1$. 
\end{lemma}

\noindent\textbf{Proof of Theorem~\ref{thm:Dual_Mult}:} First suppose that $p>1$.
Fix $u$, and consider evaluation of the worst case expectation 
\[
\sup_{P\in \mathcal W_{p,\delta}} E_P[Z-u]^+ = \inf_{\lambda}\left\{\lambda\delta^p + E_{P_0}\left[\sup_{y}(y-u)^+-\lambda|Z-y|^p\right]\right\}.
\]
The supremum above is attained by a distribution supported on the set $\{(x,y): y\in\arg\max\{(y-u)^+-\lambda^*|y-x|^p\} $ where $\lambda^*=\lambda^*(u)$ solves the infimum on the right hand side above
(see \cite{blanchet2019quantifying}, Remark 2). From Lemma~\ref{lem:Dual_Mult} this is attained at
\begin{equation}\label{eqn:wc_map_wass}
    y^* = 
x \mv{1}( x<\vartheta(u,\lambda^*)) +  
(x+ (\lambda^* p)^{\frac{-1}{p-1}} )\mv{1}(x\geq \vartheta(u,\lambda^*))
\end{equation}
if $p>1$. Observe that the worst case distribution is now the translation in \eqref{eqn:wc_map_wass} attained at $u=u^*$.
From Lemma~\ref{lem:set_of_optimisers}, $\vartheta(u^*,\lambda^*) = v_{1-\beta}(P_0)$ and $(\lambda^* p)^{-1/(p-1)} = \delta/\beta^{1/p}$. Plugging these values above completes the proof when $p>1$.

When $p=1$, $\lambda^*=1$ for any $u$, for in the region $\lambda<1$, the inner supremum becomes infinite and if $\lambda>1$, $\lambda \delta^p >\delta^p$. In this case, $\vartheta(u,\lambda) =u$, and from the second part of Lemma~\ref{lem:Dual_Mult}, the worst case coupling is supported on the set
\begin{equation}\label{eqn:opt_set}
    \bigcup_{c>0} \left\{(x,y) : y = x\mv{1}(x\leq u) + (x+c) \mv{1}(x>u)\right\}
\end{equation}
From Lemma~\ref{lem:set_of_optimisers}, $u^* = v_{1-\beta}(P_0)$. 
Substituting into \eqref{eqn:opt_set} completes the proof. \qed

\subsection{Proofs from Section~\ref{sec:divergence_dro}}
The following technical result is key to the proofs in this section. Let $\bar F_{\tt wc}(x) = \sup\{\bar F_P(x) : P\in \Dphi(P_0)\}$.
\begin{proposition}{(\textbf{Proposition 3.3 from \cite{birghila2021distributionally}})}\label{prop:sol_of_inverse}
  Let the function $\phi(\cdot)$ satisfy Assumption~\ref{assume:f-assume}. Then the worst case tail-cdf satisfies
  \[
  \bar F_{\tt wc}(x) \sim \phi^{\leftarrow}(\delta/\bar F_{P_0}(x)) \bar F_{P_0}(x) \ \ \ \ \ \text{ as }x\to\infty
  \]
\end{proposition}
We also give Karamata's theorem below for reference: this result will be useful going ahead in evaluating integrals of regularly varying functions:

\begin{theorem}[\textbf{Karamata's Theorem (\cite{deHaan}, Theorem B.1.5)}]\label{thm:Karamata}
  Let $f\in \RV(\rho)$ for some $\rho<-1$. Then, as $x\to\infty$,
  \[
  \int_{x}^\infty f(t) dt \sim \frac{xf(x)}{-\rho-1}.
  \]
\end{theorem}
\begin{lemma}\label{lem:wc_upper}
 For any ambiguity set $\mathcal P$, and $\rho_{1-\beta}$ satisfying Assumption~\ref{assume:tail_risk_measures} 
 \begin{equation}\label{eqn:wc_upper}
     \rho_{1-\beta}(\mathcal P) \leq  \int_{0}^1 w(t) v_{1-\beta t}(\mathcal P) dt.
 \end{equation}
\end{lemma}
\noindent \textbf{Proof of Proposition~\ref{prop:f_div_contents}: } Let $\tilde P$ be a distribution whose likelihood with the nominal distribution $\mathcal L(x) = \frac{d\tilde P(x)}{dP_0}$
which satisfies
\begin{equation}\label{eqn:tilde_f}
   \mathcal L(x) =  \mv{1}(x\leq m) + c_m \psi(x)\mv{1}(x\geq m)
\end{equation}
where $\psi(\cdot)$ is an almost everywhere continuously differentiable function that satisfies $\psi(m) =1$, and $c_m$ is a constant such that $E_{P_0}[\mathcal L] = 1$. Note that this implies
\begin{equation}\label{eqn:c_m}
    F_{P_0}(m) + c_{m}\int \psi(x) dP_0(x)=1 \implies  c_m = \frac{\bar F_{P_0}(m)}{E_{P_0}[\psi(Z) ; Z>m]}
\end{equation}
For each part, we demonstrate that an appropriate choice of $\psi$ and $m$ leads to the desired conclusion

\noindent \textbf{i) } Suppose that $\phi(x) \sim x\log x$, and suppose first that Assumption~\ref{assume:heavy_tailed_data} holds. Write $d\tilde P(z) = \mathcal L(z) dP_0(z)$ and note that  
\begin{align*}
    E_{\tilde P}[Z^+] &= \int_{0}^\infty \bar F_{\tilde P}(x) dx = \int_0^\infty E_{P_0}\left[\mathcal L(Z) ; Z\geq x\right]dx \\
    &=\int_{0}^\infty \left[ \int_{x}^\infty \mathcal L^\prime(t) \bar F_{P_0}(t) dt  + \bar F_{P_0}(x)\mathcal L(x)\right]dx \geq \int_{0}^\infty \left[ \int_{x}^\infty \mathcal L^\prime(t) \bar F_{P_0}(t) dt\right]dx,
\end{align*}
where the third display above is the result of integrating by parts\footnote{set $dv = dF_{P_0} = -d\bar F_{P_{0}}$ and $u =\mathcal L$ in the formula $\int udv = uv - \int vdu$. Also note that the likelihood is of bounded variation, since $\psi$ is continuously differentiable}.
In the region $x<m$, $\mathcal L(x) =1$. 
Then, setting for $x>m$,
\begin{equation}\label{eqn:DUIS}
\int_{x}^\infty \psi^\prime(t) \bar F_{P_0}(t) dt = \frac{1}{x \log x} \implies   E_{\tilde P}[Z^+]\geq  c_m \int_{m}^\infty\frac{1}{x \log x}=   \infty
\end{equation}
Differentiating \eqref{eqn:DUIS} implies that 
\[
\psi^\prime(x) \bar F_{P_0}(x) = \frac{\log x+1}{x^2\log^2x} .
\]
Choose $u_\varepsilon>e$ so that $\phi(u)\le (1+\varepsilon)u\log u$ for all $u\ge u_\varepsilon$.
Since $\psi^\prime >0$, it is possible to choose $m_0$ so large that $\psi(x) \geq u_\varepsilon$. Now,
\begin{align*}
    \int \phi(\mathcal L(x)) dP_0(x) &\leq c_m (1+\varepsilon) \int_{x>m} \psi(x) \log(\psi(x)) dP_0(x) \\
    & = c_m(1+\varepsilon) \int_{x>m} \psi^\prime(x) \bar F_{P_0}(x) (\log \psi(x)+1)dx + K_1(m)\\
    & =  c_m(1+\varepsilon) \int_{x>m} \left(\frac{\log x+1}{x^2\log ^2x}\right)(\log \psi(x) +1)dx + K_1(m).
\end{align*}
where $K_1(m) = \psi(m) \log(\psi(m)) \bar F_{P_0}(m)$. 
Note that the last line above follows upon substituting the value of $\psi^\prime(x) \bar F_P(x) $. 
To bound $c_m$, let us note that that $\psi^\prime \geq 0$, and $\psi(m) = 1$, which means that $\psi(x)\geq 1$ for all $x>m$, and from \eqref{eqn:c_m}, $c_m\leq 1$. Next,
notice that
since $\bar F_{P_0} \in \RV(-\gamma)$, Potter's bounds  (\cite[Proposition B.1.9]{deHaan}) imply that for all sufficiently large $x$,
\begin{equation}\label{eqn:psi_def}
\psi(x) = \int \frac{\log t+1}{t^2 \bar F(t) \log^2 t} dt \leq  \int \frac{1}{t^{2-\gamma-\varepsilon}}dt  \leq C^\prime t^{\gamma-1+\varepsilon}
\end{equation}
and therefore, $\log (\psi(x)) \leq (\gamma+\varepsilon)\log (x) $. Then, as $m\to\infty$, $\psi(m)\log(\psi(m))\bar F_{P_0}(m) \to 0$. 
Consequently, for a choice $m>\max\{m_0,m_1\}$ in \eqref{eqn:tilde_f}
\[
 \int_{x} \phi(\mathcal L(x)) dP_0(x) \leq (1+\varepsilon)\int_{x>m}\left(\frac{\log x+1}{x^2\log^2 x}\right)(\gamma+\varepsilon-1)\log(x)dx +K_1(m) <\delta 
\]
This implies that $\tilde P\in \Dphi(P_0)$ when $\varepsilon>0$ is chosen appropriately.

Next suppose that Assumption~\ref{assume:weibullian_tails} holds with $\gamma< 1$. Then,  $\Lambda_{P_0}\in RV(\gamma)$ and consequently from \eqref{eqn:psi_def}, $\psi(t) \leq \exp(t^{\gamma+\varepsilon})$. Thus, 
\[
\left(\frac{\log x+1}{x^2\log ^2x}\right)(\log \psi(x) +1) \leq \left(\frac{\log x+1}{x^{2-\gamma-\varepsilon}\log^2x} \right)
\]
and 
\[
 \int\phi(\mathcal L(x)) dP_0(x) \leq (1+\varepsilon) \int_{x>m} \left(\frac{\log x+1}{x^{2-\gamma-\varepsilon}\log^2x} \right) dx +K_1(m) 
\]
Since $\gamma<1$, it is possible to choose $\varepsilon>0$ such that $2-\gamma-\varepsilon>1$. Therefore, the integral above can be made arbitrarily small by an appropriate choice of $m$, and $\tilde P\in\Dphi(P_0)$.\qed

\noindent \textbf{ii) }
First, note that for every $t>m$ in \eqref{eqn:tilde_f}, integrating by parts,
\[
\tilde P(Z>t) = \int_{t}^\infty \mathcal L(x) dP_0(x) = c_m\int_{t}^\infty \psi^\prime(x) \bar F_{P_0}(x) dx  + c_m\psi(t) \bar F_{P_0}(t).
\]
Fix $\gamma^\prime$, and set $\psi^\prime(x) =  x^{\gamma-\gamma^\prime-1}$. As before, note that since $\psi(m) =1$ and $\psi^\prime>0$, $c_m\leq 1$. Then with $\bar F_{P_0}\in \RV(-\gamma)$, $\psi^\prime \bar F_{ P_0}\in \RV(-\gamma^\prime -1)$. Apply Karamata's theorem
to $g(x) = \psi^\prime(x)\bar F_{P_0}(x)$ and infer that $\tilde P(Z>t) \sim c_m(\gamma^\prime)^{-1} t \psi^\prime(t) \bar F_{P_0}(t) +c_m \psi(t) \bar F_{P_0}(t)$. Therefore $\bar F_{\tilde P }\in \RV(-\gamma^\prime)$.   

Since $\phi\in\RV(p)$ Potter's bounds imply that for an $\varepsilon>0$, there exists a $t_0>0$ and $C_\varepsilon$ such that $\phi(t)\leq C_\varepsilon t^{p+\varepsilon}$ for all $t>t_0$. Now choose $m$ so large that $\psi(x) >t_0$ for all $x>m$. Then for any $\varepsilon>0$, 
\begin{align*}
\int_x \phi\left(\mathcal{L}(x)\right) dP_0(x) &\leq  C_\varepsilon \int_{x>m} \psi^{p+\varepsilon}(x) dP_0(x)\\
    & =C_{\varepsilon} (p+\varepsilon) \int_{x>m} \psi^\prime(x) \psi^{p-1+\varepsilon}(x) \bar F_{P_0}(x)dx + C_{\varepsilon}\psi^{p+\varepsilon}(m)\bar F_{P_0}(m),
\end{align*}
where the second expression above is obtained by using integration by parts. 
By construction, $\psi(x) \sim c_1 x^{\gamma-\gamma^\prime}$ where $c_1\sim (\gamma-\gamma^\prime)^{-1}$. Substituting this in the last equation above:
\[
\int_x \phi\left(\mathcal{L}(x)\right) dP_0(x) \leq C_\varepsilon \int_{x>m} x^{(p+\varepsilon)(\gamma-\gamma^\prime)-1}\bar F_{P_0}(x)dx+ m^{(\gamma-\gamma^\prime)(p+\varepsilon)}\bar F_{P_0}(m). 
\]
Note that the  $x^{(p+\varepsilon)(\gamma-\gamma^\prime)-1}\bar F_{P_0}(x)$ is regularly varying with index $\theta=(p+\varepsilon-1)\gamma -(p+\varepsilon)\gamma^\prime-1$
and that for any $\gamma^\prime>\frac{p-1}{p}\gamma$, $\theta<-1$ whenever $\varepsilon$ is chosen small enough. Once again using Karamata's theorem,
\[
\int_{x>m} x^{(p+\varepsilon)(\gamma-\gamma^\prime)-1}\bar F_{P_0}(x)dx \sim  m^{(p+\varepsilon)(\gamma-\gamma^\prime)}\bar F_{P_0}(m)  \to 0 \text{ as }m\to\infty
\]
Hence, choosing a large enough $m$ gives $\int_x \phi\left(\mathcal{L}(x)\right) dP_0(x)\leq \delta$, and $\tilde P \in \Dphi(P_0)$.\qed

\noindent \textbf{iii)}  Let $\Lambda_{P_0}\in \RV(\gamma)$ with $\gamma\in(0,1)$. Fix $c\in\big((p-1)/p,\,1\big)$ and define on $(m,\infty)$ $\psi'(x)=\exp\!\big((1-c)\Lambda_{P_0}(x)\big)$.
Then
\[
\int_x^\infty \psi'(u)\,\bar F_{P_0}(u)\,du=\int_x^\infty e^{(1-c)\Lambda_{P_0}(u)}e^{-\Lambda_{P_0}(u)}\,du=\int_x^\infty e^{-c\,\Lambda_{P_0}(u)}\,du.
\]
By a Laplace–type (large-deviations style) asymptotic (since $\Lambda\in\RV(\gamma)$), as $x\to\infty$,
\[
\int_x^\infty e^{-c\Lambda_{P_0}(u)}\,du\ \sim\ \frac{x e^{-c\Lambda_{P_0}(x)}}{c\,\Lambda_{P_0}(x)},
\qquad
\psi(x)=1+\int_m^x e^{(1-c)\Lambda_{P_0}(u)}\,du \ \sim\ \frac{x e^{(1-c)\Lambda_{P_0}(x)}}{(1-c)\Lambda_{P_0}(x)}.
\]
Hence
\[
\bar F_{\tilde P}(x)
= c_m\!\left[\int_x^\infty \psi'(u)\bar F_{P_0}(u)\,du+\psi(x)\bar F_{P_0}(x)\right]
\sim \frac{C}{\Lambda_{P_0}(x)}\,e^{-c\Lambda_{P_0}(x)},
\]
so $\Lambda_{\tilde P}(x)\sim c\,\Lambda_{P_0}(x)$ and for any $\kappa>0$ we can choose $c\in\big((p-1)/p,\,(p-1)/p+\kappa\big)$ to get
\[
\frac{\Lambda_{\tilde P}(x)}{\Lambda_{P_0}(x)}\ \le\ \frac{p-1}{p}+\kappa
\quad\text{for all large }x.
\]

For the divergence budget, by $\phi\in \RV(p)$ there exist $C_\varepsilon,u_\varepsilon$ with $\phi(u)\le C_\varepsilon u^{p+\varepsilon}$ for $u\ge u_\varepsilon$.
Take $m$ large so $\psi(x)\ge u_\varepsilon$ for $x\ge m$. Since $c_m\le1$ and $\phi$ is increasing on $[u_\varepsilon,\infty)$,
\begin{align*}
    \int \phi(\mathcal L)\,dP_0 &\leq C_\varepsilon \int_{x>m}\psi^{p+\varepsilon}(x)\,dP_0(x)\\
&= C_\varepsilon\psi^{p+\varepsilon}(m)\bar F_{P_0}(m)
+ C_\varepsilon(p+\varepsilon)\!\int_m^\infty \psi'(x)\psi^{p-1+\varepsilon}(x)\bar F_{P_0}(x)\,dx.
\end{align*}
Using the asymptotics above,
\[
\psi'(x)\psi^{p-1+\varepsilon}(x)\bar F_{P_0}(x)
\ \asymp\ \exp\!\Big((1-c)\Lambda + (p-1+\varepsilon)(1-c)\Lambda - \Lambda\Big)
= \exp\!\big(-[\,1-p(1-c)-\varepsilon(1-c)\,]\Lambda\big).
\]
Since $c>(p-1)/p$, choose $\varepsilon>0$ small so that $1-p(1-c)-\varepsilon(1-c)>0$. Then the integral converges and can be made $\le\delta$ by taking $m$ large.
Thus $\tilde P\in\Dphi(P_0)$ and $\Lambda_{\tilde P}(x)/\Lambda_{P_0}(x)\le (p-1)/p+\kappa$ for all large $x$.\qed

\noindent \textbf{Proof of Lemma~\ref{lem:lbb_phi_div}: }\noindent\textbf{(i) KL divergence.} Suppose $\phi(x)=x\log x - x + 1$. Observe that in this case, there exists a $P_1\in \Dphi(P_0)$ such that $E_{P_1}[Z^+]=\infty$. 
Under the Lemma assumptions, there exists $m\in(0,1)$ with $w(t) > 0$ for all $t\in (0,m)$.
Since $E_{P_1}[Z_+] = \infty$,
\[
\int_0^\eta v_{1-s}(P_1)\,ds=\infty\quad\forall\,\eta>0,
\]
we have, for any $\varepsilon>0$,
\[
\int_0^\varepsilon v_{1-\beta t}(P_1)\,dt
=\frac{1}{\beta}\int_0^{\beta\varepsilon} v_{1-s}(P_1)\,ds=\infty.
\]
Therefore
\[
\rho_{1-\beta}(P_1)=\int_0^1 w(t)\,v_{1-\beta t}(P_1)\,dt
\ \geq\ \int_0^m w(t)\,v_{1-\beta t}(P_1)\,dt
\ \ge\ c \int_0^m v_{1-\beta t}(P_1)\,dt
=\infty.
\]
Thus $\rho_{1-\beta}(P_1)=\infty$ and consequently $\rho_{1-\beta}(\Dphi(P_0))=\infty$.

\noindent\textbf{(ii) Polynomial/$\RV(p)$ divergences, heavy-tailed $P_0$.}
If $\phi\in\RV(p)$ and the heavy-tail condition in (ii) of Proposition~\ref{prop:f_div_contents} holds, then there exists $P_1\in\Dphi(P_0)$ with 
$\bar F_{P_1}\in\RV(-\gamma')$ where $\gamma'=\frac{p-1}{p}\gamma+\kappa$, where $\kappa>0$ is arbitrary. Hence
\[
\rho_{1-\beta}(\Dphi(P_0)) \;\ge\; \rho_{1-\beta}(P_1).
\]
Recall that Lemma~\ref{lem:CVaR_regularity} implies $\rho_{1-\beta}(P_1) = \beta^{-1/\gamma^\prime}\ell(1/\beta)$ for a slowly varying $\ell$. Now, fix $\varsigma>0$, Taking log of the previous quantity
\[
\log \rho_{1-\beta}(\Dphi(P_0)) \geq \log\rho_{1-\beta}(P_1) \geq  \left(\frac{p}{p-1} - \delta(\kappa)\right)  \log \rho_{1-\beta}(P_0).
\]
for all $\beta$ small enough. Here, $\delta(\kappa) \downarrow 0$ when $\kappa\downarrow 0$.
Choosing $\delta(\kappa) <\varsigma$ completes the proof.

\noindent\textbf{(iii) Polynomial/$\RV(p)$ divergences, light-tailed $P_0$.}
If (iii) of Proposition~\ref{prop:f_div_contents} holds, then for any $\varepsilon>0$ there is $P_1\in\Dphi(P_0)$ and $t_0$ such that 
for all $t>t_0$,
\[
\frac{\Lambda_{P_1}(t)}{\Lambda_{P_0}(t)} \;\le\; c
\;:=\;\frac{p-1}{p}+\varepsilon \;<\;1.
\]
Let $u_\beta=\log(1/\beta)$ and take $\beta$ small enough that
$t_1:=\Lambda_{P_1}^{\leftarrow}(u_\beta)>t_0$. Then
\[
u_\beta \;=\; \Lambda_{P_1}\!\big(v_{1-\beta}(P_1)\big)
\;\le\; (c+o(1))\,\Lambda_{P_0}\!\big(v_{1-\beta}(P_1)\big),
\]
hence
\[
\Lambda_{P_0}\!\big(v_{1-\beta}(P_1)\big)\ \ge\ \frac{1+o(1)}{c}\,u_\beta.
\]
Since $\Lambda_{P_0}\in\RV(\gamma)$, inverting gives (as $\beta\downarrow0$)
\[
v_{1-\beta}(P_1)\ \ge\ c^{-1/\gamma}(1+o(1))\,v_{1-\beta}(P_0)
\;=\;\Big(\tfrac{p}{p-1}\Big)^{1/\gamma}\!(1+o(1))\,v_{1-\beta}(P_0).
\]
Because $\Lambda_{P_1}\in\RV(\gamma)$, Lemma~\ref{lem:CVaR_regularity}(ii) yields
$\rho_{1-\beta}(P_1)\sim v_{1-\beta}(P_1)$, so for any $\kappa>0$ there exists $\beta_0$
such that for all $\beta<\beta_0$,
\[
\rho_{1-\beta}(\Dphi(P_0))
\;\ge\; \rho_{1-\beta}(P_1)
\;\ge\; \Big(\Big(\tfrac{p}{p-1}\Big)^{1/\gamma}-\kappa\Big)\,\rho_{1-\beta}(P_0).
\]

\noindent \textbf{Proof of Theorem~\ref{thm:poly_divergence}:} \textbf{Upper Bound:} 
Lemma~1 of \cite{blanchet2020optimal} implies that
\[
v_{1-\beta t}(\Dphi(P_0)) = \bar F_{\tt wc}^{\leftarrow}(\beta t),
\]
where $\bar F_{\tt wc}(x) = \sup_{P\in\Dphi(P_0)}P(Z>x)$ is the worst case probability.
 Applying Proposition~\ref{prop:sol_of_inverse},
\begin{equation}\label{eqn:wc_prob_asymptotics}
    \bar F_{\tt wc}(x) \sim \bar F_{P_0}(x)\phi^{\leftarrow}(\delta/\bar F_{P_0}(x)) =  \delta^{1/p} [\bar F_{P_0}(x)]^{1-1/p+o(1)} \text{ as } x\to\infty,
\end{equation}
where the last statement follows since $\phi^\leftarrow\in \RV(1/p)$.

\noindent \textbf{i) }First, suppose that $\bar F_{P_0} \in \RV(-\gamma)$, and therefore, $\bar F_{\tt wc}\in \RV(-\gamma_0)$ where $\gamma_0=\gamma(1-1/p)$.
For $s\in [1,\infty)], $ write $U_{\tt wc}(s) = v_{1-1/s}(\Dphi(P_0))$. The previous display suggests that $U_{\tt wc}\in \RV(1/\gamma_0)$. Fix a sufficiently small $\varepsilon>0$. A change of variables $\beta t = 1/s$ in \eqref{eqn:wc_upper} yields for all $\beta$ small enough:
\begin{align*}
    \rho_{1-\beta}(\Dphi(P_0)) &\leq \beta^{-1}\int_{1/\beta}^\infty s^{-2}w\left(\frac{1}{\beta s}\right) v_{1-1/s}(\Dphi(P_0)) ds\\
    & \leq \beta^{-(\kappa+1)}\int_{1/\beta}^{\infty} s^{-(\kappa+2) + \varepsilon} U_{\tt wc}(s) ds
\end{align*}
where the last inequality follows upon using Potters bound for the regularly varying function $w(1/t) = t^{-\kappa}\ell(t)$
Since $U_{\tt wc}\in \RV(1/\gamma_0)$ the integrand above is regularly varying with index $1/\gamma_0 -\kappa-2+\varepsilon<-1$ since $\kappa > \frac{1}{\gamma} \frac{p}{p-1} -1$. Karamata's theorem therefore applies to yield (for some constant $C_{\kappa}$),
\[
\beta^{-(\kappa+1)}\int_{1/\beta}^{\infty} s^{-(\kappa+2)+\varepsilon} U_{\tt wc}(s)ds  \leq C_{\kappa}   \beta^{-\varepsilon} v_{1-\beta}(\Dphi(P_0)) \ \
 \ \text{ for all }\beta \text{ sufficiently small}. 
\]
Further, note that from Lemma~\ref{lem:CVaR_regularity}, $\rho_{1-\beta}(P_0) = (1/\beta)^{1/\gamma+o(1)}$. Therefore,
\begin{equation}\label{eqn:robust_ht_ub}
    \limsup_{\beta\to0 }\frac{\log \rho_{1-\beta}(\Dphi(P_0))}{\log \rho_{1-\beta}(P_0)}\leq \frac{p}{p-1} + \varepsilon\gamma
\end{equation}
Since $\varepsilon $ above was arbitrary, the proof of the upper bound is complete.

\noindent \textbf{ii) }Next suppose  $\Lambda_{P_0} \in \RV(\gamma)$. Then \eqref{eqn:wc_prob_asymptotics} implies that 
\[
\log \bar F_{\tt wc}(x) \sim  \log \bar F_{P_{0}}(x) + \log  \phi^{\leftarrow}(\delta/\bar F_{P_0})  \implies  - \log \bar F_{\tt wc}(x) \sim \frac{p-1}{p}\Lambda_{P_0}(x).
\]
Let $x = v_{1-s}(\Dphi(P_0))$. Note that as $s\to 0$, $x\to \infty$, since $v_{1-s}(\Dphi(P_0)) \geq v_{1-s}(P_0) \to \infty$. Therefore, as $s\to 0$,
\[
\frac{p}{p-1} \log(1/s) \sim \Lambda_{P_0}( v_{1-s}(\Dphi(P_0))).
\]
Since $\Lambda^{\leftarrow}_{P_0}\in \RV(1/\gamma)$, the above asymptotic results further in 
\[
\Lambda_{P_0}^{\leftarrow} \left(\frac{p}{p-1} \log(1/s) \right) \sim v_{1-s}(\Dphi(P_0)) \implies \left(\frac{p}{p-1}\right)^{1/\gamma} v_{1-s}(P_0) \sim v_{1-s}(\Dphi(P_0)).
\]
where the implication follows as $\Lambda^{\leftarrow}(\log(1/s)) = v_{1-s}(P_0)$.
Since  $\beta t \leq \beta$ on $t\in [0,1]$, the above implies
\[
\lim_{\beta\to 0} \sup_{t\in (0,1]} \left\vert\frac{v_{1-\beta t}(P_0)}{v_{1-\beta t}(\Dphi(P_0))} - \left(\frac{p-1}{p}\right)^{1/\gamma}\right\vert = 0.
\]
Thus,
\[
\int_{0}^1 w(t) v_{1-\beta t}(\Dphi(P_0)) dt\sim  \left(\frac{p}{p-1}\right)^{1/\gamma} \int_{0}^1 w(t) v_{1-\beta t}(P_0)dt =\left(\frac{p}{p-1}\right)^{1/\gamma}  \rho_{1-\beta}(P_0).
\]
Therefore
\begin{equation}\label{eqn:robust_lt_ub}
    \limsup_{\beta\to0 }\frac{\rho_{1-\beta}(\Dphi(P_0))}{\rho_{1-\beta}(P_0)}\leq \left(\frac{p}{p-1}\right)^{1/\gamma}.
\end{equation}

\noindent \textbf{Lower bound: }To prove the matching lower bound, fix a $\eta>0$. First suppose that $P_0$ satisfies Assumption~\ref{assume:heavy_tailed_data}. Note that for the distribution $\tilde P$ constructed in Lemma~\ref{lem:lbb_phi_div}(ii),
$\bar F_{\tilde P}\in \RV(-\gamma^\prime)$, where $\gamma^\prime = (\frac{p-1}{p}\gamma+\eta)$. Thus, 
\[
\liminf_{\beta\to 0} \frac{\log \rho_{1-\beta}(\Dphi(P_0))}{\log \rho_{1-\beta}(P_0)} \geq \liminf_{\beta\to 0}\frac{\log \rho_{1-\beta}(\tilde P)}{\log \rho_{1-\beta}(P_0)} = \frac{p}{p-1} - \delta(\eta)
\]
where $\delta(\eta) = \frac{p}{p-1}  - \frac{\gamma}{\gamma^\prime}\to 0$ as $\eta \to 0$. Since $\eta>0$ was arbitrary,
\begin{equation}\label{eqn:robust_ht_lb}
    \liminf_{\beta\to 0} \frac{\log \rho_{1-\beta}(\Dphi(P_0))}{\log \rho_{1-\beta}(P_0)} \geq \frac{p}{p-1}.
\end{equation}
Next, suppose that $P_0$ satisfies Assumption~\ref{assume:weibullian_tails}. Then for the distribution 
$\tilde P$ constructed in the proof of Proposition~\ref{prop:f_div_contents}(iii), for all $\beta<\beta_0$
\[
\frac{v_{1-\beta}(\tilde P)}{v_{1-\beta}(P_0)} \geq \left(\frac{p}{p-1}\right)^{1/\gamma} - \varepsilon
\]
Further, whenever $\Lambda_{\tilde P}\in \RV$, $\rho_{1-\beta}(\tilde P) \sim v_{1-\beta}(\tilde P)$. Then, whenever $\beta$ is sufficiently small,
\[
\frac{\rho_{1-\beta}(\tilde P)}{\rho_{1-\beta}(P_0)} \geq \left(\frac{p}{p-1}\right)^{1/\gamma} - 2\varepsilon.
\]
This implies that
\begin{equation}\label{eqn:robust_lt_lb}
    \liminf_{\beta\to 0} \frac{\rho_{1-\beta}(\Dphi(P_0))}{\rho_{1-\beta}(P_0)} \geq \left(\frac{p}{p-1}\right)^{1/\gamma} - 2\varepsilon.
\end{equation}
Since $\varepsilon$ was arbitrary,  combining \eqref{eqn:robust_ht_ub} and \eqref{eqn:robust_ht_lb} (resp. \eqref{eqn:robust_ht_ub} and \eqref{eqn:robust_lt_lb}) completes the proof. \qed

\subsection{Proofs from Section~\ref{sec:calib_uncert}} 
The following technical results are needed for proofs in this section. In both the results below $\theta\in\Theta$.

\begin{lemma}\label{lem:RVs_ht}
    Let $\log\beta_0 = \theta \log \beta$, and $\bar F_Q\in\RV(-\gamma)$. Then, for all $t\in(0,\beta]$, 
    \begin{equation}\label{eqn:RVs_ht}
        v_{1-t}(Q_{\beta}) = v_{1-\beta_0}(Q) \left(\frac{t}{\beta_0}\right)^{-1/\gamma}.
    \end{equation}
\end{lemma}

\begin{lemma}\label{lem:RVs_lt}
    Let $\log\beta_0 = \theta\log \beta$,  and $\Lambda_Q\in \RV(\gamma)$. Then, for all $t\in(0,\beta]$, 
    \begin{equation}\label{eqn:RVs_lt}
        v_{1-t}(Q_{\beta}) = v_{1-\beta_0}(Q) \left( \frac{\log (1/t)}{\log(1/\beta_0)}\right)^{1/\gamma}.
    \end{equation}
\end{lemma}

In addition, we need the following uniform version of Proposition~\ref{prop:sol_of_inverse} since in what follows, the nominal distribution is a function of the risk level $\beta$.
\begin{proposition}\label{prop:sol_of_inverse_unif}
    Let the conditions of Proposition~\ref{prop:sol_of_inverse} hold, and let $Q_\beta$ be given by \eqref{eqn:nominal_dist}. Then for any $\varepsilon<1/2$,
    \begin{equation}\label{eqn:sol_of_inverse_unif}
        \lim_{\beta\to 0}\sup_{\substack{x\geq v_{1-\beta_0}(Q)\\\theta\in[\varepsilon,1-\varepsilon],\delta\in \Delta}}\left\vert \frac{\bar F_{{\tt wc},\beta,\delta}^{(\theta)}(x) }{\phi^\leftarrow\left(\frac{\delta}{\bar F_{Q_\beta^{(\theta)}}(x)}\right)\bar F_{Q_\beta^{(\theta)}}(x)}  -1\right\vert =0
\end{equation}
where $\bar F_{{\tt wc},\beta,\delta}^{(\theta)}(x) = \sup\{ \bar F_P(x): {P\in \mathcal \Dphi(Q_\beta^{(\theta)})}\}$ is the worst-case cdf. 
\end{proposition}

\begin{lemma}\label{lem:ucc_quantiles}
    Let $U_{i:n}$ be the $i$th 
     order-statistic of $n$
    i.i.d draws from the standard uniform distribution. Let $k_{n,\theta} = \lfloor n^{1-\theta}\rfloor$. Then 
    \[
    \lim_{n\to\infty} \sup_{\theta\in [\varepsilon,1-\varepsilon]} \left|\frac{n}{k_{n,\theta}} U_{k_{n,\theta}:n} - 1\right| \quad \text{ almost surely}.
    \]
\end{lemma}

\noindent \textbf{Proof of Proposition~\ref{prop:rate_preserving_suffcient}: }  
Let us first suppose that $\bar F_Q\in \RV(-\gamma)$. 

\noindent \textbf{i) Lower Bound:} Given $\varepsilon>0$, there exists $\beta_1$ such that
\begin{align}\label{eqn:lb_robust}
    \liminf_{\beta\to 0}\inf_{\substack{t\in(0,\beta]\\\theta\in \Theta}} \frac{\log v_{1-t}(Q_\beta^{(\theta)})}{\log v_{1-t}(Q)} \geq 1 &\implies  \inf_{\substack{t\in(0,\beta]\\\theta\in \Theta}}  \frac{\log v_{1-t}(Q_\beta^{(\theta)})}{\log v_{1-t}(Q)} \geq 1-\varepsilon, \quad \forall \beta<\beta_1\nonumber\\
\end{align}
This implies that for any such $\beta$, $\log v_{1-t}(Q_\beta^{(\theta)}) \geq (1-\varepsilon) \log v_{1-t}(Q)$ for all $ t\in (0,\beta]$, $\theta\in \Theta$.
Since $Q_\beta^{(\theta)}\in\mathcal Q_{\beta,\delta}^{(\theta)}$, $\rho_{1-\beta}(\mathcal Q_{\beta,\delta}^{(\theta)}) \geq \rho_{1-\beta}(Q_\beta^{(\theta)})$. Therefore, 
\begin{align*}
    \rho_{1-\beta}(\mathcal Q_{\beta,\delta}^{(\theta)})  &\geq \int_{0}^1 w(t) v_{1-\beta t}(Q_\beta^{(\theta)})dt \geq \int_{0}^1 w(t)[v_{1-\beta t}(Q)]^{1-\varepsilon}dt\\
    &= \beta^{-\frac{1-\varepsilon}{\gamma}}\int_0^1 t^{\kappa - \frac{1-\varepsilon}{\gamma}} \ell(1/t) \ell_1(1/(\beta t)) dt
\end{align*}
In the second inequality above $\beta<\beta_1$ where $\beta_1$ is as in \eqref{eqn:lb_robust}. 
Perform a change of variables $s=1/t$ to get
\[
\int_0^1 t^{\kappa - \frac{1-\varepsilon}{\gamma}} \ell(1/t) \ell_1(1/(\beta t))dt = \int_1^\infty \ell(s) s^{\frac{1-\varepsilon}{\gamma} -\kappa- 2}\ell_1(s/\beta) ds.
\]
Let $\alpha = \frac{1-\varepsilon}{\gamma} -\kappa- 2$. Notice that $\ell_1\in \RV(0)$. Applying Potter's bounds to $\ell_1(s/\beta)$ where $s>1$, 
\[
\int_1^\infty \ell(s) s^{\frac{1-\varepsilon}{\gamma} -\kappa- 2}\ell_1(s/\beta)ds \asymp \ell_1(1/\beta) \int_1^\infty s^\alpha \ell(s) ds = C_{\alpha,\varepsilon} \ell_1(1/\beta) \footnote{By this notation, we mean that given $\varsigma>0$, for all $\beta$ sufficiently small  
\[
\ell_1(1/\beta) \int_1^\infty s^{\alpha-\varsigma} \ell(s) ds \leq \int_1^\infty \ell(s) s^{\frac{1-\varepsilon}{\gamma} -\kappa- 2}\ell_1(s/\beta)ds \leq \ell_1(1/\beta) \int_1^\infty s^{\alpha+\varsigma} \ell(s) ds.
\]}
\]
Therefore, $ \rho_{1-\beta}(\mathcal Q_{\beta,\delta}^{(\theta)}) \geq (1-\varepsilon)C_{\alpha,\varepsilon}  \beta^{-\frac{1-\varepsilon}{\gamma}}\tilde \ell(1/\beta) $
where $\tilde l = \ell_1 \ell$ is a slowly varying function. 
Taking logarithms $\log \rho_{1-\beta}(\mathcal Q_{\beta,\delta}^{(\theta)}) \geq \log C_{\alpha,\varepsilon}+ \log \tilde \ell(1/\beta) + (1-\varepsilon)/\gamma \log (1/\beta)$.  From Lemma~\ref{lem:CVaR_regularity}, $\log \rho_{1-\beta}(Q) \leq (1/\gamma+\varepsilon) \log (1/\beta)$ whenever $\beta<\beta_2$, where $\beta_2$ is chosen to be sufficiently small. 
Since $1/\gamma<\kappa+1$, $\alpha<-1$, $C_{\alpha,\varepsilon}<\infty$. Now, choosing $\beta_0 = \min\{\beta_1,\beta_2\}$, whenever $\beta<\beta_0$, for all $\theta\in \Theta$,
\[
\frac{\log \rho_{1-\beta}(\mathcal Q_{\beta,\delta}^{(\theta)}) }{\log \rho_{1-\beta}(Q)} -1 \geq -\varepsilon_0 \text{ where }\varepsilon_0\to 0\text{ whenever }\varepsilon\to 0.
\]
\noindent{\textbf{ii) Upper Bound}: }
We get the following bound, similar to \eqref{eqn:lb_robust}: for any $\varepsilon>0$, there exists a $\beta_3$ such that for all $\beta<\beta_3$, for all $(\theta,\delta)\in \Theta\times \Delta$ and $t\in (0,\beta]$,
\[
\log v_{1-t}(\mathcal Q_{\beta,\delta}^{(\theta)}) \leq (1+\varepsilon) \log v_{1-t}(Q).
\]
Recall that from Lemma~\ref{lem:wc_upper}, $\rho_{1-\beta}(\mathcal Q_{\beta,\delta}^{(\theta)}) \leq \int_{0}^1 w(t) v_{1-\beta t}(\mathcal  Q_{\beta,\delta}^{(\theta)})dt$. Repeating the steps from part (i),
\begin{align*}
    \rho_{1-\beta}(\mathcal Q_{\beta,\delta}^{(\theta)}) \leq   \beta^{-\frac{1+\varepsilon}{\gamma}}\int_0^1 t^{\kappa - \frac{1+\varepsilon}{\gamma}} \ell(1/t) \ell_1(1/(\beta t))dt
\end{align*}
Applying Potter's bounds, with $\alpha_1 = \frac{1+\varepsilon}{\gamma}- \kappa -2$,$
\int_0^1 t^{\kappa - \frac{1+\varepsilon}{\gamma}} \ell(1/t) \ell_1(1/(\beta t))dt \sim  \ell_1(1/\beta)C_{\alpha_1,\varepsilon}$. As before, with $1/\gamma<\kappa+1$, one can choose  $\varepsilon>0$ small enough such that $C_{\alpha_1,\varepsilon}<\infty$.  Further note that from Lemma~\ref{lem:CVaR_regularity}, there exists a $\beta_4$ such that $\log \rho_{1-\beta}(Q) \geq (1/\gamma-\varepsilon) \log (1/\beta)$ whenever $\beta<\beta_4$. Now, with $\tilde \beta_0 = \min\{\beta_3,\beta_4\},$ whenever $\beta<\tilde \beta_0$ for all $(\theta,\delta)\in \Theta\times \Delta$,
\[
\frac{\log \rho_{1-\beta}(\mathcal Q_{\beta,\delta}^{(\theta)}) }{\log \rho_{1-\beta}(Q)} -1 \leq \varepsilon_0 \text{ where }\varepsilon_0\to 0\text{ whenever }\varepsilon\to 0.
\]
Combining the two bounds for every $\varepsilon>0$, for all sufficiently small $\beta$,
\[
\sup_{\substack{\theta\in \Theta\\\delta\in \Delta}}\left\vert \frac{\log \rho_{1-\beta}(\mathcal Q_{\beta,\delta}^{(\theta)}) }{\log \rho_{1-\beta}(Q)} -1\right\vert  \leq C\varepsilon \quad 
\]
Since $\varepsilon$ was arbitrary, the result now follows. \qed

\noindent \textbf{Case ii) $\Lambda_Q\in \RV(\gamma)$: } \textbf{i) Lower Bound:} The first inequality in \eqref{eqn:suff_cond_lt} states that for every $q>1$, there exists $\beta_1$ such that for all $\beta<\beta_1$,
$v_{1-t}(Q_\beta^{(\theta)}) \geq (1-\varepsilon)v_{1-t}(Q)$ 
for all $t\in (\beta^{q},\beta]$ and $\theta\in \Theta$. Therefore, upon a change of variables,
\[
\rho_{1-\beta}(\mathcal Q_{\beta,\delta}^{(\theta)}) \geq \beta^{-1}\int_{0}^\beta w(t/\beta) v_{1-t}(Q_\beta^{(\theta)}) dt \geq (1-\varepsilon)\beta^{-1}\int_{\beta^{q}}^\beta w(t/\beta) v_{1-t}(Q) dt.
\]
Next, observe that in the region $(0,\beta^q]$,  $w(t/\beta) \geq (t/\beta)^{\kappa+\varepsilon}$: to see this, recall that $w(t/\beta) = (t/\beta)^{\kappa}\ell(\beta/t)$. Then
note that when $t\in (0,\beta^q]$, $\beta/t \in (0,\beta^{1-q}]$ and can be made larger than any $t_0$ for small enough $\beta$. Now, 
recall that $\ell\in \RV(0)$. Potter's bounds imply that $\ell(\beta/t) \leq (\beta/t)^{\varepsilon}$ whenever $\beta/t $ is large enough. With this, write the lower bound as
\begin{align}\label{eqn:lb_rho}
    \rho_{1-\beta}(\mathcal Q_{\beta,\delta}^{(\theta)}) &\geq \beta^{-1}\int_0^\beta w(t/\beta) v_{1-t}(Q)dt- \beta^{-1}\int_0^{\beta^q} (t/\beta)^{\kappa-\varepsilon} v_{1-t}(Q)dt\nonumber\\
    &\sim  \rho_{1-\beta}(Q) - C_{\gamma,\varepsilon,q}\,
\beta^{(q-1)(1+\kappa - \varepsilon)}\,
\bigl(\log\tfrac{1}{\beta}\bigr)^{1/\gamma+\varepsilon}
\end{align}
where to simplify the second integral above, we have applied Karamata's theorem followed by the asymptotic for the resulting Gamma function.
Further, $\rho_{1-\beta}(Q) = \log^{1/\gamma}(1/\beta) \ell_1(1/\beta)$, $\ell_1\in \RV(0)$ and thus, with  $q>1$ and $\kappa>-1$, the first term in \eqref{eqn:lb_rho} dominates. Putting everything together, whenever $\beta$ is small enough, for all $(\theta,\delta)\in \Theta\times \Delta$, 
\[
\frac{\rho_{1-\beta}(\mathcal Q_{\beta,\delta}^{(\theta)}) }{ \rho_{1-\beta}(Q)} -1 \geq -\varepsilon_0 \text{ where }\varepsilon_0\to 0\text{ whenever }\varepsilon\to 0.
\]

\noindent\textbf{ii) Upper Bound:}
First recall that from Lemma~\ref{lem:wc_upper}, $\rho_{1-\beta}(\mathcal Q_{\beta,\delta}^{(\theta)}) \leq\beta^{-1}\int_0^\beta w(t/\beta) v_{1-t}(\mathcal Q_\beta^{(\theta)}) dt$. Split the above integral as
\[
\int_0^\beta w(t/\beta) v_{1-t}(\mathcal Q_{\beta,\delta}^{(\theta)}) dt = \underbrace{\int_{\beta^q}^\beta w(t/\beta)v_{1-t}(\mathcal Q_{\beta,\delta}^{(\theta)}) dt}_{(\textbf{a)}} +  \underbrace{\int_0^{\beta^q} w(t/\beta) v_{1-t}(\mathcal Q_{\beta,\delta}^{(\theta)}) dt}_{(\textbf{b)}}  
\]
Let's analyze $\textbf{(a)}$ first. For this, we note that given $\varepsilon>0$, under \eqref{eqn:suff_cond_lt}, for all $\beta$ small enough (say less that $\beta_0$)
\[
\sup_{\substack{t\in [\beta^q,\beta]\\(\theta,\delta)\in \Theta\times \Delta}} \frac{v_{1-t}(\mathcal Q_{\beta,\delta}^{(\theta)})}{v_{1-t}(Q)}\leq (1+\varepsilon) \implies v_{1-t}(\mathcal Q_{\beta,\delta}^{(\theta}) \leq (1+\varepsilon)v_{1-t}(Q) \ \forall  t\in [\beta^q,\beta], (\theta,\delta)\in \Theta\times \Delta 
\]
Therefore,  $\beta^{-1}$ times \textbf{(a)} above can be uniformly bounded over $(\theta,\delta)\in \Theta\times \Delta$ for all $\beta<\beta_0$ by 
$(1+\varepsilon)\beta^{-1}\int_{0}^\beta w(t/\beta) v_{1-t}(Q)  dt = (1+\varepsilon)\rho_{1-\beta}(Q)$. 

To handle \textbf{(b)}, note that 
under the assumptions of the proposition,
for all $\varepsilon>0$, there exists a $\beta_1$ such that for all $\beta<\beta_1$, $v_{1-t}(\mathcal Q_{\beta,\delta}^{(\theta)}) \leq  [v_{1-t}(Q)]^{1+\varepsilon}$ for all $t<\beta^q$, $(\theta,\delta)\in \Theta\times \Delta$.  Note also that $\ell(\beta/t) \geq (\beta/t)^{-\varepsilon}$ for all $\beta<\beta_0$ and $t\in(0,\beta^q]$ where $\beta_0$ is selected such that $\beta^{1-q}>t_0$ for a suitably large $t_0$ required for Potter's bounds to hold.
Following the usual steps, $\beta^{-1}$ times \textbf{(b)} can be upper bounded by 
\begin{align*}
    \beta^{-1}\int_0^{\beta^q} (t/\beta)^{\kappa-\varepsilon} (\log(1/t))^{1/\gamma+\varepsilon} dt  &= \frac{\beta^{-(\kappa+1 -\varepsilon)}}{ (1+\kappa-\varepsilon)^{(1+1/\gamma-\varepsilon)}}
\,\Gamma\left(1+\tfrac{1}{\gamma}+\varepsilon,\ (1+\kappa+\varepsilon)\,q\log\tfrac{1}{\beta}\right) \\
&\sim C_{\gamma,\varepsilon,q}^\prime\,
\beta^{\,(q-1)(1+\kappa-\varepsilon)}\,
\bigl(\log\tfrac{1}{\beta}\bigr)^{1/\gamma+\varepsilon}
\end{align*}
where $\Gamma(s,x) = \int_x^\infty x^{s-1}e^{-x}dx\sim  x^{s-1}e^{-x}$ as $x\to\infty$ is the incomplete gamma function. 
Note that the second term above can be made negligible with respect to $\log^{1/\gamma}(1/\beta)$  since  $q>1,\kappa>-1$ and $\varepsilon>0$ is arbitrary.
Putting this observation together with the bound in \textbf{(a)}, whenever $\beta< \min\{\beta_1,\beta_2\}$, for all $(\theta,\delta)\in \Theta\times \Delta$, 
\[
\rho_{1-\beta}(\mathcal Q_{\beta,\delta}^{(\theta)}) \leq \underbrace{(1+\varepsilon) \rho_{1-\beta}(Q)}_{\text{from \textbf{(a)}}} + \underbrace{\varepsilon\rho_{1-\beta}(Q)}_{\text{from \textbf{(b)}}}. 
\]

This implies that for all $\beta<\min\{\beta_1,\beta_2\}$, for all $(\theta,\delta)\in \Theta\times \Delta$,
\[
\frac{\rho_{1-\beta}(\mathcal Q_{\beta,\delta}^{(\theta)})}{\rho_{1-\beta}(Q)} - 1 \leq 2\varepsilon
\]
Combining this with the lower bound yields that for all $\beta$ sufficiently small, 
\[
\sup_{(\theta,\delta)\in \Theta\times \Delta }\left\vert\frac{\rho_{1-\beta}(\mathcal Q_{\beta,\delta}^{(\theta)})}{\rho_{1-\beta}(Q)} - 1 \right\vert \leq 2\varepsilon
\]
Since $\varepsilon$ was arbitrary, this  completes the proof.\qed

\noindent \textbf{Proof of Lemma~\ref{lem:lb_suff}: }\textbf{Case i) $\bar F_{Q}\in \RV$: }From Lemma~\ref{lem:RVs_ht}, $v_{1-t}(Q_\beta^{(\theta)})= v_{1-\beta_0}(Q)(t/\beta_0)^{-1/\gamma}$ and $v_{1-\beta}(Q) = \beta^{1/\gamma}\ell(1/\beta)$ where $\ell \in \RV(0)$. Note that with $\beta_0 = \beta^{\theta}$ (denote this as  $\beta_0(\theta)$) for some $\theta\in \Theta$, $\sup_{\theta} \beta_0(\theta) \to 0$.  Then, apply Potter's bounds to the function $\ell(1/t)$ to infer that whenever $\beta$ is small enough (say $\beta<\beta_1)$, 
\[
\beta^{\varsigma/\varepsilon} \leq \ell(1/\beta_0) \leq \beta^{-\varsigma/\varepsilon} \text{ for all }\theta\in \Theta.
\]
Applying the above bounds to $v_{1-t}(Q)$, $v_{1-t}(Q) \geq t^{-1/\gamma +\varsigma}$ whenever $t\in(0,\beta]$ and $\beta<\beta_1$. Putting these two components together, whenever $\beta<\beta_1$, 
\[
\frac{\log v_{1-t}(Q_\beta^{(\theta)})}{\log v_{1-t}(Q)} \geq \frac{1-\gamma \varsigma/\varepsilon}{1+\gamma \varsigma}  \text{ for all }  (t,\theta) \in (0,\beta]  \times \Theta \implies \inf_{\substack{t\in(0,\beta]\\\theta\in \Theta}}\frac{\log v_{1-t}(Q_\beta^{(\theta)})}{\log v_{1-t}(Q)} \geq \frac{1-\gamma \varsigma/\varepsilon}{1+\gamma \varsigma}
\]
Since $\varsigma$ above is arbitrary, taking the $\liminf$ leads to 
\[
\liminf_{\beta\to 0} \inf_{\substack{t\in(0,\beta]\\\theta\in \Theta}}\frac{\log v_{1-t}(Q_\beta^{(\theta)})}{\log v_{1-t}(Q)}  \geq 1.
\]

\noindent \textbf{Case ii) $\Lambda_Q\in \RV$: } From Lemma~\ref{lem:RVs_lt} $v_{1-t}(Q_\beta) =v_{1-\beta_0}(Q)(\frac{\log(t)}{\log(\beta_0)})^{1/\gamma}$. Now, write
\begin{equation}\label{eqn:step_1}
    \frac{v_{1-t}(Q_\beta^{(\theta)})}{v_{1-t}(Q)}  = \frac{v_{1-\beta_0}(Q)}{v_{1-t}(Q)}\left(\frac{\log(1/\beta_0)}{\log(1/t)}\right)^{1/\gamma}
\end{equation}
Note also the following: with $t\in (\beta^q,\beta]$ and $\theta\in \Theta$, $  \log \beta_0 \leq \log(1/t) \leq q/\varepsilon \log \beta_0$
Further, with $\Lambda_Q^\leftarrow\in \RV(1/\gamma)$, Theorem B.1.9, \cite{deHaan} shows that
\[
\sup_{b\in [1,M]} \left\vert\frac{\Lambda_Q^{\leftarrow}(b\log (1/t))}{\Lambda_Q^{\leftarrow}(\log (1/t))}  -b^{1/\gamma}\right\vert \to 0,\  \text{ as }t\to 0.
\]
Substitute $M = q/\varepsilon$ above, to obtain that  as $\beta\to 0$,
\[
\sup_{\substack{t\in(\beta^q,\beta]\\ \theta\in \Theta}} \left\vert \frac{v_{1-\beta_0}(Q)}{v_{1-t}(Q)}  \left(\frac{\log(1/\beta_0)}{\log(1/t)}\right)^{1/\gamma} - 1\right\vert \to 0
\]
Substituting this observation into \eqref{eqn:step_1} completes the proof of the lemma.
\qed

\noindent \textbf{Proof of Theorem~\ref{thm:scale_preserving_dro}: }\noindent  \textbf{i)} 
To complete the proof, it is only necessary to establish the upper bound in Proposition~\ref{prop:rate_preserving_suffcient}, since the corresponding
lower bound has already been established in Lemma~\ref{lem:lb_suff}. As usual, we consider two cases:

\noindent \textbf{Case i) $\bar F_Q\in \RV$: }
From Proposition~\ref{prop:sol_of_inverse_unif} the worst case tail-cdf in $\Dphi(Q_{\beta}^{(\theta)})$ satisfies 
\begin{equation}\label{eqn:prop_asymptotics}
    \bar F_{\tt wc,\beta}(x) \sim \phi^{\leftarrow}\left(\frac{\delta}{\bar F_{Q_{\beta}^{(\theta)}}(x)}\right) \bar F_{Q_{\beta}^{(\theta)}}(x),
\end{equation}
where the $\sim$ above is to be understood as uniform convergence over $x\in[v_{1-\beta_0}(Q),\infty)$, $\theta\in \Theta$ and $\delta\in\Delta$ as in Proposition \ref{prop:sol_of_inverse_unif}. 
Observe that $x_t := x_t^{(\theta)} = v_{1-t}(\Dphi(Q_{\beta}^{(\theta)})) \geq v_{1-t}(Q_{\beta}^{(\theta)})$ for $\theta\in [\varepsilon,1-\varepsilon]$. Since $\beta_0=\beta^{\theta}$, this implies that both $x_t\to\infty$ as $t\to 0$, and  $x_t \geq  v_{1-\beta_0}(Q)$ whenever $t\in(0,\beta]$. Substituting $x=x_t$ in \eqref{eqn:prop_asymptotics}, 
\[
    \frac{1}{t}\sim \frac{1}{\phi^{\leftarrow}\left(\delta/\bar F_{Q_{\beta}^{(\theta)}}(x_t)\right) \bar F_{Q_{\beta}^{(\theta)}}(x_t)} = w_{\delta}\left(\frac{1}{\bar F_{Q_{\beta}^{(\theta)}}(x_t)}\right)
\]
uniformly over $t\in (0,\beta]$. Here $w_\delta(u) = (u/\phi^\leftarrow(\delta u))$.
Under \eqref{eqn:phi_growth}, $\phi(x) = \exp(g(x))$ where $g\in \RV(p)$. Consequently, $\phi^{\leftarrow} = g^{\leftarrow}\circ \log \in \RV(0)$ and 
uniformly over $\delta\in \Delta$ (compact set bounded away from $0$), $\phi^{\leftarrow}(\delta tx) \sim \phi^\leftarrow(\delta t)$ as $t\to\infty$.
As a consequence,
$w_{\delta}\in \RV(1)$, uniformly over $\delta\in \Delta$\footnote{a family of real valued functions $\{f_s:s\in S\}$ is said to be uniformly $\RV(p)$ if 
\[
\lim_{t\to\infty}\sup_{s\in S}\left\vert\frac{f_s(tx)}{f_s(t)} - x^p\right\vert = 0; 
\]
see \cite{deHaan}, Appendix B.3.}.
It follows from \eqref{eqn:nominal_tail} that $\bar F_{Q_\beta^{(\theta)}}(x_t) = \beta_0\left(\frac{x_t}{v_{1-\beta_0}(Q)}\right)^{-\gamma}$. Substituting these values in the above display,
\[
w^{\leftarrow}_{\delta}\left(\frac{1}{t}\right) \sim \beta_0^{-1}\left(\frac{v_{1-\beta_0}(Q)}{x_t}\right)^{-\gamma}.
\]
Upon re-arrangement,
\begin{equation}\label{eqn:re-arrange_xt}
    x_t\sim \frac{v_{1-\beta_0}(Q)}{\beta_0^{-1/\gamma}}[w^{\leftarrow}_{\delta}(1/t)]^{1/\gamma}.
\end{equation}
Recall that $\beta_0=\beta^{\theta}$, and therefore $\displaystyle\sup_{\theta\in\Theta}\beta^{\theta} \to 0$ as $\beta\to 0$. Now applying Lemma~\ref{lem:CVaR_regularity}, $\beta_0^{1/\gamma} v_{1-\beta_0}(Q) = \beta^{r^{(\theta)}(\beta_0)}$, where $\sup_{\theta\in \Theta} |r^{\theta}(\beta_0)| \to 0$ as $\beta\to 0$ (that is we can write $\beta^{o(1)}$, uniformly over $\theta\in\Theta$). To conclude, $w_{\delta}^{\leftarrow}\in \RV(1)$ (also uniformly over $\delta\in \Delta$). Applying the uniform Potter's bounds (see \cite[Theorem B.4.2]{deHaan}), for every $\varepsilon>0$, there exists a $t_1$ such that for all $t<t_1$ and $\delta\in \Delta$, $(1-\varepsilon)t^{-1/\gamma +\varepsilon} \leq [w_{\delta}^{\leftarrow}(1/t)]^{1/\gamma} \leq (1+\varepsilon)t^{-1/\gamma -\varepsilon}$. Now, substituting the above in \eqref{eqn:re-arrange_xt},  $x_t \leq t^{-1/\gamma-\varepsilon}$. Then whenever $\beta<\beta_1$, 
\[
\sup_{\substack{t\in (0,\beta]\\(\theta,\delta)\in \Theta\times \Delta}} \frac{\log v_{1-t}(\Dphi( Q_\beta^{(\theta)}))}{\log v_{1-t}(Q)} \leq \frac{1+\gamma \varepsilon}{1-\gamma \varepsilon}.
\]
Since $\varepsilon>0$ was arbitrary, the above suggests that 
\[
\limsup_{\beta\to 0}\sup_{\substack{t\in (0,\beta]\\(\theta,\delta)\in \Theta\times \Delta}}\frac{\log v_{1-t}(\Dphi( Q_\beta))}{\log v_{1-t}(Q)} \leq 1.
\]

\noindent \textbf{Case ii) $\Lambda_Q\in \RV$:}  As in part \textbf{(i)},
\[
 \frac{1}{t}\sim \frac{1}{\phi^{\leftarrow}\left(\delta/\bar F_{Q_\beta^{(\theta)}}(x_t)\right) \bar F_{Q_\beta^{(\theta)}}(x_t)}.
\]
Then,
\begin{equation}\label{eqn:disp_1}
    \log(1/t) \sim \Lambda_{Q_\beta^{(\theta)}}(x_t) -\ln \phi^{\leftarrow}\left(\delta/\bar F_{Q_\beta^{(\theta)}}(x_t)\right).
\end{equation}
Observe now that the function 
$\phi^\leftarrow_\delta(x) = \phi^\leftarrow(\delta x)$ satisfies
$\phi^\leftarrow_\delta \in \RV(0)$ uniformly over $\delta\in \Delta$. Therefore,  $\phi^\leftarrow(\delta/\bar F_{Q_\beta^{(\theta)}}(x)) = (\delta/\bar F_{Q_\beta^{(\theta)}}(x))^{o(1)}$  where the $o(1)$ is uniformly small over $x\in[v_{1-\beta_0}(Q_{\beta_{0}}),\infty)$, $\delta\in \Delta$ and $\theta\in \Theta$ as $\beta\to 0$. Then, \eqref{eqn:disp_1}  
may be simplified as
$\log(1/t) \sim \Lambda_{Q_\beta^{(\theta)}}(x_t)  - o(1)\log \bar F(x_t) \sim \Lambda_{Q_\beta^{(\theta)}}(x_t)$.
This implies that $\Lambda^{\leftarrow}_{Q_\beta^{(\theta)}}(\log(1/t)) \sim x_t$, or that $v_{1-t}(Q_{\beta^{(\theta)}}) \sim x_t$.  
As a consequence of the previous observations, the following hold uniformly over $t\in[\beta^q,\beta]$
\begin{align*}
   \frac{v_{1-t}(\Dphi( Q_\beta^{(\theta)}))}{v_{1-t}(Q)} &\sim \frac{v_{1-t}(Q_\beta^{(\theta)})}{v_{1-t}(Q)} = \frac{v_{1-\beta_0}(Q)}{v_{1-t}(Q)} \left(\frac{\log(1/t)}{\log(1/\beta_0)}\right)^{1/\gamma} \\
   &\sim \frac{\Lambda_{Q}^{\leftarrow}(\log(1/\beta_0))}{\Lambda^\leftarrow_Q(\log(1/t))} \left(\frac{\log(1/t)}{\log(1/\beta_0)}\right)^{1/\gamma}\\
    &\sim \left(\frac{\log(1/t)}{\log(1/\beta_0)}\right)^{-1/\gamma}\left(\frac{\log(1/t)}{\log(1/\beta_0)}\right)^{1/\gamma}.
\end{align*}
Here, the last statement follows since whenever $t\in[\beta^q,\beta]$, $\log(1/t)/\log(1/\beta_0) \in [1/\theta,q/\theta] $, and therefore \cite{deHaan}, Theorem B.1.4 applies to give uniform convergence  of the ratio $\frac{\Lambda_{Q}^{\leftarrow}(\log(1/\beta_0))}{\Lambda^\leftarrow_Q(\log(1/t))}$. 
Hence
\[
\lim_{\beta\to 0}\sup_{\substack{t\in [\beta^q,\beta]\\(\delta,\theta)\in (\Delta,\Theta)}} \left\vert\frac{v_{1-t}(\Dphi(Q_\beta^{(\theta)}))}{v_{1-t}(Q_{\beta}^{(\theta)})} - 1\right\vert =0.
\]

For the second part, recall that $x_t \sim  v_{1-t} (Q_\beta^{(\theta)})$ uniformly over $(\delta,\theta)$. This implies that $\log x_t = \gamma^{-1}\log\log(1/t) + O(1)$ (see Lemma~\ref{lem:RVs_lt}). From Lemma~\ref{lem:CVaR_regularity}, $\log v_{1-t}(Q) \sim \gamma^{-1} \log \log (1/t)$ as $t\to 0 $. Hence, 
\[
\sup_{\substack{t\in(0,\beta^q]\\(\theta,\delta)\in \Theta\times \Delta}} \frac{\log v_{1-t}(\Dphi( Q_\beta)))}{\log v_{1-t}(Q)}\leq (1+\varepsilon), \text{ since } \log v_{1-t}(Q) \geq (\gamma^{-1}-\varepsilon) (\log \log(1/t))
\]
whenever $\beta$ is small enough.
\qed

\noindent \textbf{Proof of Lemma~\ref{lem:intermidiate_var}:} First suppose that $\bar F_Q \in \RV(-\gamma)$. Let $U_{1},\ldots U_n$ be i.i.d. random variables which are uniformly distributed in $[0,1]$. 
Observe that $Z_i = \bar F_Q^\leftarrow(U_i)$, almost surely. Note that $v_{1-\beta_0}(\hat Q^{(n)})$ is defined as the $\lfloor n\beta_0\rfloor$th largest element of the collection $\{Z_i\}_{i\leq n}$.  
Since $\bar F_Q^\leftarrow(\cdot)$ is a decreasing function on $[0,1]$, this equals $\bar F_{Q}^{\leftarrow}\left(U_{\lfloor n\beta_0\rfloor:n}\right)$,
where $U_{i:n}$ is the $i$th order statistic of $\{U_{i}\}_{\leq n}$.  With $\beta_0 = n^{-\theta}$, where $\theta \in  [\varepsilon, 1-\varepsilon]$, $\lfloor n\beta_0\rfloor = \lfloor n^{1-\theta}\rfloor \in [n^{\varepsilon}, n^{1-\varepsilon}]$. Applying Lemma~\ref{lem:ucc_quantiles}, uniformly over $\theta\in [\varepsilon,1-\varepsilon]$,
\[
\frac{\bar F_{Q}^{\leftarrow}\left(U_{\lfloor n\beta_0\rfloor:n}\right)}{\bar F_{Q}^{\leftarrow}(\beta_0)} = \frac{\bar F_{Q}^{\leftarrow}(\beta_0 (1+o(1)))}{\bar F_{Q}^{\leftarrow}(\beta_0)} = (1+o(1))
\]
almost surely.
Here the last statement follows since $\bar F_Q^{\leftarrow}(1/t)$ is regularly varying. Thus,
\[
\frac{v_{1-\beta_0}(\hat Q^{(n)})}{\bar F_Q^\leftarrow(1/\beta_0)} -1 = r_{n,\theta}, \ \text{where almost surely} \lim_{n\to\infty}\sup_{\theta\in[\varepsilon,1-\varepsilon]} |r_{n,\theta}| =0.
\]
Conclude from the above that
\[
\lim_{n\to\infty}\sup_{\theta\in[\varepsilon,1-\varepsilon]} \left\vert\frac{v_{1-\beta_0}(\hat Q^{(n)})}{v_{1-\beta_0}(Q)} -1\right \vert = 0\quad  \text{ a.s}.
\]

\noindent \textbf{ii) }Now suppose $\Lambda_Q\in \RV(\gamma)$. Here, observe that $Z_i = \Lambda_Q^{\leftarrow}(\log(1/U_i))$. Then once again, since $\Lambda_Q^\leftarrow(1/t)$ is a decreasing function on $t\in [0,1]$, $v_{1-\beta_0}(\hat Q^{(n)})$ equals $\Lambda_{Q}^{\leftarrow}\left(-\log U_{\lfloor n\beta_0\rfloor:n}\right)$. Similar to the previous case, note that 
\[
\frac{\Lambda^{\leftarrow}_Q(-\log U_{\lfloor n\beta_0\rfloor:n})}{\Lambda^{\leftarrow}_Q(\log(1/\beta_0))} = \frac{\Lambda^{\leftarrow}_Q\left(\frac{\log (1/U_{\lfloor n\beta_0\rfloor:n})}{\log(1/\beta_0)} \log(1/\beta_0)\right)}{\Lambda^{\leftarrow}_Q(\log(1/\beta_0))} (1+o(1)).
\]
Using the uniform convergence in Lemma~\ref{lem:ucc_quantiles}:
\[
\frac{v_{1-\beta_0}(\hat Q^{(n)})}{v_{1-\beta_0}(Q)} = (1+o(1)), \  \text{ almost surely uniformly over } \theta \in [\varepsilon,1-\varepsilon]
\]
thereby completing the proof.\qed

Before proceeding with the next proof, we note down a straightforward consequence of Lemma~\ref{lem:ucc_quantiles}. Let $Y_1,\ldots Y_n$ be i.i.d. samples from a distribution with cdf $F(y) = 1-1/y$ $y\geq 1$. Then for any $\varepsilon>0,$
\begin{equation}\label{eqn:fact_lb_y}
\liminf_{n\to\infty} Y_{(\lfloor n^{\varepsilon} \rfloor)} = \infty \quad \text{ almost surely}.    
\end{equation}

\noindent\textbf{Proof of Lemma~\ref{lem:consistency_of_our_estimates}: }We prove the result in two parts

\noindent \textbf{i) Uniform Consistency of Hill Estimator: } Recall that if $\bar F_Q\in \RV(-\gamma)$, the function $U_Q(t) = \bar F_Q^{\leftarrow}(1/t)$ is regularly varying with index $1/\gamma$. Also observe that the Hill-estimator has the representation (see \cite{deHaan}, Pg. 70)
\[
\hat \gamma(k_n) = \left(\frac{1}{k_n} \sum_{i=1}^{k_n} \log\left(\frac{U_Q(Y_{(i)})}{U_Q(Y_{(k_n)})}\right)\right)^{-1}.
\]
From \eqref{eqn:fact_lb_y}, given $t_0>0$, almost everywhere, there exists an $n_0$ such that $Y_{\lfloor n^{\varepsilon} \rfloor} > t_0 $ (and consequently, $\inf_{\theta\in \Theta} Y_{\lfloor n^{\theta} \rfloor} > t_0$). Therefore, Potter's bounds apply to the function $U_Q(\cdot)$: for any $\eta>0$ almost surely, for all $\theta\in \Theta,$
\begin{equation}\label{eqn:apply_potter}
    \log(1-\eta) + \frac{(1/\gamma-\eta)}{k_n}\sum_{i=1}^{k_n}\log\left(\frac{Y_{(i)}}{Y_{(k_n)}}\right)\leq \frac{1}{\hat \gamma(k_n)} \leq \log(1+\eta) + \frac{(1/\gamma+\eta)}{k_n}\sum_{i=1}^{k_n}\log\left(\frac{Y_{(i)}}{Y_{(k_n)}}\right)  
\end{equation}
 whenever $n>n_0$.
From the Renyi representation, we have 
\[
\frac{1}{k_n}\sum_{i=1}^{k_n}\log\left(\frac{Y_{(i)}}{Y_{(k_n)}}\right) = \frac{1}{k_n} \sum_{i=1}^{k_n}E_i \text{ where } E_i \text{ are i.i.d. standard exponential}.\footnote{we assume that our probability space supports these exponential random variables. This is always possible by enlarging the underlying probability space if needed.}
\]
Let $I_n := \{k\in\mathbb N:\ n^\varepsilon \le k \le n^{1-\varepsilon}\}$ and
define $H_{k,n}$ as the right hand side above. Suppose we can demonstrate the following uniform convergence: $\sup_{k\in I_n} |H_{k,n}-1|\to 0$ almost surely as $n\to \infty$, then the uniformly of the bounds in  \eqref{eqn:apply_potter} over $\theta$ would mean that almost surely
\[
\lim_{n\to\infty}\sup_{\theta\in [\varepsilon,1-\varepsilon]}\left\vert\frac{1}{k_n} \sum_{i=1}^{k_n} \log\left(\frac{U_Q(Y_{(i)})}{U_Q(Y_{(k_n)})}\right) -\frac{1}{\gamma}\right\vert = 0 \quad \text{ almost surely}.
\]
proving the result. It then remains to show that $\sup_{k\in I_n} |H_{k,n}-1|\to 0$ almost surely.
Define the event $B_{k,n}(\eta) = \{|H_{k,n}-1|>\eta\}$. Use the Chernoff bound:
\[
\mathbb P\left(B_{k,n}(\eta) \right) \leq \exp(-C(\eta)k), \text{ for a suitable constant } C(\eta)>0.
\]
Then, for some constant $C(\eta)$,
\[
\mathbb P\left(\sup_{k\in I_n} |H_{k,n}-1| >\eta \right) \leq \sum_{k\in I_n} P\left(B_{k,n}(\eta) \right) \leq K_0(\eta) n \exp(-C(\eta) n^{\varepsilon})
\]
Importantly, the right hand side above is summable. Defining $A_{n} = \sup_{k\in I_n} |H_{k,n}-1|$,  
\[
\lim_{n\to\infty }\mathbb P( \sup_{r\geq n} |A_r| > \eta) =0 
\]
for any $\eta>0$. \cite[Theorem 1, Section 2.10]{shiryaev1996probability}  implies that $A_{n}\to 0$ almost surely, or that $\sup_{k\in I_n} |H_{k,n}-1| \to 0$ almost surely.

\noindent{\textbf{ii) Uniform consistency when $\Lambda_Q\in \RV(\gamma)$:}}
 Recall that $\beta_i$, $i \in\{0,1\}$ satisfy $\lfloor n\beta_i(n)\rfloor \in [n^{\varepsilon},n^{1-\varepsilon}]$ and, the conclusions of Lemma~\ref{lem:intermidiate_var} apply. As a result, $Z_{\lfloor n\beta_i(n)\rfloor:n} = v_{1-\beta_i}(Q)(1+r_{i,k,n})$ where $\sup_{i,k\in I_n} |r_{i,k,n}| \to 0$ a.s. as $n\to\infty$. Hence,
\begin{align*}
    \log \left(\frac{Z_{\lfloor n\beta_0(n)\rfloor:n}}{Z_{\lfloor n\beta_1(n)\rfloor:n}}\right) & = \log \left(\frac{v_{1-\beta_0}(Q)}{v_{1-\beta_1}(Q)} (1+r_{k,n}) \right) \quad \text{ where } \sup_{k\in I_n} |r_{k,n}| \xrightarrow{a.s.} 0 \\
    & =(1+o(1)) \log\left(\frac{\log^{1/\gamma}(1/\beta_0) \ell_0(\log(1/\beta_0))}{\log^{1/\gamma}(1/\beta_1) \ell_0(\log(1/\beta_1))}\right) \quad \text{since $\Lambda_Q\in \RV(\gamma)$}\\
    &=(1+o(1)) \log\left({\kappa^{-1/\gamma}_1}\right),
\end{align*}
uniformly over $k\in I_n$,
almost surely.
Here the last step is because $\ell_0\in \RV(0)$. Rearranging the last line above, $\hat \gamma_n = \gamma(1+o(1))$ which proves the lemma.\qed

In order to prove the consistency result in Theorem~\ref{thm:consistency_of_dro}, the following stochastic versions of Propositions~\ref{prop:sol_of_inverse} and \ref{prop:rate_preserving_suffcient} are required:

\begin{lemma}\label{lema:sol_of_inverse_dd}
    Let the conditions of Proposition~\ref{prop:sol_of_inverse} hold, and let $Q_\beta$ be given by \eqref{eqn:nominal_dist}. Then 
    \begin{equation}\label{eqn:sol_of_inverse_dd}
       \sup_{\substack{x\geq v_{1-\beta_0}(\hat Q^{(n,\theta)})\\\theta\in\Theta,\delta\in \Delta}}\left\vert \frac{\bar F_{{\tt wc},n}(x) }{\phi^\leftarrow\left(\frac{\delta}{\bar F_{\hat Q_\beta^{(n,\theta)}}(x)}\right)\bar F_{\hat Q_\beta^{(n,\theta)}}(x)}  -1\right\vert = o(1)
\end{equation}
almost surely, where $\bar F_{{\tt wc},n}(x) = \sup\{ \bar F_P(x): {P\in \mathcal \Dphi(\hat Q_\beta^{(n,\theta)})}\}$ is the worst-case cdf.
\end{lemma}
Recall that in the setup of Theorem~\ref{thm:consistency_of_dro}, $\beta_0(n,\theta) =n ^{-\theta} $ and $\mathcal B_n=\{\beta(n): \beta(n) = c_0 n^{-q}, \text{ for }  c_0\in (c_-,c+),  q\in [1,M]\}$

\begin{lemma}\label{lem:stochastic_sufficient}
     In the setup of Theorem~\ref{thm:consistency_of_dro} 
     the following hold almost surely as $n\to\infty$:
    \begin{enumerate}
        \item[(i)] If $\bar F_Q\in \RV$, then 
        \[
        \inf_{\substack{t\in (0,\beta]\\\theta\in \Theta, \beta(n) \in\mathcal B_n}}\frac{\log v_{1-t}(\hat Q_\beta^{(n,\theta)})}{\log v_{1-t}(Q)} = 1+o(1) \quad \text{ and }\quad  \sup_{\substack{t\in (0,\beta],\beta(n)\in \mathcal B_n \\(\theta,\delta)\in \Theta\times \Delta}}\frac{\log v_{1-t}(\Dphi(\hat Q^{(n,\theta)}_\beta))}{\log v_{1-t}(Q)} = 1+o(1). 
        \]
        \item[(ii)] If instead $\Lambda_Q\in \RV$, then for any $q>1$
        \[
        \inf_{\substack{t\in (\beta^q,\beta]\\\theta\in \Theta, \beta(n) \in\mathcal B_n}}\frac{ v_{1-t}(\hat Q_\beta^{(n,\theta)})}{ v_{1-t}(Q)} = 1+o(1) \quad \text{ and }\quad  \sup_{\substack{t\in (\beta^\kappa,\beta], \beta(n)\in \mathcal B_n \\(\theta,\delta)\in \Theta\times \Delta}}\frac{ v_{1-t}(\Dphi(\hat Q^{(n,\theta)}_\beta))}{ v_{1-t}(Q)} = 1+o(1). 
        \]
        Further, the second stochastic bound in part (i) above holds over $t\in (0,\beta^q]$.
    \end{enumerate}
\end{lemma}

\noindent \textbf{Proof of Theorem~\ref{thm:consistency_of_dro}: }
Let $\Omega_0$ be the set on which the convergence in Lemma~\ref{lem:stochastic_sufficient} occurs. 
Let us re-parametrize $\beta_0$ as $\beta_0= c_0 \beta^{\theta/q}$. Note that whenever $q\in [1,M]$ and $\theta\in[\varepsilon,1-\varepsilon],$ $\theta/q\in [\varepsilon/q,1-\varepsilon/q]$. With the above re-parametrization, for  every $\omega\in \Omega_0$, Proposition~\ref{prop:rate_preserving_suffcient} implies that so long as $\varepsilon<1/2$, for $\Theta= [\varepsilon/q,1-\varepsilon/q]$ the family of ambiguity sets $\{ \Dphi(\hat Q_{\beta}^{(n,\theta)}): (\theta,\delta)\in \Theta\times \Delta\}$ are strongly (or weakly) $Q-$rate preserving.
Since $\mathbb P(\Omega_0)=1$ and $\varepsilon$ was arbitrary,  the proof of the theorem is complete.\qed

\noindent \textbf{Proof of Proposition~\ref{prop:correctness_of_algo}: } Observe that one can write 
\[
\frac{1}{n}\sum_{i=1}^n \phi^*\left(\frac{(Z_i-u)^+ -\eta}{\lambda}\right ) \mv{1}(Z_i\leq v_{1-\beta_0}(\hat Q^{(n)}))  = \int \phi^*\left(\frac{(x-u)^+ -\eta}{\lambda}\right)\mv{1}(x\leq v_{1-\beta_0}(\hat Q^{(n)}))d\hat Q^{(n)}(x).
\]
Define the \textit{deterministic} (since here $\hat Q^{(n)}$ is fixed) and stochastic functions
\begin{align*}
    f(u, \eta,\lambda) &=  \lambda E_{\hat Q_{\beta}^{(n)}}\left[\varphi^*\left(\frac{(Z-u)^+ -\eta}{\lambda}\right)\right]\quad \text{and}\\
    f_N(u,\eta,\lambda) & =  \lambda  \int \phi^*\left(\frac{(x-u)^+ -\eta}{\lambda}\right)\mv{1}(x\leq v_{1-\beta_0}(\hat Q^{(n)}))d\hat Q^{(n)}(x)\\
    &+  \lambda  \frac{\beta_0}{N}\sum_{i=1}^N \phi^*\left(\frac{(\tilde Z_i-u)^+ -\eta}{\lambda}\right).
\end{align*}
If in addition,  $f_N\to f$ uniformly over $(u,\eta,\lambda)$. Then by \cite{shapiro2021lectures} Proposition 5.2, it can be established that $\hat C_{1-\beta}^{N}(\Dphi(\hat Q_{\beta}^{(n)})) \to  C_{1-\beta}(\Dphi(\hat Q_{\beta}^{(n)}))$ almost surely as $N\to\infty $. The rest of the proof therefore focuses on proving $f_N\to f$, uniformly over compact sets.

\noindent \textbf{i) } Recall the inverse transform method (see \cite{Glynn}, Chapter 1): for a random variable $Z$ with  distribution $P$, $F^{\leftarrow}_P(U_t)$ has the distribution of $(Z\mid Z\geq v_{1-t}(P))$ where $F$ is the cdf of $P$, and $U_t$ is uniformly distributed in $[1-t,1]$. 
Note that for $t\in [v_{1-\beta_0}(Q),\infty)$
\[
G_{\beta}(t) = \begin{cases}
    1-\beta_0 \left(\frac{t}{v_{1-\beta_0}(Q)}\right)^{-\gamma} \quad \quad\quad\quad\  \quad\text{ if $\bar F_Q\in \RV(-\gamma)$}\\
1-\exp\left(\log(\beta_0)\left(\frac{t}{v_{1-\beta_0}(Q)}\right)^\gamma\right)      \quad\text{ if } \Lambda_Q\in \RV(\gamma),
    \end{cases}  
\]
Thus whenever $y\in [1-\beta_0,1]$,
\[
[G_{\beta}]^{\leftarrow}(y) = \begin{cases}
    v_{1-\beta_0}(Q) \left(\frac{1-y}{\beta_0} \right)^{1/\gamma} \quad \quad\quad\quad\  \quad\text{ if $\bar F_Q\in \RV(-\gamma)$}\\
 v_{1-\beta_0}(Q) \left(\frac{\log(1-y)}{\log(\beta_0)} \right)^{1/\gamma}      \quad\quad\quad \quad \text{ if } \Lambda_Q\in \RV(\gamma).
    \end{cases} 
\]
Hence, by construction,
$\tilde Z = [\hat G_{\beta_0}^{(n)}]^{\leftarrow}(U_{\beta_0})$. Therefore, $\tilde Z_i$ are i.i.d. draws from the distribution of $(X_\beta \mid X_\beta \geq v_{1-\beta_0}(\hat Q^{(n)}_\beta))$ where $X_\beta$ has a cdf $\hat G_{\beta_0}^{(n)}$. 

\noindent \textbf{ii) } Now, the expected value of $\hat f_N(u,\eta,\lambda)$ equals $\lambda E_{\hat Q^{(n)}} [\phi^*(\lambda^{-1}(Z-u)^+-\eta));Z\leq v_{1-\beta_0
}(\hat Q^{(n)})]$ plus $\lambda $ times
\begin{align*}
    \beta_0 E\left[\phi^*\left(\frac{(X_\beta-u)^+-\eta}{\lambda}\right)\,\Bigg \vert\  X_{\beta}\geq v_{1-\beta_0}(\hat Q^{(n)})\right] = \int_{x\geq v_{1-\beta}(\hat Q^{(n)})} \phi^*\left(\frac{(x-u)^+-\eta}{\lambda}\right)d{\hat G_{\beta}^{(n)}}(x),  
\end{align*}
where the second statement follows since by definition $\hat G_{\beta}(v_{1-\beta_0}(\hat Q^{(n)}))  = 1-\beta_0$. Putting both the above together suggests that the expected value of $f_N(u,\eta,\lambda)$ equals $f(u,\eta,\lambda)$. Applying the law of large numbers, $f_N \to f$ point-wise in $(u,\eta,\lambda)$.  

\noindent \textbf{iii)} Finally, note that since the function $\ell(z,u) = (z-u)^+$ is convex in $u$ for all $z$, and $\phi$ is convex, $f_N(\cdot)$ is jointly convex in $(u,\eta,\lambda)$.  
Further $f(\cdot)$ is continuous and its domain has a non-empty interior. 
Coupled together with the point-wise convergence from step (ii),    \cite{shapiro2021lectures} Theorem 9.61 implies that $f_N\to f$ uniformly over $(u,\eta,\lambda)$ in compact sets. \qed

\subsection{Proofs from Section~\ref{sec:application}}
\noindent{\textbf{Proof of Proposition~\ref{prop:mv_cvar}:}}  \textbf{Part i) Heavy Tailed $\XX$: }  Let $s_t$ be such that $a_{s_t} =t^{1/\vartheta}$. Since $a_t\in \RV(1/\alpha)$, observe that $s_t\in \RV(\alpha/\vartheta)$. Write $\XX_t = \XX/s_t$ and let $L_t(\zz) = t^{-1}L(s_t\zz)$. Now, write the probability $L(\XX)\geq t$ as
\begin{align}\label{eqn:ht_ub}
    Q\{Z>t\}&= \tilde Q\left\{L(\XX) \geq t\right\} = Q\left\{\XX_t\in \Lev_1^+(L_t) \right\}\nonumber\\
    & = Q\left\{\XX_t\in \Lev_1^+(L_t) \cap B_M\right\} + Q\left\{\XX_t\in \Lev_1^+(L_t) \cap B_M^c\right\}
\end{align}
\textbf{a) Upper Bound:} Under Assumption~\ref{assume:loss_functional}, $L_t \to L^*$ continuously, and therefore epigraphically (see \cite[Theorem 7.11]{Wets}). As a result, for $\varsigma>0$, for all large enough $t$, $\left\{\XX_t\in \Lev_1^+(L_t) \cap B_M\right\} \subseteq \{\XX_t  \in [\Lev_1^+(L_t) \cap B_M]^{1+\varsigma}\}$. Substituting this into \eqref{eqn:ht_ub},
\begin{align*}
    \limsup_{t\to\infty} s_t \,Q\{Z>t\} &\leq \limsup_{t\to\infty} s_t \left( \tilde Q\left\{\XX_t \in  [\Lev_1^+(L_t) \cap B_M]^{1+\varsigma}\right\} +\tilde  Q\left\{B_M^c\right\}\right) \\
    &\leq \nu([\Lev_1^+(L^*)]^{1+\varsigma}) + \nu(B_M^c).
\end{align*}
where the last step follows from the convergence in Assumption~\ref{assume:mrv}(i). 
Since $(\varsigma,M)$ above were arbitrary, this implies that
\[
\limsup_{t\to\infty} s_t \,Q\{Z>t\} \leq \nu(\Lev_1(L^*)) \footnote{Observe that  as $(\varsigma,M) \to (0,\infty)$,  $[\Lev_1^+(L^*)]^{1+\varsigma} \downarrow \Lev_1^+(L^*)$ and $B_M^c \downarrow \emptyset$. The bound follows upon a use of continuity of measure}. 
\]

\noindent \textbf{b) Lower Bound:} From \eqref{eqn:ht_ub},
$s_t \,Q\{Z>t\} \geq s_t \tilde  Q\left\{\XX_t\in \Lev_1^+(L_t) \cap B_M\right\}.$
Using epi-graphical convergence, for any $\varsigma$, the right hand side above is lower bounded by  $ s_t \tilde  Q\left\{\XX_t\in \Lev_{1+\varsigma}^+(L^*) \cap B_M\right\}$ for all $t$ large enough. Since $(\varsigma,M)$ were arbitrary
\[
\liminf_{t\to\infty} s_t \,Q\{Z>t\} \geq \nu(\Lev_1^+(L^*))
\]
Consequently, $\displaystyle\lim_{t\to\infty} s_t \,Q\{Z>t\} = \nu(\Lev_1^+(L^*)) > 0$.  Then, $Q\{Z>t\} \in \RV(-\alpha/\nu)$. 

\noindent \textbf{Part ii) Light Tailed $\XX$: } Under the assumptions of Assumptions~\ref{assume:loss_functional} and \ref{assume:mrv}(ii), \cite[Theorem 2]{deo2023achieving} demonstrates that $\displaystyle\lim_{t\to\infty} [\Lambda(t^{1/\vartheta})]^{-1} \log \mathbb P\left(L(\XX) \geq t\right) \to I^*$ for a suitable $I^*>0$. Since $\Lambda\in \RV(\alpha)$, $\Lambda(t^{1/\vartheta}) \in \RV(\alpha/\vartheta)$, the theorem is proved.\qed

\noindent \textbf{Proof of Lemma~\ref{lem:lift_to_univariate}: }
Set $Z=L(\XX)$ and
let $P_Z=P_0\circ L^{-1}$ denote the law of $Z$ under $P_0$.

\noindent\textbf{Step 1 : data processing.}
Fix $\tilde P\in\Dphi(P_0)$.
By the data–processing inequality for $\phi$–divergences,
\[
\mathcal D_{\phi}\!\big(\tilde P\circ L^{-1}\,\big\Vert\,P_0\circ L^{-1}\big)
\;\le\;
\mathcal D_{\phi}(\tilde P\Vert P_0)
\;\le\; \delta,
\]
so $\tilde P\circ L^{-1}\in \mathcal \Dphi(P_Z)$, where $P_Z:=P_0\circ L^{-1}$.
Hence every admissible $\tilde P$ on the left produces an admissible push-forward law on $\mathbb R$:
\[
\big\{\tilde P\circ L^{-1}:\ \tilde P\in\mathcal \Dphi(P_0)\big\}
\ \subseteq\
\mathcal \Dphi(P_Z) \implies \sup_{\tilde P\in\mathcal \Dphi(P_0)} \rho(\tilde P\circ L^{-1})
\ \le\
\sup_{P\in \mathcal \Dphi(P_Z)} \rho(P),
\]
which is the desired “$\le$” inequality.

\noindent\textbf{Step 2: lift from $Z$ to $\XX$.}
Let $R\in  \Dphi(P_Z)$ and let $h:=\frac{dR}{dP_Z}$. Note that since $R\in \Dphi(P_Z)$, one must have $R\ll P_Z$, and therefore $h$ is a well defined likelihood. 
Define a probability measure $\tilde P$ on $\mathbb R^d$ by
\[
\frac{d\tilde P}{dP_0}(x)\;:=\; h\!\big(L(x)\big), \qquad x\in\mathbb R^d.
\]
Then $\tilde P\ll P_0$ and $\tilde P(\mathbb R^d)=\int h(L(x))\,dP_0(x)=\int h(z)\,dP_Z(z)=1$.
Moreover, for any bounded Borel $g:\mathbb R\to\mathbb R$,
\[
\int g(z)\,d(\tilde P\circ L^{-1})(z)
=\int g(L(x))\,h(L(x))\,dP_0(x)
=\int g(z)\,h(z)\,dP_Z(z)
=\int g(z)\,dR(z),
\]
so $\tilde P\circ L^{-1}=R$.
Finally, the divergence matches exactly:
\[
\mathcal D_\phi(\tilde P\Vert P_0)
=\int \phi\big(h(L(x))\big)\,dP_0(x)
=\int \phi\big(h(z)\big)\,dP_Z(z)
= \mathcal D_\phi(R\Vert P_Z)
\le \delta,
\]
hence $\tilde P\in \mathcal \Dphi(P_0)$.
Therefore, for this \emph{arbitrary} $R$ we have
\[
\rho(R) = \rho(\tilde P\circ L^{-1})
\ \le\ \sup_{\tilde P\in\mathcal \Dphi(P_0)} \rho(\tilde P\circ L^{-1}).
\]
Taking the supremum over all $R\in\mathcal D_\phi(P_Z,\delta)$ yields
\[
\sup_{R\in\mathcal \Dphi(P_Z)} \rho(R)
\ \le\ \sup_{\tilde P\in\mathcal \Dphi(P_0)} \rho(\tilde P\circ L^{-1}),
\]
which is the desired ``$\ge$'' inequality. Combining the two steps establishes
\[
\sup_{\tilde P \in \mathcal \Dphi(P_0)} \rho(\tilde P\circ L^{-1})
\;=\;
\sup_{P\in \mathcal \Dphi(P_Z)} \rho(P).
\]
\textbf{Proof of Lemma~\ref{lem:loss_to_factor_lift}: }
Let $X:=\mathbb R_+^d$, $Y:=\mathbb R_+$ and $S:=L(X)\subset Y$. Now, 
\[
\Gamma \;:=\; \{(x,y)\in X\times S:\ y=L(x)\}
\]
is a subset of $X\times S$ and its projection onto $S$ equals $S$. Since $L(\cdot)$ is assumed to be measurable, the Jankov-von Neumann measurable selection theorem applies and there exists a map $s:S\to X$ such that $L(s(y))=y$ for all $y\in S$.
Define $\tilde Q_\beta =Q_\beta \circ s^{-1}$ that is for all measurable $A\subset X$, $\tilde Q_\beta(A)=Q_\beta\big(s^{-1}(A)\big)$. Then, for $B\subset Y$
\begin{align*}
    (\tilde Q_\beta \circ L^{-1})(B)&\;=\;\tilde Q_\beta\!\big(L^{-1}(B)\big)
\;=\;Q_\beta\!\big(s^{-1}(L^{-1}(B))\big)
\\
&\;=\;Q_\beta\!\big(\{y\in S:\ L(s(y))\in B\}\big)
\;=\;Q_\beta(B\cap S)
\;=\;Q_\beta(B),
\end{align*}
where the last statement follows since $Q_\beta(S)=1$. Hence $Q_\beta = \tilde Q_\beta\circ L^{-1}$ as claimed.  \qed

\section{Proofs of Technical Lemmas}
\noindent \textbf{Proof of Proposition~\ref{prop:sol_of_inverse_unif}: }
Note that from \cite{blanchet2020distributionally}, Corollary 1, the following must hold true\footnote{we write $s_\beta(x)$ for $s_{\beta,\delta}^{(\theta)}(x)$ below to avoid notational clutter. For similar reasons, we write $\bar F_{\tt wc, \beta}  $ for $\bar F_{\tt wc, \beta,\delta}^{(\theta)}$}
\begin{align}\label{eqn:wc_probs}
    \delta & = (1-\bar F_{Q_\beta^{(\theta)}}(x)) \phi\left(\frac{1-s_{\beta}(x) \bar F_{Q_\beta^{(\theta)}}(x)}{1-\bar F_{Q_\beta^{(\theta)}}(x)}\right) + \bar F_{Q_\beta^{(\theta)}}(x) \phi(s_\beta(x)) \nonumber\\
    \bar F_{{\tt wc},\beta}(x)&=s_\beta(x) \bar F_{Q_\beta^{(\theta)}}(x)
\end{align}
Recall that $\phi(1) = \phi^\prime(1)=0$. Expanding the first equation of \eqref{eqn:wc_probs} about $y=1$ using a Taylor series and simplifying yields
\begin{equation}\label{eqn:tse}
    \frac{\delta}{\bar F_{Q_\beta^{(\theta)}}(x)} = \underbrace{\frac{\phi^{\prime\prime} (r_\beta(x))}{2(1-\bar F_{Q_\beta^{(\theta)}}(x))} s_\beta(x)  \bar F_{Q_\beta^{(\theta)}}(x) (s_\beta(x)-1) \left(1-\frac{1}{s_\beta(x)}\right)}_{(a)} + \phi(s_\beta(x)), 
\end{equation}
where $r_\beta(x) \in (1-y(x),1)$ where 
\[
y(x) = \frac{1-s_\beta(x) \bar F(x)}{1-\bar F(x)}.
\]
Since $\phi^{\prime\prime}$  is continuous (note that $\phi$ is twice continuously differentiable), it is bounded in the vicinity of $1$. Therefore, for $\beta$ small enough, $\phi^{\prime\prime}(r_\beta(x)) <\infty$
 Note that from \eqref{eqn:nominal_tail} in the region $x\geq v_{1-\beta_0}(Q)$, $\bar F_{Q_\beta^{(\theta)}}(x) \leq  \beta^{\varepsilon+o(1)}_0$ for all $\theta\in \Theta$. Recall that $\Delta $ is a compact subset of $(0,\infty)$. 
Hence, $\displaystyle \inf_{\substack{x\geq v_{1-\beta_0}(Q)\\\theta\in \Theta,\delta\in \Delta}} s_{\beta}(x) \to \infty$ as $\beta\to 0$, 
for otherwise the
left hand side of \eqref{eqn:tse} will become infinite, but the right hand side will remain finite as $\beta\to 0$. 
Next, since $\phi(x) /x\to \infty$,  
$\phi^\leftarrow(u)/u\to 0$ as $u\to\infty$. From \eqref{eqn:tse},
\[
\frac{\delta}{\bar F_{Q_\beta}^{(\theta)}(x)} \geq \phi(s_\beta(x)) \implies \bar F_{Q_\beta}^{(\theta)} s_\beta(x) \leq \phi^{\leftarrow}\left(\frac{\delta}{\bar F_{Q_\beta}^{(\theta)}(x)}\right) \bar F_{Q_\beta}^{(\theta)}.
\]
Since $\displaystyle \sup_{\substack{x\geq v_{1-\beta_0}(Q)\\\theta\in \Theta}} \bar F_{Q_\beta^{(\theta)}}(x) \to 0$ as $\beta\to 0$, the above display implies that
$\displaystyle\sup_{\substack{x\geq v_{1-\beta_0}(Q)\\\theta\in \Theta,\delta\in \Delta}} \bar F_{Q_\beta}(x) s_\beta(x) \to 0$ as $\beta$ to $0$. Putting these two facts together,  the term labeled \textbf{(a)} in \eqref{eqn:tse} can be made arbitrarily small with respect to $\phi(s_\beta(x))$, uniformly over $x\geq v_{1-\beta}(Q)$, $\delta\in \Delta$ and $\theta\in \Theta$ as $\beta\to 0$. Thus, 
\[
\frac{\delta}{\bar F_{Q_\beta^{(\theta)}}(x)}\sim \phi(s_\beta(x)) \implies s_\beta(x) \sim \phi^\leftarrow\left(\frac{\delta}{\bar F_{Q_\beta^{(\theta)}}(x)}\right)
\]
where once again $\sim$ is to be read in the context of Proposition~\ref{prop:sol_of_inverse_unif}. This further implies that $F_{{\tt wc},\beta}(x) \sim \bar F_{Q_\beta}(x) \phi^\leftarrow(\delta/F_{Q_\beta}(x))$, thereby establishing Proposition~\ref{prop:sol_of_inverse_unif}.\qed

\noindent \textbf{Proof of Lemma~\ref{lem:Dual_Mult}: }
We complete the proof in two steps. 

\noindent \textbf{Part 1 - $p>1$:} Fix $z$ and note that the first order condition implies that any maximizer of $g_{\lambda}(y) = (y-u)^+ -\lambda |y-z|^p$ satisfies
\begin{equation}\label{eqn:foc}
\mv{1}(y^*\geq u) = \lambda p |y^*-z|^{p-1} {\tt sgn}(y^*-z),    
\end{equation}
where ${\tt sgn}(x)$ is the sign function which takes the value $1$ if $x\geq0 $ and $-1$ otherwise.
To solve \eqref{eqn:foc}, either $y^*<u$ and $y^* =z$ (so that both sides of \eqref{eqn:foc} equal $0$), or one must have that $y^*\geq u$, and
\begin{equation}\label{eqn:case_2}
    |y^*-z|^{p-1}{\tt sgn}(y^*-z) = \frac{1}{\lambda p}  \implies y^* = z+  (\lambda p )^{\frac{-1}{p-1}}.
\end{equation}
Let us analyze the solution case by case. 

\noindent \textbf{Case 1:} Let us first suppose that $z<\vartheta(u,\lambda)$. Then, if in addition, $z< u - (\lambda p)^{-\frac{1}{p-1}}$, $y_1^*= z$ is the only solution to \eqref{eqn:foc}, since then $ z+ (\lambda p )^{\frac{-1}{p-1}} < u$ and so \eqref{eqn:case_2} cannot hold. Next, suppose $z>u- (\lambda p )^{\frac{-1}{p-1}}$. In this case, both $y_1^*=z$ and $y_2^*=z+ (\lambda p )^{\frac{-1}{p-1}}$ solve \eqref{eqn:foc}. To evaluate the maximizer, we find that $g_\lambda(y_1^*)=0$ and 
\begin{align*}
    g_\lambda(y_2^*) &= \left(z-u +(\lambda p)^{\frac{-1}{p-1}}\right)^+  -\lambda (\lambda p )^{-\frac{p}{p-1}}\\
    & <  \left(\frac{1}{(\lambda p)^{\frac{1}{p-1}}} - \frac{1-1/p}{(\lambda p)^{\frac{1}{p-1}}}\right) - \frac{1}{p (\lambda p)^{\frac{1}{p-1}}} = 0
\end{align*}
where the last statement follows, since additionally $z< \vartheta(u,\lambda)$.
Thus  $y^*=z$ if $z<\vartheta(u,\lambda)$. Therefore, when $z<\vartheta(u,\lambda)$, $g(y^*)=0$.

\noindent \textbf{Case 2:} Now, suppose that $z> \vartheta(u,\lambda)$. Suppose in addition, $z\geq u$. Then the only solution to \eqref{eqn:foc} is $y^* = z+ {(\lambda p)^{-\frac{1}{p-1}}}$, since both
$y^*<u$ and $y^*=z$ is cannot hold. 
Finally, suppose $z< u$. In this case $y^*_1=z$ and $y^*_2 =z + (\lambda p)^{\frac{-1}{p-1}}$ \textit{both} satisfy \eqref{eqn:foc}. To find the supremum, we compare the function evaluation at each of
these values. Note that $g_\lambda(y_1^*) =0$ and 
 \begin{align*}
      g_{\lambda}(y^*_2) &= (z-u+(\lambda p)^{\frac{-1}{p-1}})^+ - \lambda(\lambda p)^{\frac{-p}{p-1}}\\
      & = z -\left(u - \frac{1-1/p}{(\lambda p)^{\frac{1}{p-1}}}\right) = (z-\vartheta(u,\lambda)) > 0
 \end{align*}
Therefore, $g(y^*) = (z-\vartheta(u,\lambda))$ when $z>\vartheta(u,\lambda)$.

\noindent  \textbf{Case 3:} Finally, suppose that $z =\vartheta(u,\lambda)$. Then, one can check that both $y^*_1 =z$ and $y_2^* = z+(\lambda p)^\frac{-1}{p-1}$ are optimal since $g(y^*_1) = g(y_2^*) = 0$.\\

Putting the above three cases together, $g(y^*) = (z-\vartheta(u,\lambda))^+$ 
\[
E_{P_0}\left[\sup_y\left\{(y-u)^+ -\lambda |y-Z|^p\right\}\right] = E_{P_0}\left[Z-\vartheta(u,\lambda)\right]^+.
\]
Substituting the above expression into \eqref{eqn:wass_duality},
\begin{align}\label{eqn:cvar_wass_expansion}
    C_{1-\beta}(\mathcal W_{p,\delta}) &= \inf_u\left\{u+ \beta^{-1}\inf_{\lambda}\left\{\lambda\delta^p + E_{P_0}[Z-\vartheta(u,\lambda)]^+ \right\} \right\}\nonumber\\
    & = \inf_{u,\lambda}\left\{u+ \beta^{-1}\lambda\delta^p + \beta^{-1} E_{P_0}[Z-\vartheta(u,\lambda)]^+\right\},
\end{align}
From the previous arguments, \begin{enumerate}
    \item[(i)]  $y^* = z$ when $z<\vartheta(u,\lambda)$ and
    \item[(ii)] $y^* =z +(\lambda p)^{-1/(p-1)}$ when $z\geq \vartheta(u,\lambda)$.
\end{enumerate}
Putting  the above two together $y^* = z \mv{1}\{z<\vartheta(u,\lambda)\} +(z +(\lambda p)^{-1/(p-1)}) \mv{1}\{z>\vartheta(u,\lambda)\}$.

\noindent \textbf{Part - II:} Suppose now $p=1$. Here, the inner optimization problem becomes 
\[
\sup_{y}\left\{(y-u)^+ - \lambda |y-z|\right\} =\sup_{y}g_\lambda(y).
\]
First, suppose that $\lambda<1$. Then for every large enough $y$, $(y-u)^+ -|y-z| = (1-\lambda)y -(u+\lambda z)$, and so the supremum above is $+\infty$. Next, let $\lambda > 1 $. Then observe that a piece-wise linear analysis suggests that the only maximizer of $g_\lambda$ is $y^*=z$, since for $y\neq z$, one can verify that $g_\lambda(y) < g_\lambda(z)$. If instead $\lambda=1$, note that for any $c>0$, $y^* = z + c$ is a maximizer so long as $z>u$, else the above expression gets maximized at $y^*=z$. Putting all of the previous observations together, for $\lambda\geq 1$, $g_\lambda(y^*) = [z-u]^+$. Substituting this information into \eqref{eqn:cvar_wass_expansion} and recalling that $\vartheta(u,\lambda)=u$ if $\lambda\geq 1$ and $-\infty$ otherwise,
\[
\sup_{y}\left\{(y-u)^+ - \lambda |z-y|\right\} = [z-\vartheta(u,\lambda)]^+.
\]
Repeating the steps leading up to \eqref{eqn:cvar_wass_expansion} completes the proof.\qed

\noindent \textbf{Proof of Lemma~\ref{lem:set_of_optimisers}: }
Suppose that $p>1$. Differentiate \eqref{eqn:cvar_wass_expansion} and set derivatives to $0$ to obtain that the optimal $(u^*,\lambda^*)$:
\begin{align*}
    \frac{\partial \varphi_\beta(u,\lambda)}{\partial u} = 0 &\implies \beta = \bar F_{P_0}(\vartheta(u^*,\lambda^*)), \\
    \frac{\partial \varphi_\beta(u,\lambda)}{\partial \lambda} = 0&\implies \beta^{-1} \delta^p = \beta^{-1}\bar F_{P_0}(\vartheta(u^*,\lambda^*))\left(\frac{1}{\lambda^* p}\right)^{p/(p-1)}.
\end{align*}
The first of the above equations implies that $\vartheta(u^*,\lambda^*) = v_{1-\beta}(P_0)$.  Substitute the value of $\bar F_{P_0}(\vartheta(u^*,\lambda^*) = v_{1-\beta}(P_0)) =\beta$ into the second equation to obtain $(\lambda^*p)^{-1/(p-1)} = \delta/\beta^{1/p}$.

Next, suppose that $p=1$. Here, 
note that $\nu(u,\lambda) = u$ if $\lambda\geq 1$ and equals $-\infty$ otherwise. Thus $E_{P_0}[Z-\vartheta(u,\lambda)]^+ = \infty$ whenever $\lambda<1$. Since $\varphi_\beta(u,\lambda)$ is monotone in $\lambda$ for every $u$, $\lambda^*=1$  and $u^* =v_{1-\beta}(P_0)$.\qed

\noindent \textbf{Proof of Lemma~\ref{lem:wc_upper}: }
By definition, for any $P\in \mathcal P$ and $t\in(0,1]$, $v_{1-t}(P)\leq v_{1-t}(\mathcal P)$. Therefore,
\[
\rho_{1-\beta}(P) \leq \int_{0}^1 w(t) v_{1-\beta t}(\mathcal P) dt  \ \ \ \ \ \text{ for all } P\in\mathcal P.
\]
Since the right hand side above is a constant, taking the supremum over all $P\in \mathcal P$ yields $\rho_{1-\beta}(\mathcal P) \leq \int_0^1 w(t) v_{1-\beta t}(\mathcal P) dt$.\qed

\noindent \textbf{Proof of Lemmas~\ref{lem:RVs_ht} and \ref{lem:RVs_lt}: }Note that the tail cdf of $Q_\beta$ is given by 
\[
\bar F_{Q_\beta} (x) = \bar F_Q(x) \mv{1}(x\leq v_{1-\beta_0}(x)) + G_\beta(x) \mv{1}(x> v_{1-\beta_0}(x)).
\]
Fix $t<\beta_0$ above, and note that by construction, $v_{1-t}(Q_\beta) = G_\beta^{\leftarrow}(t)$. With $\log \beta  \sim \theta \log\beta$ for $0<\theta<1$, one has that 
\[
G_{\beta_0}(x) = \begin{cases}
    \beta_0 \left(\frac{x}{v_{1-\beta_0}(Q)}\right)^{-\gamma} \quad \quad\quad\quad\  \quad\text{ if $\bar F_Q\in \RV(\gamma)$}\\
\exp\left(\log(\beta_0)\left(\frac{x}{v_{1-\beta_0}(Q)}\right)^\gamma\right)      \quad\text{ if } \Lambda_Q\in \RV(\gamma)
    \end{cases}  
\]
Taking the inverse of the above expression completes the proof.
\qed 

\noindent\textbf{Proof of Lemma~\ref{lem:ucc_quantiles}: } Let $U_1,\ldots$ be uniform random variables. For each $n$, define the empirical count process
\[
N_n(x) := \sum_{i=1}^n \mathbf 1_{\{U_i \le x\}}, \qquad x\in[0,1].
\]
Then $N_n(x)\sim\mathrm{Bin}(n,x)$ and for each $k\in\{1,\dots,n\}$,
\[
\{U_{k:n} \le x\} = \{N_n(x)\ge k\}, 
\qquad
\{U_{k:n} \ge x\} = \{N_n(x)\le k-1\}.
\]

\medskip\noindent
\textbf{Step 1: Exponential tail bounds for each fixed $(n,k)$.}
Fix $0<\eta\le 1/2$, $n\in\mathbb N$ and $k\in\{1,\dots,n\}$. Set
\[
x_+ := (1+\eta)\frac{k}{n},
\qquad
x_- := (1-\eta)\frac{k}{n}.
\]
Note that $x_\pm\in(0,1)$ for $k\le n$ and $\eta\le 1/2$. We have
$\{U_{k:n} \ge x_+\}
= \{N_n(x_+) \le k-1\},$
with
$\mu_+ := \mathbb E[N_n(x_+)] = n x_+ = (1+\eta)k$.
Choose 
$\delta_+ := \frac{\eta}{2(1+\eta)} \in (0,1).$
Then
\[
(1-\delta_+)\mu_+
= (1-\tfrac{\eta}{2(1+\eta)})(1+\eta)k
= (1+\tfrac{\eta}{2})k
\ge k
\ge k-1.
\]
Hence $
\{N_n(x_+) \le k-1\}
\subseteq
\{N_n(x_+) \le (1-\delta_+)\mu_+\}.$ 
By the 
Chernoff bound for a binomial random variable $Y$ with mean $\mu$,
\[
\mathbb P(Y \le (1-\delta)\mu) \le \exp\!\Big(-\frac{\delta^2}{2}\,\mu\Big)
\qquad (0<\delta<1),
\]
we obtain
\[
\mathbb P\bigl(U_{k:n} \ge (1+\eta)\tfrac{k}{n}\bigr)
\le \exp\!\Big(-\frac{\delta_+^2}{2}\,\mu_+\Big)
= \exp\!\Big(-\frac{\eta^2}{8(1+\eta)}\,k\Big)
\le \exp(-c_1\eta^2 k),
\]
for some absolute constant $c_1>0$.
A similar Chernoff bound yields
\[
\mathbb P(Y \ge (1+\delta)\mu) \le \exp\!\Big(-\frac{\delta^2}{3}\,\mu\Big),
\qquad 0<\delta\le 1.
\]
Then
\[
\mathbb P\bigl(U_{k:n} \le (1-\eta)\tfrac{k}{n}\bigr)
\le \exp\!\Big(-\frac{\delta_-^2}{3}\,\mu_-\Big)
= \exp\!\Big(-\frac{\eta^2}{12(1-\eta)}\,k\Big)
\le \exp(-c_2\eta^2 k),
\]
for some absolute constant $c_2>0$. 
Combining both tails, there exists an absolute constant $c>0$ such that for all $n$, all $k\in\{1,\dots,n\}$ and all $0<\eta\le 1/2$,
\begin{equation}\label{eqn:BCL_prelim}
\mathbb P\!\left(
\left|U_{k:n} - \frac{k}{n}\right|
> \eta\,\frac{k}{n}
\right)
\le 2\exp(-c\eta^2 k).    
\end{equation}

\medskip\noindent
\textbf{Step 2: Apply Borel-Cantelli.}
Fix $\eta\in(0,1/2]$. For the intermediate index set
$I_n := \{k\in\mathbb N:\ n^\varepsilon \le k \le n^{1-\varepsilon}\},$
we apply a union bound using \eqref{eqn:BCL_prelim}:
\[
\begin{aligned}
\mathbb P\Bigg(
\sup_{k\in I_n}
\left|
\frac{U_{k:n}}{k/n} - 1
\right| > \eta
\Bigg)
&= \mathbb P\Bigg(
\exists\,k\in I_n:\ \left|U_{k:n} - \frac{k}{n}\right| > \eta\frac{k}{n}
\Bigg)\\
&\leq \sum_{k\in I_n}
\mathbb P\!\left(
\left|U_{k:n} - \frac{k}{n}\right| > \eta\frac{k}{n}
\right)\\
&\leq \sum_{k\in I_n} 2\exp(-c\eta^2 k)\leq 2\,|I_n|\,\exp(-c\eta^2 n^\varepsilon)\\
&\leq 2n\,\exp(-c\eta^2 n^\varepsilon).
\end{aligned}
\]
Since $n^\varepsilon/\log n\to\infty$, the series
$\sum_{n=1}^\infty 2n\,\exp(-c\eta^2 n^\varepsilon)$
converges. Denote the random variable $A_{n} = \sup_{k\in I_n}|(n/k)U_{k:n} -1|$. Then, we have 
\[
\limsup_{n\to\infty}\mathbb P\left(\sup_{m\geq n}A_m > \eta\right) \leq \limsup_{n\to\infty}\sum_{m>n}\sum_{m=n}^\infty 2m\,\exp(-c\eta^2 m^\varepsilon) =0.
\]
As a consequence, \cite[Theorem 1, Section 2.10]{shiryaev1996probability} implies that $A_{n}\to0 $ almost surely, concluding the proof.\qed
\end{appendices}


\noindent\textbf{{Proof of Lemma~\ref{lema:sol_of_inverse_dd}: }}From Lemma~\ref{lem:intermidiate_var}, $v_{1-\beta_0}(\hat Q^{(n)}) = v_{1-\beta_0}(Q) (1+o(1))$, uniformly over $\theta\in \Theta$
on a set $\Omega_0$ whose probability is $1$.  Further, since $\sup_{\theta\in \Theta}\beta_0 = \sup_{\theta\in \Theta} n^{-\theta}\to 0$ as $n\to\infty$, $\inf_{\theta\in \Theta} v_{1-\beta_0}(Q) \to\infty$.
Consequently, with probability 1, $\inf_{\theta\in \Theta
}v_{1-\beta_0}(\hat Q^{(n)}) \to \infty$.
Applying the proof of Proposition~\ref{prop:sol_of_inverse_unif} point-wise on $\Omega_0$ shows that for every $\omega\in \Omega_0$,
\begin{equation}\label{eqn:dd_uc}
    \sup_{\substack{x\geq v_{1-\beta_0}(\hat Q^{(n)})\\\theta\in \Theta,\delta\in \Delta}}\left\vert \frac{\bar F_{{\tt wc},n}(x) }{\phi^\leftarrow\left(\frac{\delta}{\bar F_{\hat Q_\beta^{(n)}}(x)}\right)\bar F_{\hat Q_\beta^{(n)}}(x)}  -1\right\vert = o(1),
\end{equation}
which establishes the conclusion of the lemma. \qed

\noindent \textbf{Proof of Lemma~\ref{lem:stochastic_sufficient}: } Throughout this proof, we let $\beta_0 = \beta_0(n,\theta) = n^{-\theta}$ denote the intermediate level and $\beta \in \mathcal B_n$ to be the target level. 

\noindent \textbf{Case i) $\bar F_Q\in \RV$:} \textbf{i) Lower Bound:} Let  $\Omega_1$ be the set on which $\hat \gamma_n(k_{n,\theta}) \to \gamma$ uniformly over $\theta \in \Theta$. 
Note that for $\omega\in \Omega_1$,  
\[
v_{1-t}(\hat Q_{\beta}^{(n,\theta)}) = \left(\frac{t}{\beta_0}\right)^{1/\gamma+r_{n,1}(\theta)}  v_{1-\beta_0}(Q) (1+r_{n,2}(\theta))
\]
where the the uniform consistency of $\hat \gamma(k_n),$ $\sup_{\theta\in \Theta}|\max \{r_{n,1}(\theta),r_{n,2}(\theta)\}| \to 0$ almost surely. 
For any $\beta(\theta) = n^{-\theta}$ and $\beta(n) \in \mathcal B_n$, observe that $\beta_0 \geq  \beta^{M/\varepsilon}$.
Apply Potter's bounds to $v_{1-\beta_0}(Q)$ as in the proof of Lemma~\ref{lem:lb_suff}:
\[
\beta^{\varsigma M /\varepsilon} \leq \ell(1/\beta_0) \leq \beta^{-\varsigma M/\varepsilon} \quad \quad  \forall (\beta,\theta) \in \mathcal B(n) \times \Theta
\]
whenever $n$ is large enough ($\beta_0$ is small enough). Proceeding as in the proof of Lemma~\ref{lem:lb_suff} shows that for every $\omega\in \Omega_1$ (which has a probability $1$),
\[
\inf_{\substack{t\in(0,\beta]\\\theta\in \Theta,\beta\in \mathcal B_n}} \frac{\log v_{1-t}(\hat Q_\beta^{(n,\theta)})}{\log v_{1-t}(Q)} \geq  \frac{1-\gamma M\varsigma/\varepsilon}{1+\gamma \varsigma}.
\]
Since $M<\infty$, and $\varsigma >0$ was arbitrary,  conclude that with probability 1, 
\[
\liminf_{n\to\infty} \inf_{\substack{t\in(0,\beta]\\\theta\in \Theta,\beta\in \mathcal B(n)}} \frac{\log v_{1-t}(\hat Q_\beta^{(n,\theta)})}{\log v_{1-t}(Q)}  \geq 1.
\]
\noindent \textbf{i) Upper Bound: } Set $\Omega^\prime = \Omega_0\cap \Omega_1$. 
Then, for every $\omega\in \Omega^\prime$,
\[
\bar F_{{\tt wc},n}(x) \sim \phi^\leftarrow\left(\frac{\delta}{\bar F_{\hat Q_\beta^{(n)}}(x)}\right)\bar F_{\hat Q_\beta^{(n)}}(x) 
\]
uniformly over $x\geq v_{1-\beta_0}(\hat Q^{(n)})$, $(\theta,\delta)\in \Theta\times \Delta$ in the sense given by Lemma~\ref{lema:sol_of_inverse_dd}. Set $x_t = v_{1-t}(\Dphi(\hat Q_{\beta}^{(n)}))$ and note that whenever $t\in(0,\beta]$, $v_{1-t}(\hat Q^{(n)}) \geq  v_{1-\beta_0}(\hat Q^{(n)})  =   v_{1-\beta_0}(\hat Q^{(n)}_\beta) $.  Thus, $x_t\geq v_{1-\beta_0}(\hat Q^{(n)}_\beta)  $. Then, uniformly over $t\in(0,\beta]$, $(\theta,\delta)\in \Theta\times\Delta$, 
\[
\frac{1}{t }\sim  w_\delta\left(\frac{1}{\bar F_{\hat Q_{\beta}^{(n)}}(x_t)}\right), \text{ where } w_\delta(x) = \frac{1}{x \phi^\leftarrow(\delta/x) }.
\]
With $\hat \gamma_n= \gamma+o(1)$ on $\Omega^\prime$, re-arranging the above and substituting the value of $\bar F_{\hat Q_{\beta}^{(n)}}$ yields that 
\begin{equation}\label{eqn:crit_ub_h}
    w^{\leftarrow}\left(\frac{1}{t}\right) \sim \beta_0^{-1}\left(\frac{v_{1-\beta_0}(Q)}{x_t}\right)^{-\gamma+o(1)} \implies x_t \sim \frac{v_{1-\beta_0}(Q)}{\beta_0^{-1/\gamma+o(1)}} [w^\leftarrow(1/t)]^{1/\gamma+o(1)} = t^{-1/\gamma+o(1)}.
\end{equation}
Finally, with $\beta_0 = n^{-\theta}$ for $\theta\in \Theta$, $\beta\in \mathcal B_n$ and $t\in (0,\beta]$, one can further write $\beta_0^{o(1)} = t^{o(1)}$, uniformly over $(t,\theta)$. Then,
\[
\sup_{\substack{t\in (0,\beta],\beta\in \mathcal B_n\\(\theta,\delta)\in \Theta \times \Delta}}\frac{\log v_{1-t}(\Dphi(\hat Q_\beta^{(n)}))}{\log v_{1-t}(Q)} = 1+o(1). 
\]
This completes the proof in the case where $\bar F_Q\in \RV$. 

\noindent \textbf{Case ii) $\Lambda_Q\in \RV$: }\textbf{i) Lower Bound:}  On the set $\Omega_0$, $v_{1-\beta_0}(\hat Q_\beta^{(n)}) = v_{1-\beta_0}(Q_\beta)(1+r_{n,3}(\theta))$, where by Lemma~\ref{lem:consistency_of_our_estimates}, $r_{n,3}(\theta)\to 0$ uniformly over $\theta\in \Theta$. Therefore, repeating calculations from the proof  Lemma~\ref{lem:lb_suff}(ii), for any $q>1$, uniformly over $t\in[\beta^q,\beta]$
\begin{align}
     \frac{v_{1-t}(\hat Q_\beta^{(n,\theta)})}{v_{1-t}(Q)}
     & = \frac{v_{1-\beta_0}(Q)}{v_{1-t}(Q)} \left(\frac{\log(1/t)}{\log(1/\beta_0)}\right)^{1/\gamma+r_{n,4}(\theta)}(1+r_{n,3}(\theta))\label{eqn:key_expr}
\end{align}
Since $\log(1/\beta_0) \leq \log (1/t)\leq (qM/\varepsilon)\log(1/\beta_0)$ whenever $\beta = n^{-\theta}$ for $\theta\in \Theta$, $\beta \in \mathcal B_n$, and $t\in (\beta^q,\beta]$, note that on $\omega\in \Omega_0 \cap \Omega_1$ (still a probability 1 set), the right hand side of \eqref{eqn:key_expr} equals $1+\tilde r_n(t,\theta)$, where $\sup_{t\in (\beta^q,\beta], \theta\in \Theta}|\tilde r_n(t,\theta)|\to 0$, thereby establishing the lower bound. 

\noindent \textbf{ii) Upper Bound:} Following the proof of Case ii) from Theorem~\ref{thm:scale_preserving_dro}, observe that on the set $\Omega_0$ where the uniform convergence \eqref{eqn:dd_uc} occurs, one has that 
$\log(1/t) = \Lambda_{\hat Q_\beta^{(n)}}(x_t)(1+o(1))$ uniformly over $t\in(0,\beta]$, $(\theta,\delta)\in \Theta\times \Delta$ where recall that $x_t = v_{1-t}(\Dphi(\hat Q_\beta^{(n)}))$ is the worst case VaR. Then, proceeding as in that case, $v_{1-t}(\hat Q_{\beta}^{(n)}) \sim x_t$. Further on $\Omega_0$, from the previous part, $v_{1-t}(\hat Q^{(n)}_\beta) = v_{1-t}(Q_\beta) (1+o(1))$. Putting these two together, on $\Omega^\prime$, we get 
\begin{align*}
    \sup_{t\in[\beta^\kappa,\beta]}\frac{v_{1-t}(\Dphi( \hat Q_\beta^{(n)}))}{v_{1-t}(Q)} &\sim \sup_{t\in [\beta^\kappa,\beta]}\frac{\Lambda_{Q_{\beta}^{(n)}}^{\leftarrow}(\log(1/t))}{\Lambda^\leftarrow_Q(\log(1/t))}(1+o(1))  \\
    &\sim \sup_{t\in [\beta^\kappa,\beta]} \frac{\Lambda_{Q_\beta}(\log(1/t))}{\Lambda_Q^\leftarrow(\log 1/t)} \ \ \  \text{ since } v_{1-t}(\hat Q^{(n)}_\beta) = v_{1-t}(Q_\beta)\\
    &=(1+o(1))
\end{align*}
where the last bound follows from the proof of Theorem~\ref{thm:scale_preserving_dro}. The second statement follows because $v_{1-\beta}(Q) = [\log(1/\beta)]^{1/\gamma+o(1)}$ and on the set $\Omega^\prime$, with $\hat \gamma = \gamma +o(1)$ 
\[
v_{1-t}(\hat Q_{\beta}^{(n)}) = v_{1-\beta_0}\left(\frac{\log(1/t)}{\log(1/\beta_0)}\right)^{1/\gamma+o(1)}.
\]
Noting that the above bound is uniform over $t\in (0,\beta]$, taking logs and simplifying
\[
\sup_{t\in (0,\beta^\kappa]} \frac{\log v_{1-t}(\hat Q_\beta^{(n)})}{\log v_{1-t}(Q)} = 1+o(1)
\]
thereby completing the proof.\qed
\end{document}